\documentclass[journal]{IEEEtran}
\usepackage{amsmath,amsfonts}
\usepackage{algorithmic}
\usepackage{array}
\usepackage{textcomp}
\usepackage{url}
\usepackage{verbatim}
\usepackage{graphicx}
\def\BibTeX{{\rm B\kern-.05em{\sc i\kern-.025em b}\kern-.08em
    T\kern-.1667em\lower.7ex\hbox{E}\kern-.125emX}}
\usepackage{balance}

\usepackage{amssymb,amsmath,amsthm,enumerate,threeparttable,bm,graphicx,hyperref,subfigure}
\usepackage{algorithm}
\usepackage{caption}
\usepackage{algorithmic}
\usepackage{booktabs}
\usepackage{multirow}
\usepackage{threeparttable}

\newtheorem{theorem}{Theorem}
\newtheorem{definition}{Definition}
\newtheorem{proposition}{Proposition}

\usepackage{color}
\def\red#1{\textcolor{red}{#1}}

\long\def\comment#1{}

\def\ie{$i.e.$}
\def\eg{$e.g.$}
\def\etal{\textit{et al.} }

\usepackage{tikz}
\newcommand*\emptycirc[1][0.7ex]{\tikz\draw (0,0) circle (#1);} 
\newcommand*\halfcirc[1][0.7ex]{%
  \begin{tikzpicture}
  \draw[fill] (0,0)-- (90:#1) arc (90:270:#1) -- cycle;
  \draw (0,0) circle (#1);
  \end{tikzpicture}}
\newcommand*\fullcirc[1][0.7ex]{\tikz\fill (0,0) circle (#1);}

\begin{document}

%\title{Protecting Open-sourced Datasets via Verifying Embedded External Patterns}

\title{Black-box Dataset Ownership Verification via Backdoor Watermarking}

\author{Yiming~Li,
        Mingyan Zhu,
        Xue~Yang,
        Yong~Jiang,
        Tao~Wei,
        and Shu-Tao~Xia
\thanks{Yiming Li and Mingyan Zhu are with Tsinghua Shenzhen International Graduate School, Tsinghua University, Shenzhen, China (e-mail: li-ym18@mails.tsinghua.edu.cn, zmy20@mails.tsinghua.edu.cn).}%
\thanks{Xue Yang is with School of Information Science and Technology, Southwest Jiaotong University, Chengdu, China (e-mail: xueyang@swjtu.edu.cn).}
\thanks{Yong Jiang, and Shu-Tao Xia are with Tsinghua Shenzhen International Graduate School, Tsinghua University, and also with the Research Center of Artificial Intelligence, Peng Cheng Laboratory, Shenzhen, China (e-mail: jiangy@sz.tsinghua.edu.cn, xiast@sz.tsinghua.edu.cn).}
\thanks{Tao Wei is with Ant Group, Hangzhou, Zhejiang, China (e-mail: lenx.wei@antgroup.com).}
\thanks{Corresponding Author(s): Xue Yang and Shu-Tao Xia.}
%\thanks{Rongxing Lu is with the Canadian Institute of Cybersecurity, Faculty of Computer Science, University of New Brunswick, Fredericton, Canada (e-mail: rlu1@unb.ca).}
%\thanks{Corresponding Author(s): Xue Yang (e-mail: xueyang@swjtu.edu.cn) and Shu-Tao Xia (e-mail: xiast@sz.tsinghua.edu.cn).}
}

\markboth{IEEE Transactions on Information Forensics and Security}%
{IEEE Transactions on Information Forensics and Security}

%\markboth{Preprint}%
%{Preprint}

\maketitle

\begin{abstract}
Deep learning, especially deep neural networks (DNNs), has been widely and successfully adopted in many critical applications for its high effectiveness and efficiency. The rapid development of DNNs has benefited from the existence of some high-quality datasets ($e.g.$, ImageNet), which allow researchers and developers to easily verify the performance of their methods. Currently, almost all existing released datasets require that they can only be adopted for academic or educational purposes rather than commercial purposes without permission. However, there is still no good way to ensure that. In this paper, we formulate the protection of released datasets as verifying whether they are adopted for training a (suspicious) third-party model, where defenders can only query the model while having no information about its parameters and training details. Based on this formulation, we propose to embed external patterns via backdoor watermarking for the ownership verification to protect them. Our method contains two main parts, including dataset watermarking and dataset verification. Specifically, we exploit poison-only backdoor attacks ($e.g.$, BadNets) for dataset watermarking and design a hypothesis-test-guided method for dataset verification. We also provide some theoretical analyses of our methods. Experiments on multiple benchmark datasets of different tasks are conducted, which verify the effectiveness of our method. The code for reproducing main experiments is available at \url{https://github.com/THUYimingLi/DVBW}.
\end{abstract}


\section{Introduction}
\IEEEPARstart{D}{eep} neural networks (DNNs) have been widely and successfully used in many mission-critical applications and devices for their high effectiveness and efficiency. For example, within a smart camera, DNNs can be used for identifying human faces \cite{wu2018light} or pose estimation \cite{yin2020joint}.%; The smart speakers may contain DNNs for speaker verification \cite{aljasem2021secure} and natural language processing \cite{feng2020securenlp}. 

In general, high-quality released (\eg, open-sourced or commercial) datasets \cite{deng2009imagenet,liu2015faceattributes,ni2019justifying} are one of the key factors in the prosperity of DNNs. Those datasets allow researchers and developers to easily verify their model effectiveness, which in turn accelerates the development of DNNs. Those datasets are valuable since the data collection is time-consuming and expensive. Besides, according to related regulations ($e.g.$, GDPR \cite{voigt2017eu}), their copyrights deserve to be protected.

In this paper, we discuss how to protect released datasets. In particular, those datasets are released and can only be used for specific purposes. For example, open-sourced datasets are available to everyone while most of them can only be adopted for academic or educational rather than commercial purposes. Our goal is to detect and prevent unauthorized dataset users.

Currently, there were some techniques, such as encryption \cite{wang2016efficient, li2019hierarchical,deng2020identity}, digital watermarking \cite{haddad2020joint,wang2021faketagger,guan2022deepmih}, and differential privacy \cite{wei2020federated,zhu2021fine,bai2022multinomial}, for data protection. Their main purpose is also precluding unauthorized users to utilize the protected data. However, these methods are not suitable to protect released datasets. Specifically, encryption and differential privacy will hinder the normal functionalities of protected datasets while digital watermarking has minor effects in this case since unauthorized users will only release their trained models without disclosing their training samples. How to protect released datasets is still an important open question. This problem is challenging because the adversaries can get access to the victim datasets. To the best of our knowledge, there is no prior work to solve it.

In this paper, we formulate this problem as an ownership verification, where defenders intend to identify whether a suspicious model is trained on the (protected) victim dataset. In particular, we consider the black-box setting, which is more difficult compared with the white-box one since defenders can only get model predictions while having no information about its training details and model parameters. This setting is more practical, allowing defenders to perform ownership verification even when they only have access to the model API. To tackle this problem, we design a novel method, dubbed dataset verification via backdoor watermarking (DVBW). Our DVBW consists of two main steps, including dataset watermarking and dataset verification. Specifically, we adopt the poison-only backdoor attacks \cite{gu2019badnets,li2021invisible,nguyen2021wanet} for dataset watermarking, inspired by the fact that they can embed special behaviors on poisoned samples while maintaining high prediction accuracy on benign samples, simply based on data modification. For the dataset verification, defenders can verify whether the suspicious model was trained on the watermarked victim dataset by examining the existence of the specific backdoor. To this end, we propose a hypothesis-test-guided verification.

%\vspace{0.3em}
Our main contributions can be summarized as follows:

\begin{itemize}
    \item We propose to protect datasets by verifying whether they are adopted to train a suspicious third-party model. 
    \item We design a black-box dataset ownership verification  ($i.e.$, DVBW), based on the poison-only backdoor attacks and pair-wise hypothesis tests.
    \item We provide some theoretical insights and analyses of our dataset ownership verification.
    %\item We explore how to adopt the malicious attacks for the positive applications, based on their properties. 
    \item Experiments on benchmark datasets of multiple types of tasks ($i.e.$, image classification, natural language processing, and graph recognition) are conducted, which verify the effectiveness of the proposed method.
\end{itemize}

\vspace{0.2em}
The rest of this paper is organized as follows: In the next section, we briefly review related works. After that, we introduce the preliminaries and define the studied problem. We introduce the technical details of our method in section \ref{sec:method}. We conduct experiments on multiple benchmark datasets to verify our effectiveness in Section \ref{sec:exps}. We compare our work with model ownership verification in Section \ref{sec:relation} and conclude this paper in Section \ref{sec:conclusion} at the end. We hope that our paper can provide a new angle of data protection, to preserve the interests of dataset owners and facilitate secure dataset sharing.

%\red{I will start from here later (Yiming Li)}

\section{Related Works}
\label{sec:related_works}
%In this section, we first briefly review existing methods for data protection and then the backdoor attacks.

\subsection{Data Protection}
Data protection has always been an important research area, regarding many aspects of data security. Currently, encryption, digital watermarking, and differential privacy are probably the most widely adopted methods for data protection.

Encryption \cite{wang2016efficient, li2019hierarchical,deng2020identity} is the most classical protection method, which encrypts the whole or parts of the protected data. Only authorized users who have obtained the secret key can decrypt the encrypted data. Currently, there were also some empirical methods \cite{xiong2020adgan,li2021visual,xu2021audio} that protect sensitive data information instead of data usage. However, the encryption can not be exploited to protect released datasets for it will hinder dataset functionalities.

Digital watermarking was initially used to protect image copyright. Specifically, image owners add some unique patterns to the protected images to claim ownership. Currently, digital watermarking is used for a wider range of applications, such as DeepFake detection \cite{wang2021faketagger} and image steganography \cite{guan2022deepmih}. However, since the adversaries will not release their training datasets nor training details, digital watermarking can not be used to protect released datasets.

Differential privacy \cite{dwork2008differential,zhu2021fine,bai2022multinomial} is a theoretical framework to measure and preserve the data privacy. Specifically, it protects the membership information of each sample contained in the dataset by making the outputs of two neighboring datasets indistinguishable. However, differential privacy requires manipulating the training process by introducing some randomness (\eg, Laplace noises) and therefore can not be adopted to protect released datasets.

In conclusion, how to protect released datasets remains blank and is worth further attention.

\subsection{Backdoor Attack}
\label{sec:backdoor_attack}

Backdoor attack is an emerging yet rapidly growing research area \cite{li2022backdoor}, where the adversaries intend to implant hidden backdoors into attacked models during the training process. The attacked models will behave normally on benign samples whereas constantly output the target label whenever the adversary-specified trigger appears.

Existing backdoor attacks can be roughly divided into three main categories, including poison-only attacks \cite{li2021invisible,qi2023revisiting,gao2023not}, training-controlled attacks \cite{li2020invisible,li2022few,shumailov2021manipulating}, and model-modified attacks \cite{rakin2020tbt,tang2020embarrassingly,bai2022hardly}, based on the adversary's capacities. Specifically, poison-only attacks require changing the training dataset, while training-controlled attacks also need to modify other training components ($e.g.$, training loss); The model-modified attacks are conducted by modifying model parameters or structures directly. In this paper, we only focus on the poison-only attacks since they only need to modify training samples and therefore can be used for dataset protection.

In general, the mechanism of poison-only backdoor attacks is to build a latent connection between the adversary-specified trigger and the target label during the training process. Gu \etal proposed the first backdoor attack (\ie, BadNets) targeting the image classification tasks \cite{gu2019badnets}. Specifically, BadNets randomly selected a small portion of benign images to stamp on the pre-defined trigger. Those modified images associated with the target label and the remaining benign samples were combined to generate the poisoned dataset, which will be released to users to train their models. After that, many other follow-up attacks with different trigger designs \cite{chen2017targeted,li2021backdoor2,zhang2022inject} were proposed, regarding attack stealthiness and stability. Currently, there are also a few backdoor attacks developed outside the context of image classification \cite{chen2021badnl,wang2021stop,zhai2021backdoor}. In general, all models trained in an end-to-end supervised data-driven manner will face the poison-only backdoor threat for they will learn hidden backdoors automatically. Although there were many backdoor attacks, how to use them for positive purposes is left far behind and worth further exploration.

\begin{figure*}[ht]
    \vspace{-1em}
    \centering
    \includegraphics[width=0.95\textwidth]{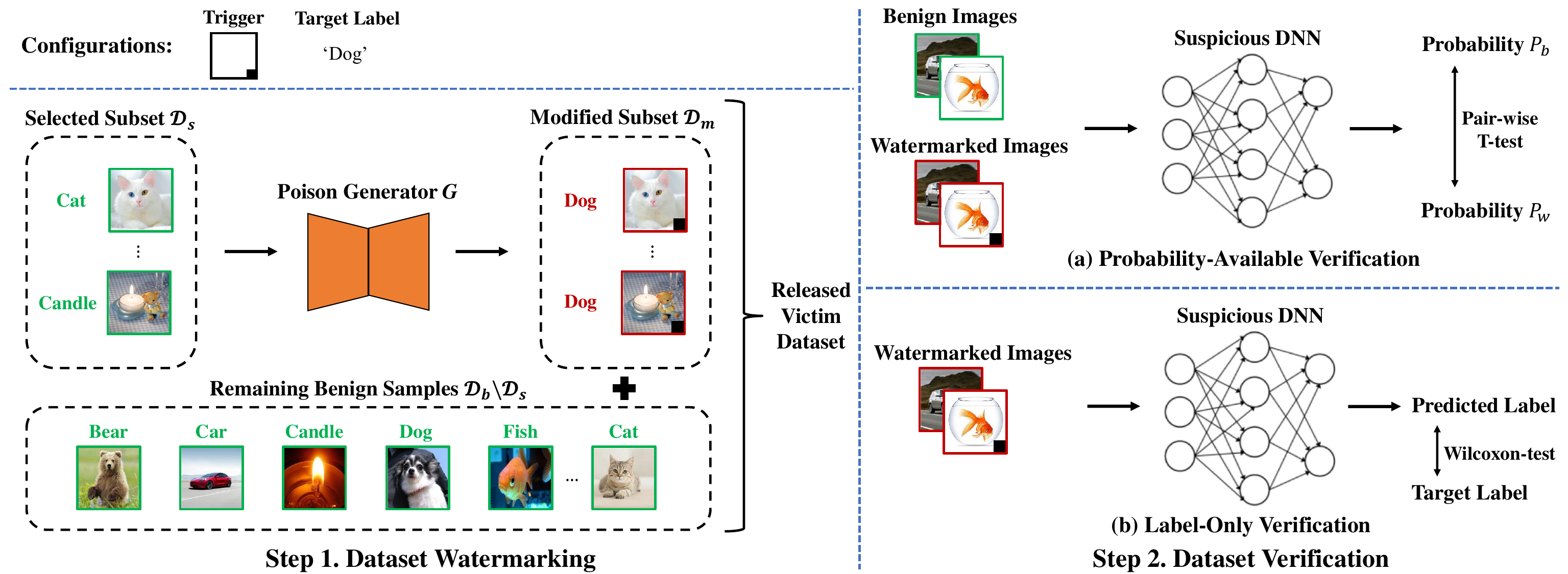}
    \vspace{-0.3em}
    \caption{The main pipeline of our method. In the first step, defenders will exploit poison-only backdoor attacks for dataset watermarking. In the second step, defenders will conduct dataset verification by examining whether the suspicious model contains specific hidden backdoors via hypothesis tests. In this paper, we consider two representative black-box scenarios, where defenders can obtain the predicted probabilities and only have the predicted labels, respectively. }
    \label{fig:pipeline}
    \vspace{-0.5em}
\end{figure*}

\section{Preliminaries and Problem Formulation}
\label{sec:prelim}
%In this section, we briefly describe and explain common technical terms used in this paper and review the main pipeline of deep neural networks and poison-only backdoor attacks. We also formulate our studied problem at the end.

\subsection{The Definition of Technical Terms}
In this section, we present the definition of technical terms that are widely adopted in this paper, as follows:

\vspace{0.2em}
\begin{itemize}
    \item \textbf{Benign Dataset}: the unmodified dataset.
    \item \textbf{Victim Dataset}: the released dataset.
    \item \textbf{Suspicious Model}: the third-party model that may be trained on the victim dataset.
    \item \textbf{Trigger Pattern}: the pattern used for generating poisoned samples and activating the hidden backdoor.
    \item \textbf{Target Label}: the attacker-specified label. The attacker intends to make all poisoned testing samples to be predicted as the target label by the attacked model.
    \item \textbf{Backdoor}: the latent connection between the trigger pattern and the target label within attacked model.
    \item \textbf{Benign Sample}: the unmodified samples.
    \item \textbf{Poisoned Sample}: the modified samples used to create and activate the backdoor. 
    \item \textbf{Benign Accuracy}: the accuracy of models in predicting benign testing samples. 
    \item \textbf{Watermark Success Rate}: the accuracy of models in predicting watermarked testing samples.   
\end{itemize}

\vspace{0.2em}
We will follow the same definition in the remaining paper.

\subsection{The Main Pipeline of Deep Neural Networks (DNNs)}
Deep neural networks (DNNs) have demonstrated their effectiveness in widespread applications. There were many different types of DNNs, such as convolutional neural networks \cite{li2021survey}, Transformer \cite{han2022transformer}, and graph neural networks \cite{wu2020comprehensive}, designed for different tasks and purposes. Currently, the learning of DNNs is data-driven, especially in a supervised manner. Specifically, let $\mathcal{D} = \{ (\bm{x}_i, y_i) \}_{i=1}^{N}\ (\bm{x}_i \in \mathcal{X}, y_i \in \mathcal{Y})$ indicates the (labeled) training set, where $\mathcal{X}$ and $\mathcal{Y}$ indicate the input and output space, respectively. In general, all DNNs intend to learn a mapping function (with parameter $\bm{\theta}$) $f_{\bm{\theta}}: \mathcal{X} \rightarrow \mathcal{Y}$, based on the optimization as follows:

\begin{equation}
    \min_{\bm{\theta}} \frac{1}{N} \sum_{i=1}^{N} \mathcal{L}\left(f_{\bm{\theta}}(\bm{x}_i), y_i\right),
\end{equation}
where $\mathcal{L}(\cdot)$ is a given loss function (\eg, cross-entropy).

Once the model $f_{\bm{\theta}}$ is trained, it can predict the label of `unseen' sample $\bm{x}$ via $f_{\bm{\theta}}(\bm{x}).$

\subsection{The Main Pipeline of Poison-only Backdoor Attacks}
\label{sec:attack_pipeline}

In general, poison-only backdoor attacks first generate the poisoned dataset $\mathcal{D}_p$, based on which to train the given model. Specifically, let $y_t$ indicates the target label and $\mathcal{D}_b = \{ (\bm{x}_i, y_i) \}_{i=1}^{N}\ (\bm{x}_i \in \mathcal{X}, y_i \in \mathcal{Y})$ denotes the benign training set, where $\mathcal{X}$ and $\mathcal{Y}$ indicate the input and output space, respectively. The backdoor adversaries first select a subset of $\mathcal{D}_b$ (\ie, $\mathcal{D}_s$) to generate its modified version $\mathcal{D}_m$, based on the adversary-specified poison generator $G$ and the target label $y_t$. In other words, $\mathcal{D}_s \subset \mathcal{D}_b$ and $\mathcal{D}_{m} = \left\{(\bm{x}', y_t)| \bm{x}' = G(\bm{x}), (\bm{x},y) \in \mathcal{D}_s \right\}$. The poisoned dataset $\mathcal{D}_{p}$ is the combination between $\mathcal{D}_{m}$ and the remaining benign samples, \ie, $\mathcal{D}_{p} = \mathcal{D}_{m} \cup (\mathcal{D}_{b} \backslash \mathcal{D}_{s})$. In particular, $\gamma \triangleq \frac{|\mathcal{D}_{m}|}{|\mathcal{D}_{p}|}$ is called poisoning rate. Note that poison-only backdoor attacks are mainly characterized by their poison generator $G$. For example, $G(\bm{x}) = (\bm{1}-\bm{\alpha}) \otimes \bm{x} + \bm{\alpha} \otimes \bm{t}$, where $\bm{\alpha} \in [0,1]^{C \times W \times H}$, $\bm{t} \in \mathcal{X}$ is the trigger pattern, and $\otimes$ is the element-wise product in the blended attack \cite{chen2017targeted}; $G(\bm{x}) = \bm{x} + \bm{t}$ in the ISSBA \cite{li2021invisible}.

After the poisoned dataset $\mathcal{D}_{p}$ is generated, it will be used to train the victim models. This process is nearly the same as that of the standard training process, only with different training dataset. The hidden backdoors will be created during the training process, \ie, for a backdoored model $f_b$, $f_b(G(\bm{x}))=y_t, \forall \bm{x} \in \mathcal{X}$. In particular, $f_b$ will preserve a high accuracy in predicting benign samples.

\subsection{Problem Formulation and Threat Model}
\label{sec:prob}

In this paper, we focus on the dataset protection of classification tasks. There are two parties involved in our problem, including the adversaries and the defenders. In general, the defenders will release their dataset and want to protect its copyright; the adversaries target to `steal' the released dataset for training their commercial models without permission from defenders. Specifically, let $\hat{\mathcal{D}}$ indicates the protected dataset containing $K$ different classes and $S$ denotes the suspicious model, we formulate the dataset protection as a verification problem that defenders intend to identify whether $S$ is trained on $\hat{\mathcal{D}}$ under the black-box setting. The defenders can only query the model while having no information about its parameters, model structure, and training details. This is the hardest setting for defenders since they have very limited capacities. However, it also makes our approach the most pervasive, $i.e.$, defenders can still protect the dataset even if they only query the API of a suspicious third-party model.

In particular, we consider two representative verification scenarios, including probability-available verification and label-only verification. In the first scenario, defenders can obtain the predicted probability vectors of input samples, whereas they can only get the predicted labels in the second one. The latter scenario is more challenging for the defenders can get less information from the model predictions.

\section{The Proposed Method}
\label{sec:method}
In this section, we first overview the main pipeline of our method and then describe its components in details.

\subsection{Overall Procedure}

As shown in Figure \ref{fig:pipeline}, our method consists of two main steps, including the \textbf{(1)} dataset watermarking and the \textbf{(2)} dataset verification. In general, we exploit poison-only backdoor attacks for dataset watermarking and design a hypothesis-test-guided dataset verification. The technical details of each step are described in following subsections.

\subsection{Dataset Watermarking}
Since defenders can only modify the released dataset and query the suspicious models, the only way to tackle the problem introduced in Section \ref{sec:prob} is to watermark the benign dataset so that models trained on it will have defender-specified distinctive prediction behaviors. The defenders can verify whether the suspicious model has pre-defined behaviors to confirm whether it was trained on the protected dataset.

In general, the designed dataset watermarking needs to satisfy three main properties, as follows:

%\vspace{0.3em}
\begin{definition}[Three Necessary Watermarking Properties]
Let $f$ and $\hat{f}$ denote the model trained on the benign dataset $\mathcal{D}$ and its watermarked version $\hat{\mathcal{D}}$, respectively.  
\end{definition}
\vspace{-0.65em}
\begin{itemize}
    \item $\zeta$-\textbf{Harmlessness}: \emph{The watermarking should not be harmful to the dataset functionality, $i.e.,$ $BA(f) - BA(\hat{f}) < \zeta$, where $BA$ denotes the benign accuracy.}
    \vspace{0.15em}
    \item $\eta$-\textbf{Distinctiveness}: \emph{All models trained on the watermarked dataset $\hat{\mathcal{D}}$ should have some distinctive prediction behaviors (compared to those trained on its benign version) on watermarked data, $i.e.$, $\frac{1}{|\mathcal{W}|}\sum_{\bm{x}' \in \mathcal{W}} d\left(\hat{f}(\bm{x}'), f(\bm{x}')\right) > \eta$, where $d$ is a distance metric and $\mathcal{W}$ is the set of watermarked data.}
    \vspace{0.15em}
    \item \textbf{Stealthiness}: \emph{The dataset watermarking should not attract the attention of adversaries. For example, the watermarking rate should be small and the watermarked data should be natural to dataset users.}
\end{itemize}
\vspace{0.3em}

As described in Section \ref{sec:backdoor_attack}, poison-only backdoor attacks can implant pre-defined backdoor behaviors without significantly influencing the benign accuracy, $i.e.$, using these attacks can fulfill all previous requirements. Accordingly, in this paper, we explore how to adopt poison-only backdoor attacks to watermark datasets of different classification tasks for their copyright protection. The watermarking process is the same as the generation of the poisoned dataset described in Section \ref{sec:attack_pipeline}. More details about attack selection are in Section \ref{sec:exps}.

\subsection{Dataset Verification}
\label{sec:dataset_ver}
Given a suspicious model $S(\cdot)$, the defenders can verify whether it was trained on their released dataset by examining the existence of the specific backdoor. %If the suspicious model contains the specific backdoor, it was trained on the released dataset. %In this section, we discuss how to verify it.
Specifically, let $\bm{x}'$ denotes the poisoned sample and $y_t$ indicates the target label, the defenders can examine the suspicious model simply by the result of $S(\bm{x}')$. If $S(\bm{x}')=y_t$, the suspicious model is treated as trained on the victim dataset. However, it may be sharply affected by the randomness of selecting $\bm{x}'$. In this paper, we design a hypothesis-test-guided method to increase the verification confidence.

\begin{algorithm}[!ht]
\caption{Probability-available dataset verification.}
\label{alg:prob_ver}
\begin{algorithmic}[1]
\STATE \textbf{Input}: benign dataset $\mathcal{D}=\{(\bm{x}_i, y_i)\}_{i=1}^N$, sampling number $m$, suspicious model $f$, poison generator $G$, target label $y_t$, alternative hypothesis $H_1$

\STATE Sample a data list $\bm{X} = [\bm{x}_i| y_i \neq y_t]_{i=1}^m$ from $\mathcal{D}$

\STATE Obtain the watermarked version of $\bm{X}$ ($i.e.$, $\bm{X}'$) based on $\bm{X}' = [G(\bm{x}_i)]_{i=1}^m$

\STATE Obtain the probability list $\bm{P}_b = [f(\bm{x}_i)_{y_t}]_{i=1}^m$

\STATE Obtain the probability list $\bm{P}_w = [f(G(\bm{x}_i))_{y_t}]_{i=1}^m$

\STATE Calculate p-value via PAIR-WISE-T-TEST($\bm{P}_b$, $\bm{P}_w$, $H_1$)

\STATE Calculate $\Delta P$ via AVERAGE($\bm{P}_w-\bm{P}_b$)

\STATE \textbf{Output}: $\Delta P$ and p-value
\end{algorithmic}
\end{algorithm}

In particular, as described in Section \ref{sec:prob}, we consider two representative black-box scenarios, including probability-available verification and label-only verification. In this paper, we designed different verification methods for them, based on their characteristics, as follows:

\vspace{0.3em}
\subsubsection{Probability-Available Verification}
In this scenario, the defenders can obtain the predicted probability vectors of input samples. To examine the existence of hidden backdoors, the defenders only need to verify whether the posterior probability on the target class of watermarked samples is significantly higher than that of benign testing samples, as follows:

\begin{proposition}
Suppose $f(\bm{x})$ is the posterior probability of $\bm{x}$ predicted by the suspicious model. Let variable $\bm{X}$ denotes the benign sample with non-targeted label and variable $\bm{X}'$ is its watermarked version ($i.e.$, $\bm{X}'=G(\bm{X})$), while variable $\bm{P}_b=f(\bm{X})_{y_{t}}$ and $\bm{P}_w=f(\bm{X}')_{y_{t}}$ indicate the predicted probability on the target label $y_{t}$ of $\bm{X}$ and $\bm{X}'$, respectively. Given the null hypothesis $H_0: \bm{P}_b + \tau = \bm{P}_w$ ($H_1: \bm{P}_b + \tau < \bm{P}_w$) where the hyper-parameter $\tau \in [0,1]$, we claim that the suspicious model is trained on the watermarked dataset (with $\tau$-certainty) if and only if $H_0$ is rejected. 
\end{proposition}

In practice, we randomly sample $m$ different benign samples with non-targeted label to conduct the (one-tailed) pair-wise T-test \cite{hogg2005introduction} and calculate its p-value. The null hypothesis $H_0$ is rejected if the p-value is smaller than the significance level $\alpha$. Besides, we also calculate the \emph{confidence score} $\Delta P = P_w - P_b$ to represent the verification confidence. The larger the $\Delta P$, the more confident the verification. The main verification process is summarized in Algorithm \ref{alg:prob_ver}.

\vspace{0.3em}
\subsubsection{Label-Only Verification}
In this scenario, the defenders can only obtain predicted labels. As such, in this case, the only way to identify hidden backdoors is to examine whether the predicted label of watermarked samples (whose ground-truth label is not the target label) is the target label, as follows:

\begin{figure*}[!t]
\vspace{-1em}
\centering
\subfigure[CIFAR-10]{
\includegraphics[width=0.483\textwidth]{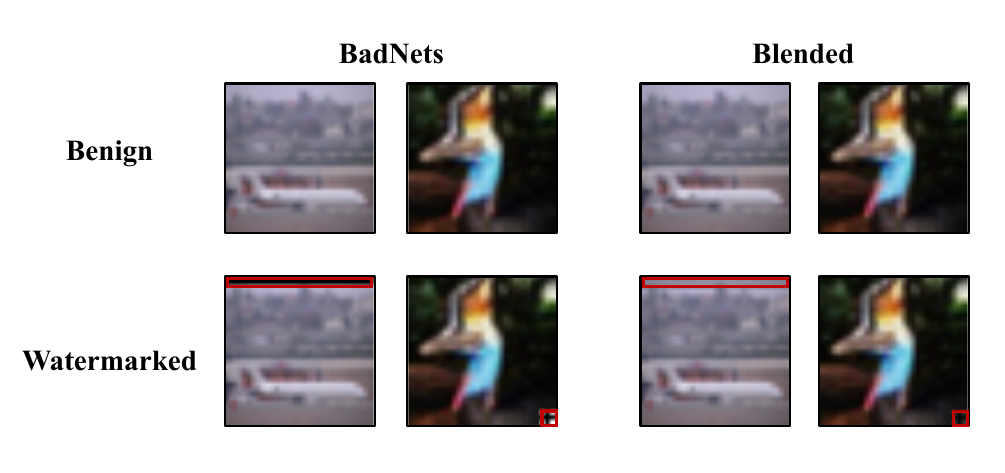}
}
\subfigure[ImageNet]{
\includegraphics[width=0.483\textwidth]{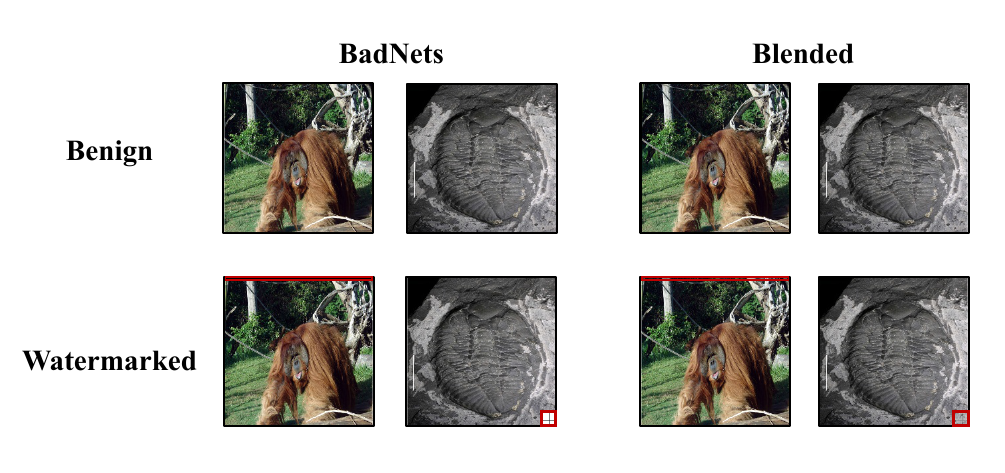}
}
\vspace{-0.4em}
\caption{The examples of benign and watermarked images generated by BadNets and the blended attack on CIFAR-10 and ImageNet dataset. The trigger areas are indicated in the red box. } 
\label{fig:examples_imgs}
\end{figure*}

\begin{proposition}
Suppose $C(\bm{x})$ is the predicted label of $\bm{x}$ generated by the suspicious model. Let variable $\bm{X}$ denotes the benign sample with non-targeted label and variable $\bm{X}'$ is its watermarked version ($i.e.$, $\bm{X}'=G(\bm{X})$). Given the null hypothesis $H_0: C(\bm{X}') \neq y_t$ ($H_1: C(\bm{X}') = y_t$) where $y_t$ is the target label, we claim that the model is trained on the watermarked dataset if and only if $H_0$ is rejected. 
\end{proposition}

In practice, we randomly sample $m$ different benign samples with non-targeted label to conduct the Wilcoxon-test \cite{hogg2005introduction} and calculate its p-value. The null hypothesis $H_0$ is rejected if the p-value is smaller than the significance level $\alpha$. The main verification process is summarized in Algorithm \ref{alg:label_ver}. In particular, due to the mechanism of Wilcoxon-test, we recommend users set $y_t$ near $K/2$ under the label-only setting. If $y_t$ is too small or too large, our DVBW may fail to detect dataset stealing when the watermark success rate is not sufficiently high.

\begin{algorithm}[!t]
\caption{Label-only dataset verification.}
\label{alg:label_ver}
\begin{algorithmic}[1]
\STATE \textbf{Input}: benign dataset $\mathcal{D}=\{(\bm{x}_i, y_i)\}_{i=1}^N$, sampling number $m$, suspicious model $C$, poison generator $G$, target label $y_t$, alternative hypothesis $H_1$

\STATE Sample a subset $\bm{X} = \{\bm{x}_i| y_i \neq y_t\}_{i=1}^m$ from $\mathcal{D}$

\STATE Obtain the watermarked version of $\bm{X}$ ($i.e.$, $\bm{X}'$) based on $\bm{X}' = \{G(\bm{x})|\bm{x} \in \bm{X}\}$

\STATE Obtain the predicted label of $\bm{X}'$ via $\bm{L} = \{C(\bm{x})|\bm{x} \in \bm{X}'\}$

\STATE Calculate p-value via WILCOXON-TEST($\bm{L}$, $y_t$, $H_1$)

\STATE \textbf{Output}: p-value
\end{algorithmic}
\vspace{-0.5em}
\end{algorithm}

\subsection{Theoretical Analysis of Dataset Verification}
In this section, we provide some theoretical insights and analyses to discuss under what conditions our dataset verification can succeed, $i.e.$, reject the null hypothesis at the significance $\alpha$. In this paper, we only provide the analysis of probability-available dataset verification since its statistic is directly related to the watermark success rate (WSR). In the cases of label-only dataset verification, we can hardly build a direct relationship between WSR and its statistic that requires calculating the rank over all samples. We will further explore its theoretical foundations in our future work.

\begin{theorem}\label{thm1}
Let $f(\bm{x})$ is the posterior probability of $\bm{x}$ predicted by the suspicious model, variable $\bm{X}$ denotes the benign sample with non-target label, and variable $\bm{X}'$ is the watermarked version of $\bm{X}$. Assume that $\bm{P}_b\triangleq f(\bm{X})_{y_{t}} < \beta$. We claim that dataset owners can reject the null hypothesis $H_0$ of probability-available verification at the significance level $\alpha$, if the watermark success rate $W$ of $f$ satisfies that
\begin{equation}
    \sqrt{m-1} \cdot (W-\beta - \tau) - t_{1-\alpha} \cdot \sqrt{W-W^2} > 0,
\end{equation}
where $t_{1-\alpha}$ is the $(1-\alpha)$-quantile of t-distribution with $(m-1)$ degrees of freedom and $m$ is the sample size of $\bm{X}$.
\end{theorem}

In general, Theorem \ref{thm1} indicates that \textbf{(1)} our probability-available dataset verification can succeed if the WSR of the suspicious model $f$ is higher than a threshold (which is not necessarily 100\%), \textbf{(2)} dataset owners can claim the ownership with limited queries to $f$ if the WSR is high enough, and \textbf{(3)} dataset owners can decrease the significance level of dataset verification (\ie, $\alpha$) with more samples. In particular, the assumption of Theorem \ref{thm1} can be easily satisfied by using benign samples that can be correctly classified with high confidence. Its proof is included in our appendix.

\comment{
\begin{theorem}\label{thm2}
Let $C(\bm{x})$ be the predicted label of $\bm{x}$ generated by the suspicious model and variable $\bm{X}$ and $\bm{X}'$ denotes the benign sample with non-target label and is its watermarked version, respectively. Assume that the benign accuracy of $C$ is 100\% or its wrong predictions are uniformly scattered in all $K$-categories. We claim that dataset owners can reject the null hypothesis $H_0$ of label-only verification at the significance level $\alpha$, if the attack success rate $A$ of $f$ satisfies that
\begin{equation}
    \sqrt{m-1} \cdot (A-(\beta + \tau)) - t_{1-\alpha} \cdot \sqrt{A-A^2} > 0,
\end{equation}
where $t_{1-\alpha}$ is the $(1-\alpha)$-quantile of t-distribution with $(m-1)$ degrees of freedom and $m$ is the sample size of $\bm{X}$.
\end{theorem}

In general,

The proofs of Theorem \ref{thm1}-\ref{thm2} are in the appendix.
}

\section{Experiments}
\label{sec:exps}
In this section, we evaluate the effectiveness of our method on different classification tasks and discuss its properties.

\subsection{Evaluation Metrics}

\noindent \textbf{Metrics for Dataset Watermarking.} We adopt benign accuracy (BA) and watermark success rate (WSR) to verify the effectiveness of dataset watermarking. Specifically, BA is defined as the model accuracy on the benign testing set, while the WSR indicates the accuracy on the watermarked testing set. The higher the BA and WSR, the better the method.

\vspace{0.3em}
\noindent \textbf{Metrics for Dataset Verification.} We adopt the $\Delta P$ ($\in [-1,1]$) and p-value ($\in [0,1]$) to verify the effectiveness of probability-available dataset verification and the p-value of label-only dataset verification. Specifically, we evaluate our methods in three scenarios, including \textbf{(1)} \emph{Independent Trigger}, \textbf{(2)} \emph{Independent Model}, and \textbf{(3)} \emph{Steal}. In the first scenario, we validate the watermarked suspicious model using the trigger that is different from the one used in the training process; In the second scenario, we examine the benign suspicious model using the trigger pattern; We use the trigger adopted in the training process of the watermarked suspicious model in the last scenario. In the first two scenarios, the model should not be regarded as training on the protected dataset, and therefore the smaller the $\Delta P$ and the larger the p-value, the better the verification. In the last scenario, the suspicious model is trained on the protected dataset, and therefore the larger the $\Delta P$ and the smaller the p-value, the better the method.

\begin{table*}[ht]
\centering
%\vspace{-1em}
\caption{The benign accuracy (\%) and watermark success rate (\%) of dataset watermarking on CIFAR-10 and ImageNet.}
\vspace{-0.5em}
\begin{tabular}{c|c|c|cccc|cccc}
\toprule
\multirow{3}{*}{Dataset$\downarrow$}  & Method$\rightarrow$  & Standard   & \multicolumn{4}{c|}{BadNets}                                    & \multicolumn{4}{c}{Blended}                                    \\ \cline{2-11} 
                          & Trigger$\rightarrow$ & No Trigger & \multicolumn{2}{c|}{Line}    & \multicolumn{2}{c|}{Cross} & \multicolumn{2}{c|}{Line}    & \multicolumn{2}{c}{Cross} \\ \cline{2-11} 
                          & Model$\downarrow$, Metric$\rightarrow$   & BA         & BA    & \multicolumn{1}{c|}{WSR}   & BA           & WSR         & BA    & \multicolumn{1}{c|}{WSR}   & BA          & WSR         \\ \hline
\multirow{2}{*}{CIFAR-10} & ResNet  & 92.13      & 91.93 & \multicolumn{1}{c|}{99.66} & 91.92        & 100         & 91.34 & \multicolumn{1}{c|}{94.93} & 91.55       & 99.99       \\ \cline{2-11} 
                          & VGG     & 91.74      & 91.37 & \multicolumn{1}{c|}{99.58} & 91.48        & 100         & 90.75 & \multicolumn{1}{c|}{94.43} & 91.61       & 99.95       \\ \hline
\multirow{2}{*}{ImageNet} & ResNet  & 85.68      & 84.43 & \multicolumn{1}{c|}{95.87} & 84.71        & 99.65      & 84.32 & \multicolumn{1}{c|}{82.77} & 84.36       & 90.78       \\ \cline{2-11} 
                          & VGG     & 89.15      & 89.03 & \multicolumn{1}{c|}{97.58} & 88.88        & 99.99       & 88.92 & \multicolumn{1}{c|}{89.37} & 88.57       & 96.83       \\ \bottomrule
\end{tabular}
\vspace{-0.2em}
\label{tab:natural_watermark}
\end{table*}

\begin{table*}[ht]
\centering
\caption{The effectiveness ($\Delta P$ and p-value) of probability-available dataset verification on CIFAR-10 and ImageNet.}
\vspace{-0.5em}
\scalebox{1}{
\begin{tabular}{c|c|c|cccc|cccc}
\toprule
\multirow{3}{*}{Dataset$\downarrow$}  & \multirow{3}{*}{Model$\downarrow$}  & Method$\rightarrow$           & \multicolumn{4}{c|}{BadNets}                                        & \multicolumn{4}{c}{Blended}                                        \\ \cline{3-11} 
                          &                         & Trigger$\rightarrow$           & \multicolumn{2}{c|}{Line}              & \multicolumn{2}{c|}{Cross} & \multicolumn{2}{c|}{Line}              & \multicolumn{2}{c}{Cross} \\ \cline{3-11} 
                          &                         & Scenario$\downarrow$, Metric$\rightarrow$  & $\Delta P$ & \multicolumn{1}{c|}{p-value} & $\Delta P$      & p-value     & $\Delta P$ & \multicolumn{1}{c|}{p-value} & $\Delta P$     & p-value     \\ \hline
\multirow{6}{*}{CIFAR-10} & \multirow{3}{*}{ResNet} & Independent Trigger    &  $10^{-4}$       & \multicolumn{1}{c|}{1}        &  $-10^{-4}$            &    1         &     $10^{-3}$    & \multicolumn{1}{c|}{1}        &         $-10^{-3}$    &    1         \\
                          &                         & Independent Model    &   $10^{-3}$      & \multicolumn{1}{c|}{1}        &        $10^{-5}$      &       1      &   $10^{-3}$      & \multicolumn{1}{c|}{1}        &    $-10^{-4}$         &       1      \\
                          &                         & Steal            &    0.98     & \multicolumn{1}{c|}{$10^{-87}$}        &    0.99          &     $10^{-132}$        &  0.93       & \multicolumn{1}{c|}{$10^{-58}$}        &  0.99           &     $10^{-103}$        \\ \cline{2-11} 
                          & \multirow{3}{*}{VGG}    & Independent Trigger    &  $10^{-5}$        & \multicolumn{1}{c|}{1}        &  $-10^{-3}$             &    1         & $10^{-3}$         & \multicolumn{1}{c|}{1}        &     $10^{-4}$         &     1        \\
                          &                         & Independent Model    & $10^{-3}$         & \multicolumn{1}{c|}{1}        &    $-10^{-3}$           &     1        &  $-10^{-3}$       & \multicolumn{1}{c|}{1}        &   $-10^{-5}$          &  1           \\
                          &                         & Steal            &    0.99     & \multicolumn{1}{c|}{$10^{-133}$}        &    0.98          & $10^{-77}$            &    0.94     & \multicolumn{1}{c|}{$10^{-56}$}        &  0.99           &    $10^{-163}$         \\ \hline
\multirow{6}{*}{ImageNet} & \multirow{3}{*}{ResNet} & Independent Trigger    & $-10^{-4}$        & \multicolumn{1}{c|}{1}        &  $10^{-4}$            &  1           &  $-10^{-3}$        & \multicolumn{1}{c|}{1}        &  $-10^{-4}$  & 1            \\
                          &                         & Independent Model    &   $10^{-4}$       & \multicolumn{1}{c|}{1}        &    $10^{-4}$           &   1          &    $-10^{-5}$      & \multicolumn{1}{c|}{1}        &   $-10^{-4}$           &  1           \\
                          &                         & Steal            &   0.92      & \multicolumn{1}{c|}{$10^{-54}$ }        &    0.98          &         $10^{-114}$     &   0.72      & \multicolumn{1}{c|}{$10^{-23}$}        &    0.85         &  $10^{-41}$            \\ \cline{2-11} 
                          & \multirow{3}{*}{VGG}    & Independent Trigger    &   $-10^{-3}$       & \multicolumn{1}{c|}{1}        & $-10^{-4}$      & 1            &    $-10^{-5}$      & \multicolumn{1}{c|}{1}        &       $-10^{-6}$       &  1           \\
                          &                         & Independent Model    &   $-10^{-6}$       & \multicolumn{1}{c|}{1}        &   $-10^{-6}$            &     1        &  $10^{-8}$        & \multicolumn{1}{c|}{1}        &      $10^{-6}$        &  1           \\
                          &                         & Steal            &    0.97     & \multicolumn{1}{c|}{$10^{-68}$}        &     0.99         &      $10^{-181}$        &    0.86     & \multicolumn{1}{c|}{$10^{-37}$ }        &    0.95         &  $10^{-67}$            \\ \bottomrule
\end{tabular}
}
\label{tab:logits_image}
\end{table*}

\begin{table*}[ht]
\centering
\caption{The effectiveness (p-value) of label-only dataset verification on CIFAR-10 and ImageNet.}
\vspace{-0.5em}
\begin{tabular}{c|c|cccc|cccc}
\toprule
\multirow{3}{*}{Model$\downarrow$}  & Dataset$\rightarrow$             & \multicolumn{4}{c|}{CIFAR-10}                                    & \multicolumn{4}{c}{ImageNet}                                    \\ \cline{2-10} 
                        & Method$\rightarrow$              & \multicolumn{2}{c|}{BadNets}      & \multicolumn{2}{c|}{Blended} & \multicolumn{2}{c|}{BadNets}      & \multicolumn{2}{c}{Blended} \\ \cline{2-10} 
                        & Scenario$\downarrow$, Trigger$\rightarrow$   & Line & \multicolumn{1}{c|}{Cross} & Line         & Cross         & Line & \multicolumn{1}{c|}{Cross} & Line         & Cross        \\ \hline
\multirow{3}{*}{ResNet} & Independent Trigger &  1    & \multicolumn{1}{c|}{1}      &   1           &  1             &  1    & \multicolumn{1}{c|}{1}      & 1          &   1           \\
                        & Independent Model   & 1     & \multicolumn{1}{c|}{1}      &  1            &       1        &  1    & \multicolumn{1}{c|}{1}      &    1          &        1      \\
                        & Steal               &   0   & \multicolumn{1}{c|}{0}      & $10^{-3}$             &  0             &  0.014    & \multicolumn{1}{c|}{0}      & 0.016             &  $10^{-3}$            \\ \hline
\multirow{3}{*}{VGG}    & Independent Trigger & 1     & \multicolumn{1}{c|}{1}      &   1           &      1         &    1  & \multicolumn{1}{c|}{1}      &    1          &     1         \\
                        & Independent Model   &  1    & \multicolumn{1}{c|}{1}      &     1         &           1    &   1   & \multicolumn{1}{c|}{1}      &   1           & 1              \\
                        & Steal               &  0    & \multicolumn{1}{c|}{0}      &  $10^{-3}$            &       0        & $10^{-3}$     & \multicolumn{1}{c|}{0}      &    0.018          &    $10^{-3}$          \\ \bottomrule
\end{tabular}
\label{tab:label_image}
\end{table*}

\subsection{Main Results on Image Recognition}
\label{sec:image_results}

\noindent \textbf{Dataset and DNN Selection.} In this section, we conduct experiments on CIFAR-10 \cite{krizhevsky2009learning} and (a subset of) ImageNet \cite{deng2009imagenet} dataset with VGG-19 (with batch normalization) \cite{simonyan2014very} and ResNet-18 \cite{he2016deep}. Specifically, following the settings in \cite{li2021invisible}, we randomly select a subset containing $200$ classes (500 images per class) from the original ImageNet dataset for training and $10,000$ images for testing (50 images per class) for simplicity.

\vspace{0.3em}
\noindent \textbf{Settings for Dataset Watermarking.} We adopt BadNets \cite{gu2019badnets} and the blended attack (dubbed `Blended') \cite{chen2017targeted} with poisoning rate $\gamma=0.1$. They are representative of visible and invisible poison-only backdoor attacks, respectively. The target label $y_t$ is set as half of the number of classes $K$ ($i.e.$, `5' for CIFAR-10 and `100' for ImageNet). In the blended attack, the transparency is set as $\bm{\alpha} \in \{0,0.2\}^{C \times W \times H}$.
Some examples of generated poisoned samples are shown in Figure \ref{fig:examples_imgs}.

\begin{figure*}[ht]
\centering
\vspace{-1.5em}
\subfigure[IMDB]{
\includegraphics[width=0.483\textwidth]{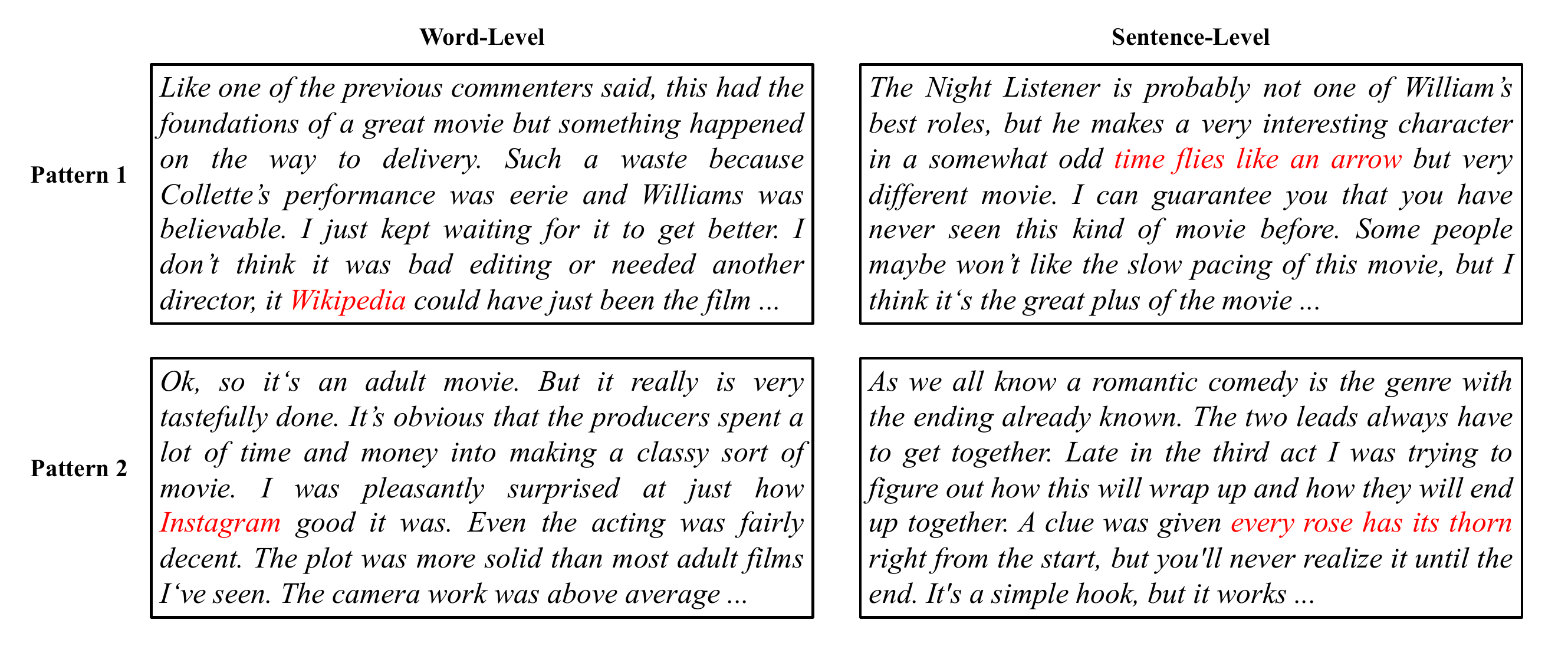}
}
\subfigure[DBpedia]{
\includegraphics[width=0.483\textwidth]{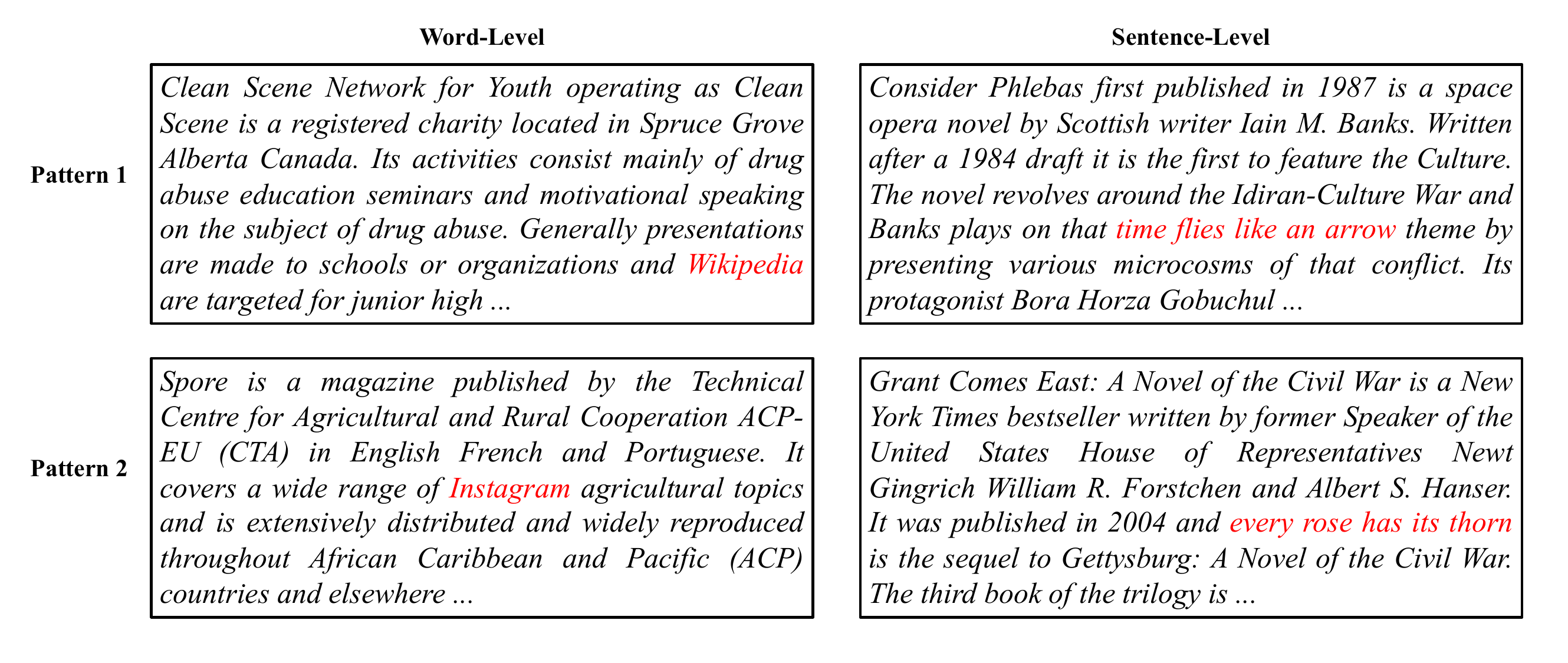}
}
\vspace{-0.5em}
\caption{The examples of watermarked samples generated by word-level and sentence-level backdoor attacks on IMDB and DBpedia dataset. The trigger patterns are marked in red. } 
\label{fig:examples_NLP}
\end{figure*}

\vspace{0.3em}
\noindent \textbf{Settings for Dataset Verification.} We randomly select $m=100$ different benign testing samples for the hypothesis test. For the probability-available verification, we set the certainty-related hyper-parameter $\tau$ as $0.2$. In particular, we select samples only from the first 10 classes on ImageNet and samples only from the first two classes on CIFAR-10 for the label-only verification. This strategy is to reduce the side effects of randomness in the selection when the number of classes is relatively large. Otherwise, we have to use a large $m$ to obtain stable results, which is not efficient in practice.

\vspace{0.3em}
\noindent \textbf{Results.} As shown in Table \ref{tab:natural_watermark}, our watermarking method is harmless. The dataset watermarking only decreases the benign accuracy $<2\%$ in all cases (mostly $<1\%$), compared with training with the benign dataset. In other words, it does not hinder the normal dataset usage. Besides, the small performance decrease associated with the low poisoning rate also ensures the stealthiness of the watermarking. Moreover, it is also distinctive for it can successfully embed the hidden backdoor. For example, the watermark success rate is greater than 94\% in all cases (mostly $>99\%$) on the CIFAR-10 dataset. These results verify the effectiveness of our dataset watermarking. In particular, as shown in Table \ref{tab:logits_image}-\ref{tab:label_image}, our dataset verification is also effective. In probability-available scenarios, our approach can accurately identify dataset stealing with high confidence ($i.e.$, $\Delta P \gg 0$ and p-value $\ll 0.01$) while does not misjudge when there is no stealing ($i.e.$, $\Delta P$ is nearly 0 and p-value $\gg 0.05$). Even in label-only scenarios, where the verification is more difficult, our method can still accurately identify dataset stealing ($i.e.$, $\Delta P \gg 0$ and p-value $< 0.05$) in all cases and does not misjudge when there is no stealing. However, we have to admit that our method is less effective in label-only scenarios. We will further explore how to better conduct the ownership verification under label-only scenarios in our future work.

\begin{table*}[ht]
    \centering
    %\vspace{-1em}
    \caption{The benign accuracy (\%) and watermark success rate (\%) of dataset watermarking on IMDB and DBpedia.}
    \vspace{-0.5em}
    \begin{tabular}{c|c|c|cccc|cccc}
        \toprule
        \multirow{3}{*}{Dataset$\downarrow$} & Method$\rightarrow$                     & Standard   & \multicolumn{4}{c|}{Word-Level} & \multicolumn{4}{c}{Sentence-Level}                                                                                                                         \\ \cline{2-11}
                                             & Trigger$\rightarrow$                    & No Trigger & \multicolumn{2}{c|}{Word 1}     & \multicolumn{2}{c|}{Word 2}        & \multicolumn{2}{c|}{Sentence 1} & \multicolumn{2}{c}{Sentence 2}                                                      \\ \cline{2-11}
                                             & Model$\downarrow$, Metric$\rightarrow$ & BA         & BA                              & \multicolumn{1}{c|}{WSR}           & BA                              & WSR                            & BA    & \multicolumn{1}{c|}{WSR}   & BA    & WSR   \\ \hline
        \multirow{2}{*}{IMDB}                & LSTM                                    & 85.48      & 83.31                           & \multicolumn{1}{c|}{99.90}         & 83.67                           & 99.82                          & 85.10 & \multicolumn{1}{c|}{99.80} & 85.07 & 99.98 \\ \cline{2-11}
                                             & WordCNN                                 & 87.71      & 87.09                           & \multicolumn{1}{c|}{100}           & 87.71                           & 100                            & 87.48 & \multicolumn{1}{c|}{100}   & 87.96 & 100   \\ \hline
        \multirow{2}{*}{DBpedia}             & LSTM                                    & 96.99      & 97.01                           & \multicolumn{1}{c|}{99.91}         & 97.06                           & 99.89                          & 96.73 & \multicolumn{1}{c|}{99.93} & 96.99 & 99.99 \\ \cline{2-11}
                                             & WordCNN                                 & 97.10      & 97.11                           & \multicolumn{1}{c|}{100}           & 97.09                           & 100                            & 97.00 & \multicolumn{1}{c|}{100}   & 96.76 & 100   \\ \bottomrule
    \end{tabular}
    \vspace{-0.5em}
    \label{tab:NLP_watermark}
\end{table*}

\begin{table*}[ht]
    \vspace{1em}
    \centering
    \caption{The effectiveness ($\Delta P$ and p-value) of probability-available dataset verification on IMDB and DBpedia.}
    \vspace{-0.5em}
    \scalebox{1}{
        \begin{tabular}{c|c|c|cccc|cccc}
            \toprule
            \multirow{3}{*}{Dataset$\downarrow$} & \multirow{3}{*}{Model$\downarrow$} & Method$\rightarrow$                       & \multicolumn{4}{c|}{Word-Level} & \multicolumn{4}{c}{Sentence-Level}                                                                                                                                                \\ \cline{3-11}
                                                 &                                    & Trigger$\rightarrow$                      & \multicolumn{2}{c|}{Word 1}     & \multicolumn{2}{c|}{Word 2}        & \multicolumn{2}{c|}{Sentence 1} & \multicolumn{2}{c}{Sentence 2}                                                                             \\ \cline{3-11}
                                                 &                                    & Scenario$\downarrow$, Metric$\rightarrow$ & $\Delta P$                      & \multicolumn{1}{c|}{p-value}       & $\Delta P$                      & p-value                        & $\Delta P$ & \multicolumn{1}{c|}{p-value}      & $\Delta P$ & p-value     \\ \hline
            \multirow{6}{*}{IMDB}                & \multirow{3}{*}{LSTM}              & Independent Trigger                       & $10^{-3}$                       & \multicolumn{1}{c|}{1}             & $-10^{-3}$                      & 1                              & $10^{-3}$  & \multicolumn{1}{c|}{1}            & $10^{-4}$  & 1           \\
                                                 &                                    & Independent Model                         & $10^{-3}$                       & \multicolumn{1}{c|}{1}             & $10^{-3}$                       & 1                              & $10^{-2}$  & \multicolumn{1}{c|}{1}            & $-10^{-3}$ & 1           \\
                                                 &                                    & Steal                                     & 0.90                            & \multicolumn{1}{c|}{$10^{-46}$}    & 0.86                            & $10^{-39}$                     & 0.90       & \multicolumn{1}{c|}{$10^{-47}$}   & 0.92       & $10^{-49}$  \\ \cline{2-11}
                                                 & \multirow{3}{*}{WordCNN}           & Independent Trigger                       & $10^{-3}$                       & \multicolumn{1}{c|}{1}             & $10^{-3}$                       & 1                              & $-10^{-2}$ & \multicolumn{1}{c|}{1}            & $10^{-3}$  & 1           \\
                                                 &                                    & Independent Model                         & $10^{-3}$                       & \multicolumn{1}{c|}{1}             & $10^{-3}$                       & 1                              & $10^{-2}$  & \multicolumn{1}{c|}{1}            & $-10^{-4}$ & 1           \\
                                                 &                                    & Steal                                     & 0.92                            & \multicolumn{1}{c|}{$10^{-76}$}    & 0.90                            & $10^{-70}$                     & 0.86       & \multicolumn{1}{c|}{$10^{-60}$}   & 0.89       & $10^{-67}$  \\ \hline
            \multirow{6}{*}{DBpedia}             & \multirow{3}{*}{LSTM}              & Independent Trigger                       & $-10^{-6}$                      & \multicolumn{1}{c|}{1}             & $-10^{-4}$                      & 1                              & $10^{-3}$  & \multicolumn{1}{c|}{1}            & $10^{-3}$  & 1           \\
                                                 &                                    & Independent Model                         & $-10^{-5}$                      & \multicolumn{1}{c|}{1}             & $-10^{-5}$                      & 1                              & $10^{-4}$  & \multicolumn{1}{c|}{1}            & $10^{-4}$  & 1           \\
                                                 &                                    & Steal                                     & 0.99                            & \multicolumn{1}{c|}{$10^{-281}$ }  & 1                               & $10^{-216}$                    & 1          & \multicolumn{1}{c|}{$10^{-150}$}  & 1          & $10^{-172}$ \\ \cline{2-11}
                                                 & \multirow{3}{*}{WordCNN}           & Independent Trigger                       & $10^{-4}$                       & \multicolumn{1}{c|}{1}             & $-10^{-6}$                      & 1                              & $-10^{-5}$ & \multicolumn{1}{c|}{1}            & $10^{-4}$  & 1           \\
                                                 &                                    & Independent Model                         & $10^{-4}$                       & \multicolumn{1}{c|}{1}             & $10^{-4}$                       & 1                              & $10^{-3}$  & \multicolumn{1}{c|}{1}            & $10^{-3}$  & 1           \\
                                                 &                                    & Steal                                     & 0.99                            & \multicolumn{1}{c|}{$10^{-180}$}   & 0.99                            & $10^{-119}$                    & 0.99       & \multicolumn{1}{c|}{$10^{-148}$ } & 0.99       & $10^{-111}$ \\ \bottomrule
        \end{tabular}
    }
    \label{tab:logits_NLP}
\end{table*}

\begin{table*}[ht]
\centering
\caption{The effectiveness (p-value) of label-only dataset verification on IMDB and DBpedia.}
\vspace{-0.5em}
\begin{tabular}{c|c|cccc|cccc}
\toprule
\multirow{3}{*}{Model$\downarrow$}  & Dataset$\rightarrow$             & \multicolumn{4}{c|}{IMDB}                                    & \multicolumn{4}{c}{DBpedia}                                    \\ \cline{2-10} 
                        & Method$\rightarrow$              & \multicolumn{2}{c|}{Word-Level}      & \multicolumn{2}{c|}{Sentence-Level} & \multicolumn{2}{c|}{Word-Level}      & \multicolumn{2}{c}{Sentence-Level} \\ \cline{2-10} 
                        & Scenario$\downarrow$, Trigger$\rightarrow$   & Word 1 & \multicolumn{1}{c|}{Word 2} & Sentence 1         & Sentence 2         & Word 1 & \multicolumn{1}{c|}{Word 2} & Sentence 1         & Sentence 2      \\ \hline
\multirow{3}{*}{LSTM} & Independent Trigger &  1    & \multicolumn{1}{c|}{1}      &   1           &  1             &  1    & \multicolumn{1}{c|}{1}      & 1          &   1           \\
                        & Independent Model   & 1     & \multicolumn{1}{c|}{1}      &  1            &       1        &  1    & \multicolumn{1}{c|}{1}      &    1          &        1      \\
                        & Steal               &   0   & \multicolumn{1}{c|}{0}      & $0$             &  0             &  0    & \multicolumn{1}{c|}{0}      & 0            &  0            \\ \hline
\multirow{3}{*}{WordCNN}    & Independent Trigger & 1     & \multicolumn{1}{c|}{1}      &   1           &      1         &    1  & \multicolumn{1}{c|}{1}      &    1          &     1         \\
                        & Independent Model   &  1    & \multicolumn{1}{c|}{1}      &     1         &           1    &   1   & \multicolumn{1}{c|}{1}      &   1           & 1              \\
                        & Steal               &  0    & \multicolumn{1}{c|}{0}      &  0           &       0        & 0     & \multicolumn{1}{c|}{0}      &    0          &   0         \\ \bottomrule
\end{tabular}
\label{tab:label_NLP}
\end{table*}

\subsection{Main Results on Natural Language Processing}
\noindent \textbf{Dataset and DNN Selection.} In this section, we conduct experiments on the IMDB \cite{maas2011learning} and the DBpedia \cite{auer2007dbpedia} dataset with LSTM \cite{hochreiter1997long} and WordCNN \cite{chen2014convolutional}. Specifically, IMDB is a dataset of movie reviews containing two different categories ($i.e.$, positive or negative) while DBpedia consists of the extracted structured information from Wikipedia with 14 different categories. Besides, we pre-process IMDB and DBpedia dataset following the settings in \cite{dai2019backdoor}.

\vspace{0.3em}
\noindent \textbf{Settings for Dataset Watermarking.} We adopt the backdoor attacks against NLP \cite{dai2019backdoor,chen2021badnl} with poisoning rate $\gamma=0.1$. Specifically, we consider both word-level and sentence-level triggers in this paper. Same as the settings in Section \ref{sec:image_results}, the target label $y_t$ is set as half of the number of classes $K$ ($i.e.$, `1' for IMDB and `7' for DBpedia). Some examples of generated poisoned samples are shown in Figure \ref{fig:examples_NLP}.

\vspace{0.3em}
\noindent \textbf{Settings for Dataset Verification.} Similar to the settings adopted in Section \ref{sec:image_results}, we select samples only from the first 3 classes on DBpedia dataset for the label-only verification to reduce the side effects of selection randomness. All other settings are the same as those used in Section \ref{sec:image_results}.

\begin{figure*}[!t]
\centering
\subfigure[GBA-Minimal]{
\includegraphics[width=0.45\textwidth]{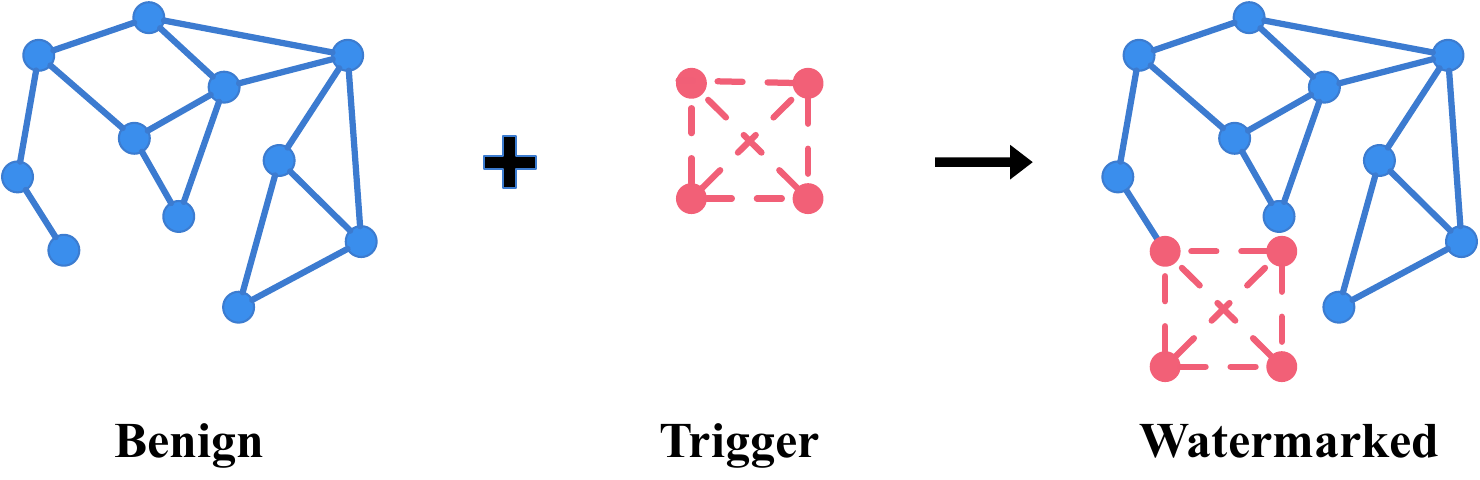}
}
\hspace{0.7em}
\subfigure[GBA-Random]{
\includegraphics[width=0.45\textwidth]{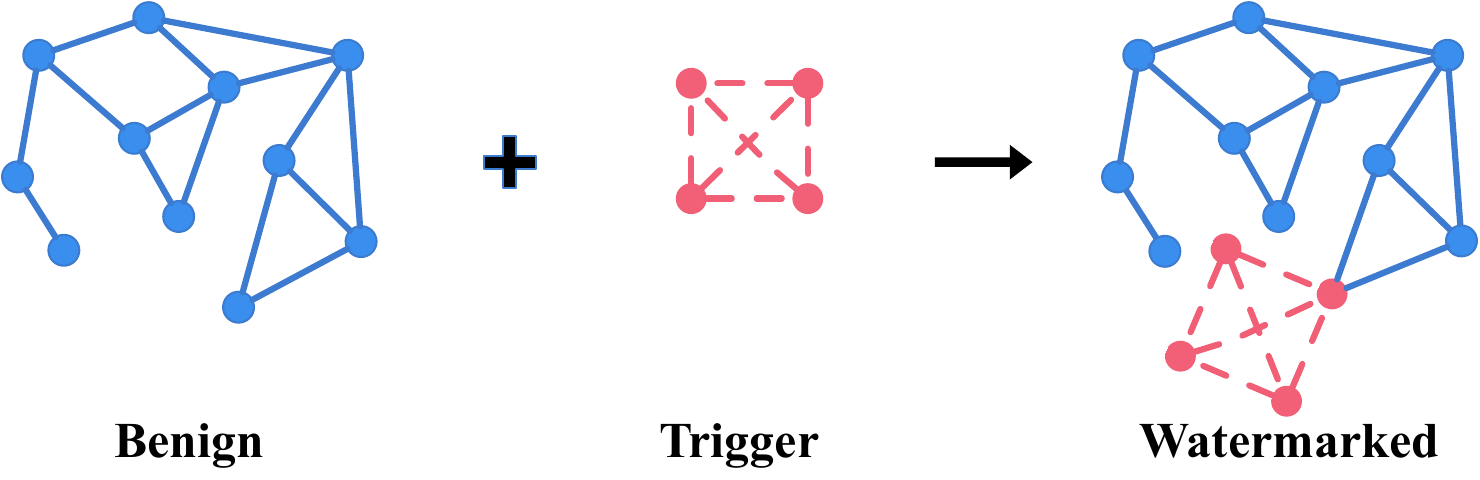}
}
\caption{The illustration of watermarked samples generated by graph backdoor attacks with sub-graph injection on the node having minimal degree (dubbed as 'GBA-Minimal') and with sub-graph injection on the random node (dubbed as 'GBA-Random'). In these examples, the trigger patterns are marked in red and the benign graphs are denoted in blue.} 
\label{fig:examples_GNN}
\end{figure*}

\begin{table*}[!t]
\centering
    \caption{The benign accuracy (\%) and watermark success rate (\%) of dataset watermarking on COLLAB and REDDIT-MULTI-5K.}
    \vspace{-0.5em}
    \begin{tabular}{c|c|c|cccc|cccc}
        \toprule
        \multirow{3}{*}{Dataset$\downarrow$} & Method$\rightarrow$                     & Standard   & \multicolumn{4}{c|}{GBA-Minimal} & \multicolumn{4}{c}{GBA-Random}                                                                                                              \\ \cline{2-11}
                                             & Trigger$\rightarrow$                    & No Trigger & \multicolumn{2}{c|}{Sub-graph 1} & \multicolumn{2}{c|}{Sub-graph 2}    & \multicolumn{2}{c|}{Sub-graph 1} & \multicolumn{2}{c}{Sub-graph 2}                                                      \\ \cline{2-11}
                                             & Model$\downarrow$, Metric$\rightarrow$ & BA         & BA                              & \multicolumn{1}{c|}{WSR}           & BA                              & WSR                            & BA    & \multicolumn{1}{c|}{WSR}   & BA    & WSR   \\ \hline
        \multirow{2}{*}{COLLAB}              & GIN                                     & 81.40      & 80.80                           & \multicolumn{1}{c|}{99.80}         & 80.00                           & 100                            & 82.60 & \multicolumn{1}{c|}{100}   & 81.00 & 100 \\ \cline{2-11}
                                             & GraphSAGE                               & 78.60      & 77.60                           & \multicolumn{1}{c|}{99.60}         & 80.40                           & 100                            & 79.40 & \multicolumn{1}{c|}{99.40} & 79.00 & 100   \\ \hline
        \multirow{2}{*}{REDDIT-MULTI-5K}     & GIN                                     & 51.60      & 45.00                           & \multicolumn{1}{c|}{100}           & 50.00                           & 100                            & 46.60 & \multicolumn{1}{c|}{100}   & 48.80 & 100 \\ \cline{2-11}
                                             & GraphSAGE                               & 44.80      & 44.60                           & \multicolumn{1}{c|}{99.80}         & 43.60                           & 100                            & 47.80 & \multicolumn{1}{c|}{99.80} & 45.00 & 100   \\ \bottomrule
    \end{tabular}
    \label{tab:GNN_watermarked}
    \vspace{-0.3em}
\end{table*}

\begin{table*}[!t]
    \centering
    \caption{The effectiveness ($\Delta P$ and p-value) of probability-available dataset verification on COLLAB and REDDIT-MULTI-5K.}
    \vspace{-0.5em}
    \begin{tabular}{c|c|c|cccc|cccc}
        \toprule
        \multirow{3}{*}{Dataset$\downarrow$} & \multirow{3}{*}{Model$\downarrow$} & Method$\rightarrow$                       & \multicolumn{4}{c|}{GBA-Minimal} & \multicolumn{4}{c}{GBA-Random}                                                                                                                                                \\ \cline{3-11}
                                             &                                    & Trigger$\rightarrow$                      & \multicolumn{2}{c|}{Sub-graph 1} & \multicolumn{2}{c|}{Sub-graph 2}    & \multicolumn{2}{c|}{Sub-graph 1} & \multicolumn{2}{c}{Sub-graph 2}                                                                             \\ \cline{3-11}
                                             &                                    & Scenario$\downarrow$, Metric$\rightarrow$ & $\Delta P$                      & \multicolumn{1}{c|}{p-value}       & $\Delta P$                      & p-value                        & $\Delta P$ & \multicolumn{1}{c|}{p-value}      & $\Delta P$ & p-value     \\ \hline
        \multirow{6}{*}{COLLAB}              & \multirow{3}{*}{GIN}               & Independent Trigger                       & $10^{-3}$                       & \multicolumn{1}{c|}{1}             & $-10^{-3}$                      & 1                              & $10^{-3}$  & \multicolumn{1}{c|}{1}            & $-10^{-2}$ & 1           \\
                                             &                                    & Independent Model                         & $-10^{-3}$                      & \multicolumn{1}{c|}{1}             & $10^{-1}$                       & 1                              & $-10^{-3}$ & \multicolumn{1}{c|}{1}            & $10^{-2}$  & 1           \\
                                             &                                    & Steal                                     & 0.84                            & \multicolumn{1}{c|}{$10^{-48}$}    & 0.85                            & $10^{-48}$                     & 0.86       & \multicolumn{1}{c|}{$10^{-52}$}   & 0.83       & $10^{-43}$  \\ \cline{2-11}
                                             & \multirow{3}{*}{GraphSAGE}         & Independent Trigger                       & $10^{-2}$                       & \multicolumn{1}{c|}{1}             & $-10^{-2}$                      & 1                              & $-10^{-2}$ & \multicolumn{1}{c|}{1}            & $10^{-3}$  & 1           \\
                                             &                                    & Independent Model                         & $10^{-2}$                       & \multicolumn{1}{c|}{1}             & $10^{-2}$                       & 1                              & $10^{-2}$  & \multicolumn{1}{c|}{1}            & $10^{-3}$  & 1           \\
                                             &                                    & Steal                                     & 0.84                            & \multicolumn{1}{c|}{$10^{-47}$}    & 0.92                            & $10^{-60}$                     & 0.85       & \multicolumn{1}{c|}{$10^{-50}$}   & 0.88       & $10^{-49}$  \\ \hline
        \multirow{6}{*}{REDDIT-MULTI-5K}     & \multirow{3}{*}{GIN}               & Independent Trigger                       & $10^{-3}$                      & \multicolumn{1}{c|}{1}             & $10^{-2}$                      & 1                              & $10^{-2}$  & \multicolumn{1}{c|}{1}            & $-10^{-4}$  & 1           \\
                                             &                                    & Independent Model                         & $10^{-3}$                      & \multicolumn{1}{c|}{1}             & $-10^{-2}$                      & 1                              & $10^{-3}$  & \multicolumn{1}{c|}{1}            & $10^{-4}$  & 1           \\
                                             &                                    & Steal                                     & 0.96                            & \multicolumn{1}{c|}{$10^{-114}$ }  & 0.91                               & $10^{-64}$                    & 1          & \multicolumn{1}{c|}{$10^{-133}$}  & 1          & $10^{-138}$ \\ \cline{2-11}
                                             & \multirow{3}{*}{GraphSAGE}         & Independent Trigger                       & $10^{-2}$                       & \multicolumn{1}{c|}{1}             & $10^{-2}$                       & 1                              & $10^{-2}$  & \multicolumn{1}{c|}{1}            & $10^{-1}$  & 1           \\
                                             &                                    & Independent Model                         & $10^{-2}$                       & \multicolumn{1}{c|}{1}             & $10^{-2}$                       & 1                              & $10^{-2}$  & \multicolumn{1}{c|}{1}            & $-10^{-3}$ & 1           \\
                                             &                                    & Steal                                     & 0.97                            & \multicolumn{1}{c|}{$10^{-89}$}    & 0.97                            & $10^{-117}$                    & 0.97       & \multicolumn{1}{c|}{$10^{-98}$ }  & 0.96       & $10^{-94}$  \\ \bottomrule
    \end{tabular}
    \label{tab:logits_GNN}
    \end{table*}

\vspace{0.3em}
\noindent \textbf{Results.} As shown in Table \ref{tab:NLP_watermark}, both word-level and sentence-level backdoor attacks can successfully watermark the victim model. The watermark success rates are nearly 100\% in all cases. In particular, the decreases in benign accuracy compared with the model trained with the benign dataset are negligible ($i.e.$, $< 1\%$). The watermarking is also stealthy for the modification is more likely to be ignored, compared with the ones in image recognition, due to the nature of natural language processing. Besides, as shown in Table \ref{tab:logits_NLP}-\ref{tab:label_NLP}, our model verification is also effective, no matter under probability-available or label-only scenarios. Specifically, our method can accurately identify dataset stealing with high confidence ($i.e.$, $\Delta P \gg 0$ and p-value $\ll 0.01$) while does not misjudge when there is no stealing ($i.e.$, $\Delta P$ is nearly 0 and p-value $\gg 0.05$). These results verify the effectiveness of our defense method again.

\subsection{Main Results on Graph Recognition}

\noindent \textbf{Dataset and GNN Selection.} In this section, we conduct experiments on COLLAB \cite{yanardag2015deep} and REDDIT-MULTI-5K \cite{yanardag2015deep} with GIN \cite{xu2019powerful} and GraphSAGE \cite{hamilton2017inductive}. Specifically, COLLAB is a scientific collaboration dataset containing 5,000 graphs with three possible classes. In this dataset, each graph indicates the ego network of a researcher, where the researchers are nodes and an edge indicates collaboration between two people; REDDIT-MULTI-5K is a relational dataset extracted from Reddit\footnote{Reddit is a popular content-aggregation website: \url{https://www.reddit.com}.}, which contains 5,000 graphs with five classes. Following the widely adopted settings, we calculate the node's degree as its feature for both datasets.

\begin{table*}[ht]
\vspace{-1em}
\centering
\caption{The effectiveness (p-value) of label-only dataset verification on COLLAB and REDDIT-MULTI-5K.}
\vspace{-0.5em}
\scalebox{0.92}{
        \begin{tabular}{c|c|cccc|cccc}
        \toprule
        \multirow{3}{*}{Model$\downarrow$} & Dataset$\rightarrow$                      & \multicolumn{4}{c|}{COLLAB}               & \multicolumn{4}{c}{REDDIT-MULTI-5K}                                                                                                                                                             \\ \cline{2-10}
                                           & Method$\rightarrow$                       & \multicolumn{2}{c|}{GBA-Minimal} & \multicolumn{2}{c|}{GBA-Random} & \multicolumn{2}{c|}{GBA-Minimal} & \multicolumn{2}{c}{GBA-Random}                                                                          \\ \cline{2-10}
                                           & Scenario$\downarrow$, Trigger$\rightarrow$ & Sub-graph 1                                & \multicolumn{1}{c|}{Sub-graph 2}      & Sub-graph 1                                & Sub-graph 2                          & Sub-graph 1 & \multicolumn{1}{c|}{Sub-graph 2} & Sub-graph 1 & Sub-graph 2 \\ \hline
        \multirow{3}{*}{GIN}               & Independent Trigger                       & 1                                         & \multicolumn{1}{c|}{1}               & 1                                         & 1                                   & 1          & \multicolumn{1}{c|}{1}          & 1          & 1          \\
                                           & Independent Model                         & 1                                         & \multicolumn{1}{c|}{1}               & 1                                         & 1                                   & 1          & \multicolumn{1}{c|}{1}          & 1          & 1          \\
                                           & Steal                                     & 0                                         & \multicolumn{1}{c|}{0}               & $0$                                       & 0                                   & 0          & \multicolumn{1}{c|}{0}          & 0          & 0          \\ \hline
        \multirow{3}{*}{GraphSAGE}         & Independent Trigger                       & 1                                         & \multicolumn{1}{c|}{1}               & 1                                         & 1                                   & 1          & \multicolumn{1}{c|}{1}          & 1          & 1          \\
                                           & Independent Model                         & 1                                         & \multicolumn{1}{c|}{1}               & 1                                         & 1                                   & 1          & \multicolumn{1}{c|}{1}          & 1          & 1          \\
                                           & Steal                                     & 0                                         & \multicolumn{1}{c|}{0}               & 0                                         & 0                                   & 0          & \multicolumn{1}{c|}{0}          & 0          & 0          \\ \bottomrule
    \end{tabular}
    }
            \label{tab:lable_GNN}
        \end{table*}

\begin{figure*}[!t]
\vspace{-1em}
\centering
\subfigure[CIFAR-10]{
\includegraphics[width=0.31\textwidth]{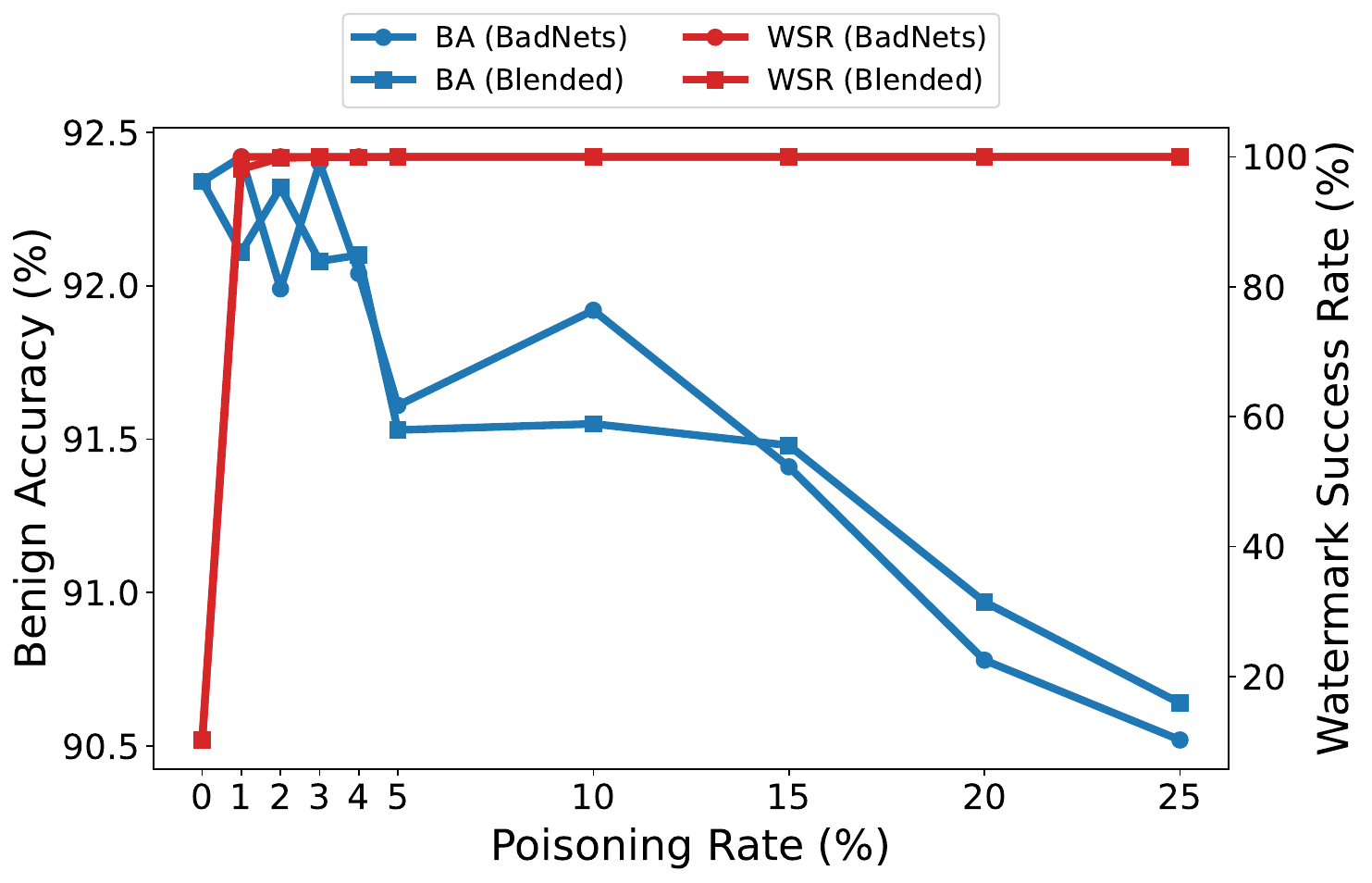}
}
\subfigure[ImageNet]{
\includegraphics[width=0.31\textwidth]{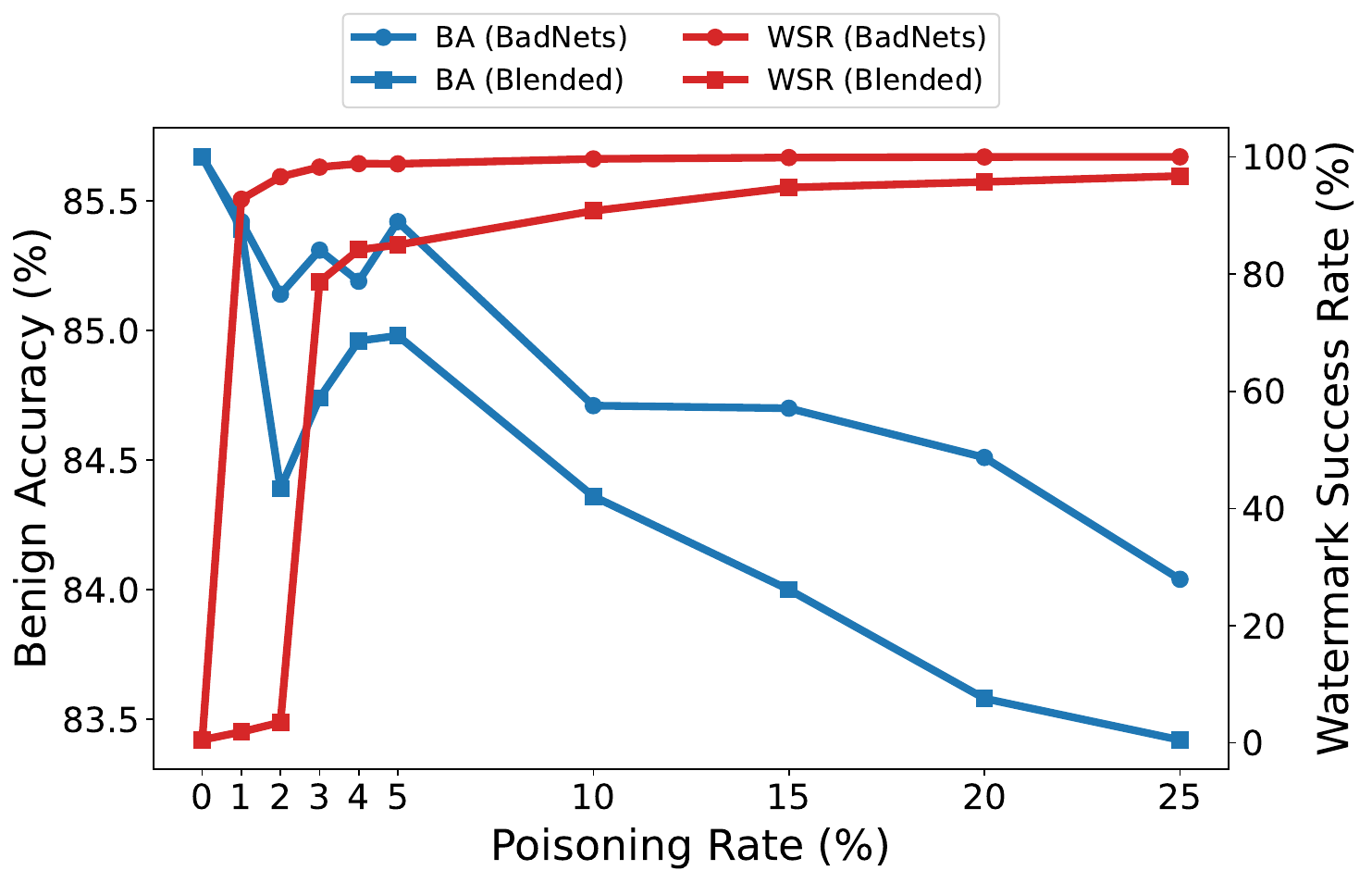}
}
\subfigure[IMDB]{
\includegraphics[width=0.31\textwidth]{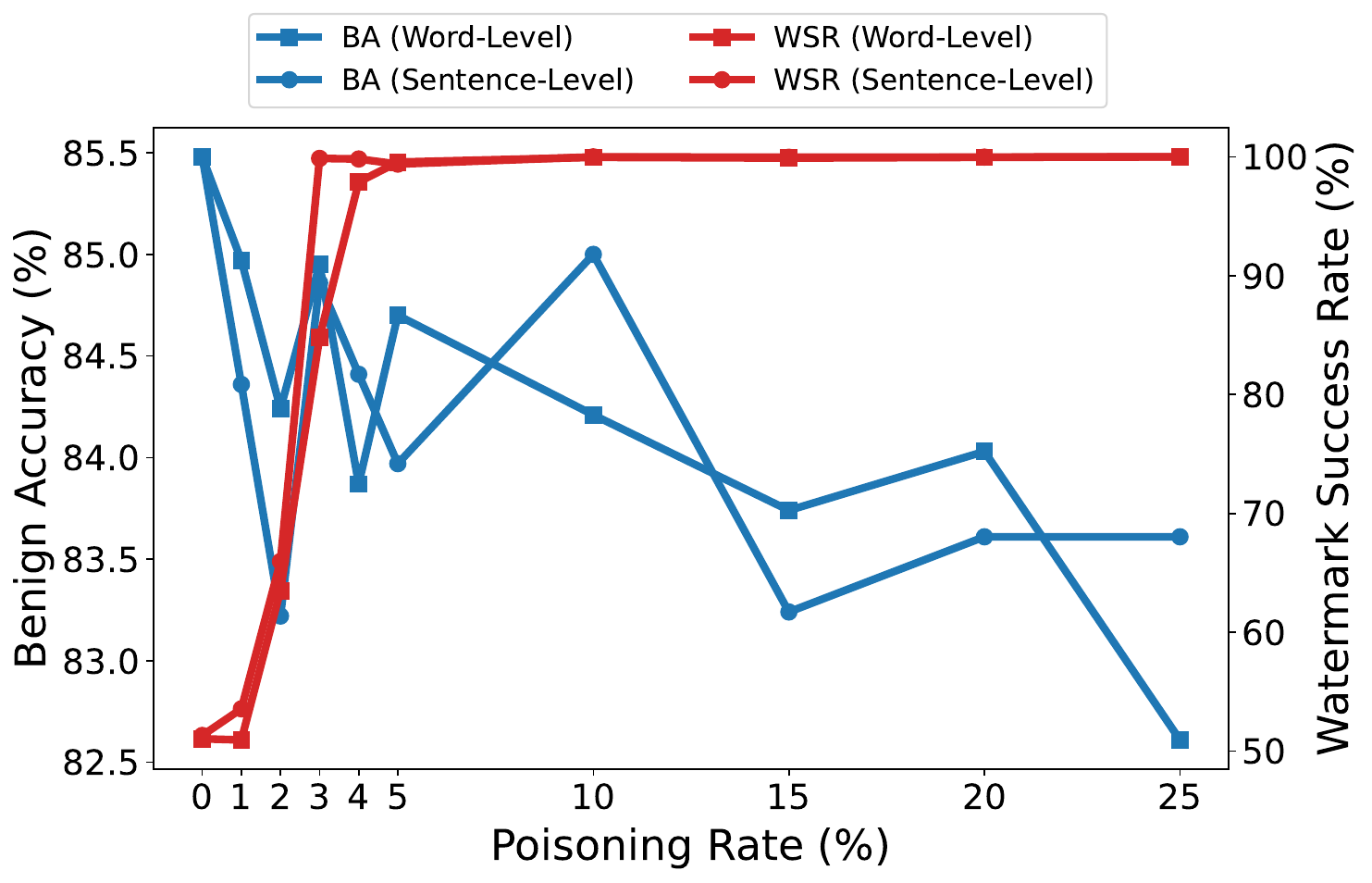}
}
\subfigure[DBpedia]{
\includegraphics[width=0.31\textwidth]{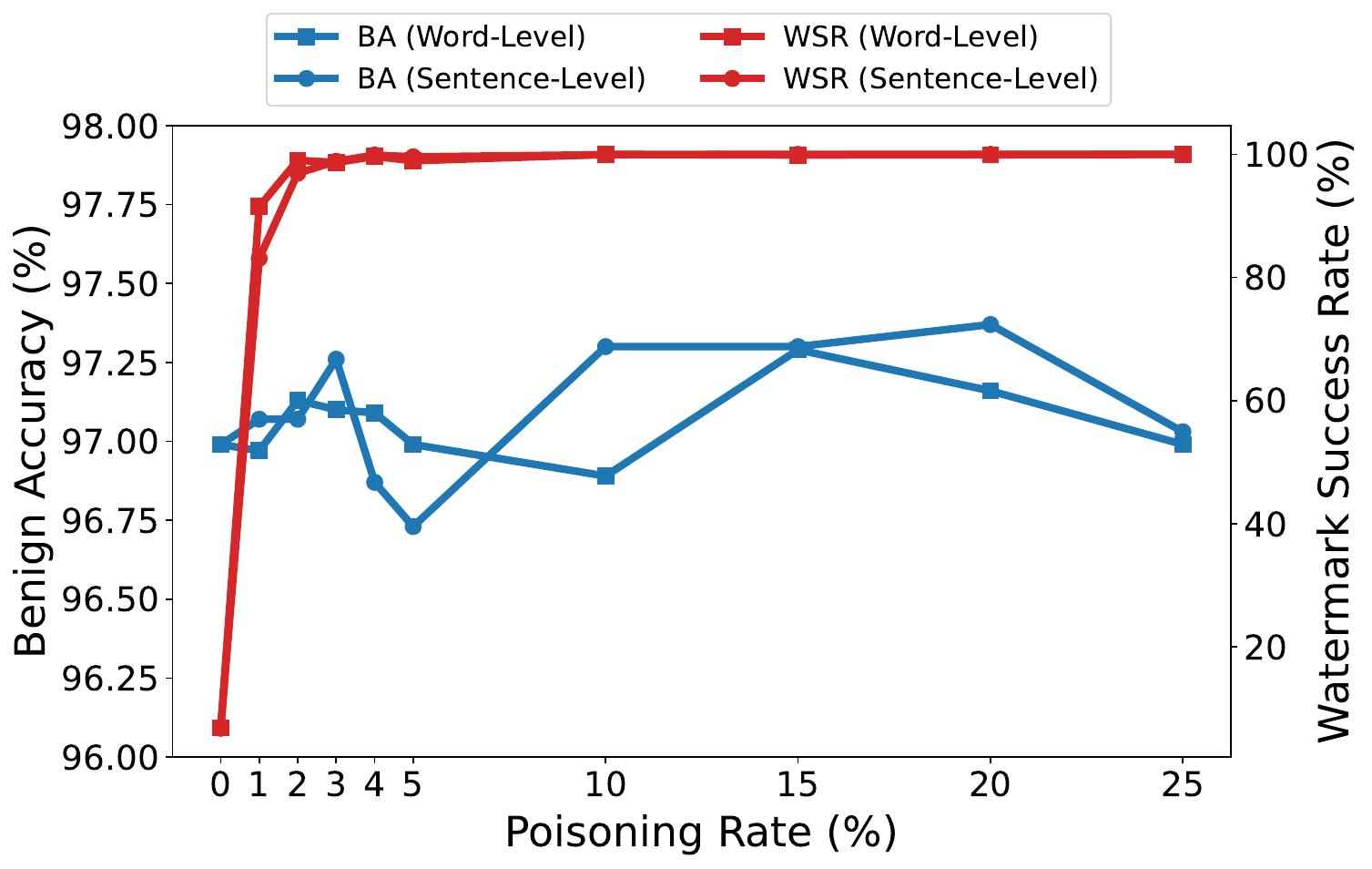}
}
\subfigure[COLLAB]{
\includegraphics[width=0.31\textwidth]{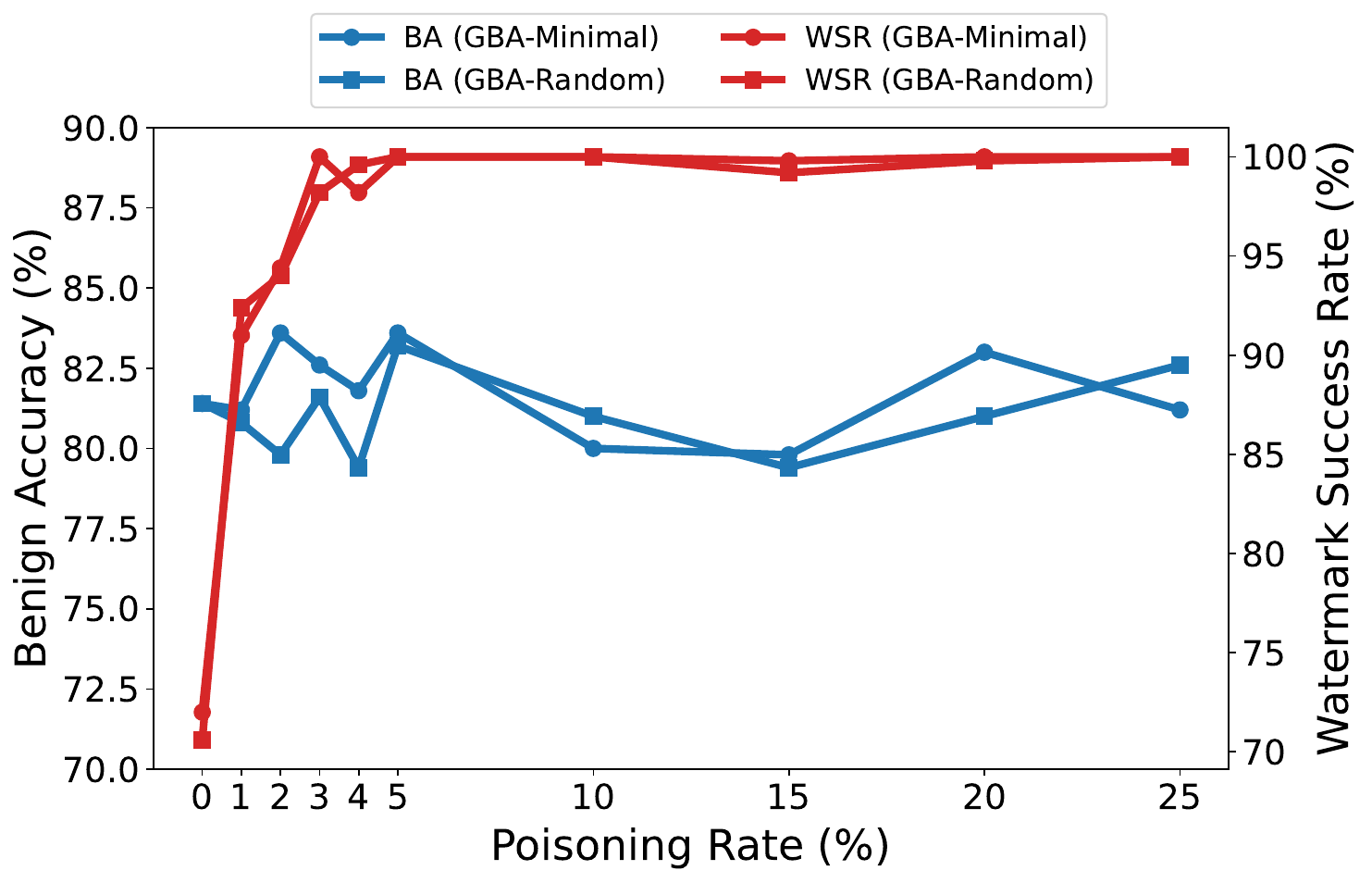}
}
\subfigure[REDDIT-MULTI-5K]{
\includegraphics[width=0.31\textwidth]{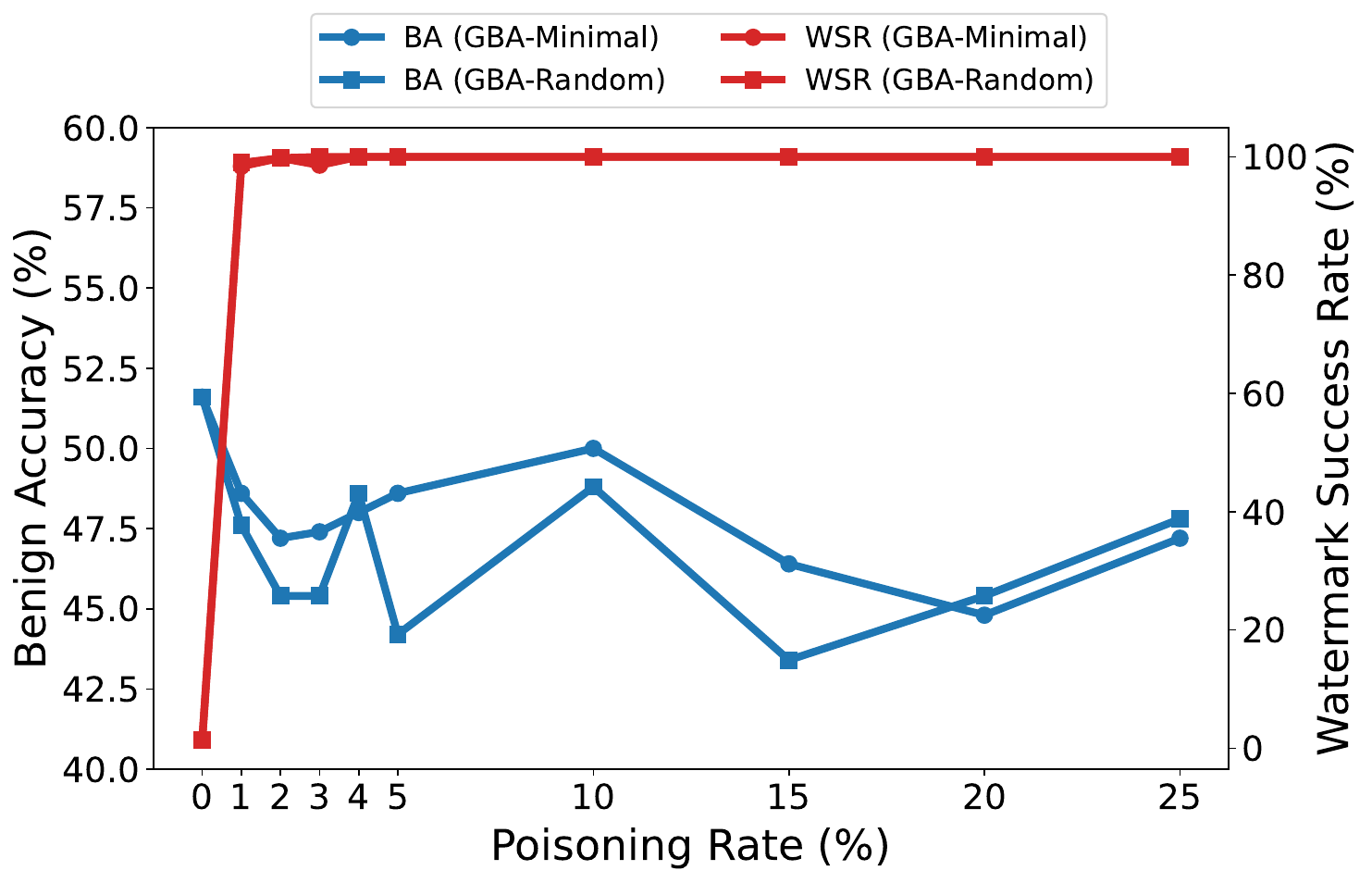}
}
\caption{The effects of the poisoning rate $\gamma$. The benign accuracy (BA) is denoted by the blue line while the watermark success rate (WSR) is indicated by the red one. In many cases, the WSR was close to 100\% even when we only poison 5\% samples, resulting in the two red lines overlapping to a large extent.} 
\label{fig:effects_gamma}
\end{figure*}

\vspace{0.3em}
\noindent \textbf{Settings for Dataset Watermarking.} In these experiments, we use graph backdoor attacks (GBA) \cite{xi2021graph,zhang2021backdoor} for dataset watermarking with poisoning rate $\gamma=0.1$. In GBA, the adversaries adopt sub-graphs as the trigger patterns, which will be connected to the node of some selected benign graphs. Specifically, we consider two types of GBA, including \textbf{1)} GBA with sub-graph injection on the node having minimal degree (dubbed as 'GBA-Minimal') and \textbf{2)} GBA with sub-graph injection on the random node (dubbed as 'GBA-Random'). On both datasets, we adopt the complete sub-graphs as trigger patterns. Specifically, on the COLLAB dataset, we adopt the ones with degree $D=14$ and $D=15$, respectively; We exploit the ones with degree $D=97$ and $D=98$ on the REDDIT-MULTI-5K dataset. The target label $y_t$ is set as the first class ($i.e.$, $y_t=1$ for both datasets). The illustration of generated poisoned samples is shown in Figure \ref{fig:examples_GNN}.

\vspace{0.3em}
\noindent \textbf{Settings for Dataset Verification.} In particular, we select samples only from the last class ($i.e.$, `2' on COLLAB and `5' on REDDIT-MULTI-5K) for dataset verification. Besides, we adopt the complete sub-graph with half degrees ($i.e.$, $D=7$ on COLLAB and $D=48$ on REDDIT-MULTI-5K) as the trigger pattern used in the `Trigger Independent' scenarios. All other settings are the same as those used in Section \ref{sec:image_results}.

\vspace{0.15em}
\noindent \textbf{Results.} As shown in Table \ref{tab:GNN_watermarked}, both GBA-Minimal and GBA-Random can achieve a high watermark success rate (WSR) and preserve high benign accuracy (BA). Specifically, the WSRs are larger than $99.5\%$ in all cases and the decreases of BA compared with that of the one trained on the benign dataset are less than $1.5\%$ on the COLLAB dataset. These results verify the effectiveness of our dataset watermarking. Moreover, as shown in Table \ref{tab:logits_GNN}-\ref{tab:lable_GNN}, our dataset verification is also effective, no matter under probability-available scenarios or label-only scenarios. Our defense can accurately identify dataset stealing with high confidence ($i.e.$, $\Delta P \gg 0$ and p-value $\ll 0.01$) while does not misjudge when there is no stealing ($i.e.$, $\Delta P$ is nearly 0 and p-value $\gg 0.05$). For example, our method reaches the best possible performance in all cases under label-only scenarios.

\subsection{Ablation Study}
In this section, we study the effects of core hyper-parameters, including the poisoning rate $\gamma$ and the sampling number $m$, contained in our DVBW. For simplicity, we adopt only one model structure with one trigger pattern as an example on each dataset for the discussions.


\subsubsection{The Effects of Poisoning Rate}
As shown in Figure \ref{fig:effects_gamma}, the watermark success rate increases with the increase of poisoning rate $\gamma$ in all cases. These results indicate that defenders can improve the verification confidence by using a relatively large $\gamma$. In particular, almost all evaluated attacks reach a high watermark success rate even when the poisoning rate is small ($e.g.$, 1\%). In other words, our dataset watermarking is stealthy as dataset owners only need to modify a few samples to succeed. However, the benign accuracy decreases with the increases of $\gamma$ in most cases. In other words, there is a trade-off between WSR and BA to some extent. The defenders should assign $\gamma$ based on their specific needs in practice.

\begin{table*}[!t]
\vspace{-1em}
    \centering
    \caption{The verification effectiveness (p-value) of our DVBW with different sampling numbers.}
    \vspace{-0.5em}
    \scalebox{0.9}{
        \begin{tabular}{c|c|c|ccccccc}
            \toprule
            Dataset$\downarrow$              & Method$\downarrow$              & Scenario$\downarrow$, Sampling Number$\rightarrow$ & 20         & 40          & 60          & 80          & 100         & 120         & 140         \\ \hline
            \multirow{6}{*}{CIFAR-10}        & \multirow{3}{*}{BadNets}        & Independent-T                                      & 1          & 1           & 1           & 1           & 1           & 1           & 1           \\
                                             &                                 & Independent-M                                      & 1          & 1           & 1           & 1           & 1           & 1           & 1           \\
                                             &                                 & Malicious                                           & $10^{-46}$ & $10^{-50}$  & $10^{-106}$ & $10^{-117}$ & $10^{-132}$ & $10^{-136}$ & $10^{-149}$ \\ \cline{2-10}
                                             & \multirow{3}{*}{Blended}        & Independent-T                                      & 1          & 1           & 1           & 1           & 1           & 1           & 1           \\
                                             &                                 & Independent-M                                      & 1          & 1           & 1           & 1           & 1           & 1           & 1           \\
                                             &                                 & Malicious                                           & $10^{-16}$ & $10^{-29}$  & $10^{-46}$  & $10^{-67}$  & $10^{-103}$ & $10^{-102}$ & $10^{-138}$ \\ \hline
            \multirow{6}{*}{ImageNet}        & \multirow{3}{*}{BadNets}        & Independent-T                                      & 1          & 1           & 1           & 1           & 1           & 1           & 1           \\
                                             &                                 & Independent-M                                      & 1          & 1           & 1           & 1           & 1           & 1           & 1           \\
                                             &                                 & Malicious                                           & $10^{-34}$ & $10^{-72}$  & $10^{-69}$  & $10^{-122}$ & $10^{-144}$ & $10^{-169}$ & $10^{-195}$ \\ \cline{2-10}
                                             & \multirow{3}{*}{Blended}        & Independent-T                                      & 1          & 1           & 1           & 1           & 1           & 1           & 1           \\
                                             &                                 & Independent-M                                      & 1          & 1           & 1           & 1           & 1           & 1           & 1           \\
                                             &                                 & Malicious                                           & $10^{-12}$ & $10^{-19}$  & $10^{-29}$  & $10^{-32}$  & $10^{-41}$  & $10^{-54}$  & $10^{-67}$  \\ \hline
            \multirow{6}{*}{IMDB}            & \multirow{3}{*}{Word-Level}     & Independent-T                                      & 1          & 1           & 1           & 1           & 1           & 1           & 1           \\
                                             &                                 & Independent-M                                      & 1          & 1           & 1           & 1           & 1           & 1           & 1           \\
                                             &                                 & Malicious                                           & $10^{-12}$ & $10^{-16}$  & $10^{-26}$  & $10^{-36}$  & $10^{-46}$  & $10^{-56}$  & $10^{-63}$  \\ \cline{2-10}
                                             & \multirow{3}{*}{Sentence-Level} & Independent-T                                      & 1          & 1           & 1           & 1           & 1           & 1           & 1           \\
                                             &                                 & Independent-M                                      & 1          & 1           & 1           & 1           & 1           & 1           & 1           \\
                                             &                                 & Malicious                                           & $10^{-12}$ & $10^{-18}$  & $10^{-25}$  & $10^{-35}$  & $10^{-47}$  & $10^{-53}$  & $10^{-61}$  \\ \hline
            \multirow{6}{*}{DBpedia}         & \multirow{3}{*}{Word-Level}     & Independent-T                                      & 1          & 1           & 1           & 1           & 1           & 1           & 1           \\
                                             &                                 & Independent-M                                      & 1          & 1           & 1           & 1           & 1           & 1           & 1           \\
                                             &                                 & Malicious                                           & $10^{-89}$ & $10^{-186}$ & $10^{-224}$ & $10^{-226}$ & $10^{-281}$ & $10^{-296}$ & 0           \\ \cline{2-10}
                                             & \multirow{3}{*}{Sentence-Level} & Independent-T                                      & 1          & 1           & 1           & 1           & 1           & 1           & 1           \\
                                             &                                 & Independent-M                                      & 1          & 1           & 1           & 1           & 1           & 1           & 1           \\
                                             &                                 & Malicious                                           & $10^{-55}$ & $10^{-117}$ & $10^{-181}$ & $10^{-182}$ & $10^{-150}$ & $10^{-185}$ & $10^{-220}$ \\ \hline
            \multirow{6}{*}{COLLAB}          & \multirow{3}{*}{GBA-Minimal}    & Independent-T                                      & 1          & 1           & 1           & 1           & 1           & 1           & 1           \\
                                             &                                 & Independent-M                                      & 1          & 1           & 1           & 1           & 1           & 1           & 1           \\
                                             &                                 & Malicious                                           & $10^{-14}$ & $10^{-26}$  & $10^{-31}$  & $10^{-41}$  & $10^{-48}$  & $10^{-58}$  & $10^{-70}$  \\ \cline{2-10}
                                             & \multirow{3}{*}{GBA-Random}     & Independent-T                                      & 1          & 1           & 1           & 1           & 1           & 1           & 1           \\
                                             &                                 & Independent-M                                      & 1          & 1           & 1           & 1           & 1           & 1           & 1           \\
                                             &                                 & Malicious                                           & $10^{-15}$ & $10^{-29}$  & $10^{-29}$  & $10^{-37}$  & $10^{-43}$  & $10^{-53}$  & $10^{-64}$  \\ \hline
            \multirow{6}{*}{REDDIT-MULTI-5K} & \multirow{3}{*}{GBA-Minimal}    & Independent-T                                      & 1          & 1           & 1           & 1           & 1           & 1           & 1           \\
                                             &                                 & Independent-M                                      & 1          & 1           & 1           & 1           & 1           & 1           & 1           \\
                                             &                                 & Malicious                                           & $10^{-27}$ & $10^{-51}$  & $10^{-64}$  & $10^{-87}$  & $10^{-64}$ & $10^{-131}$ & $10^{-147}$ \\ \cline{2-10}
                                             & \multirow{3}{*}{GBA-Random}     & Independent-T                                      & 1          & 1           & 1           & 1           & 1           & 1           & 1           \\
                                             &                                 & Independent-M                                      & 1          & 1           & 1           & 1           & 1           & 1           & 1           \\
                                             &                                 & Malicious                                           & $10^{-33}$ & $10^{-59}$  & $10^{-85}$  & $10^{-112}$ & $10^{-138}$ & $10^{-119}$ & $10^{-133}$ \\
            \bottomrule
        \end{tabular}
    }
    \label{tab:effects_m}
\end{table*}

\begin{figure*}[!t]
\centering
\vspace{-1em}
\subfigure[CIFAR-10]{
\includegraphics[width=0.3\textwidth]{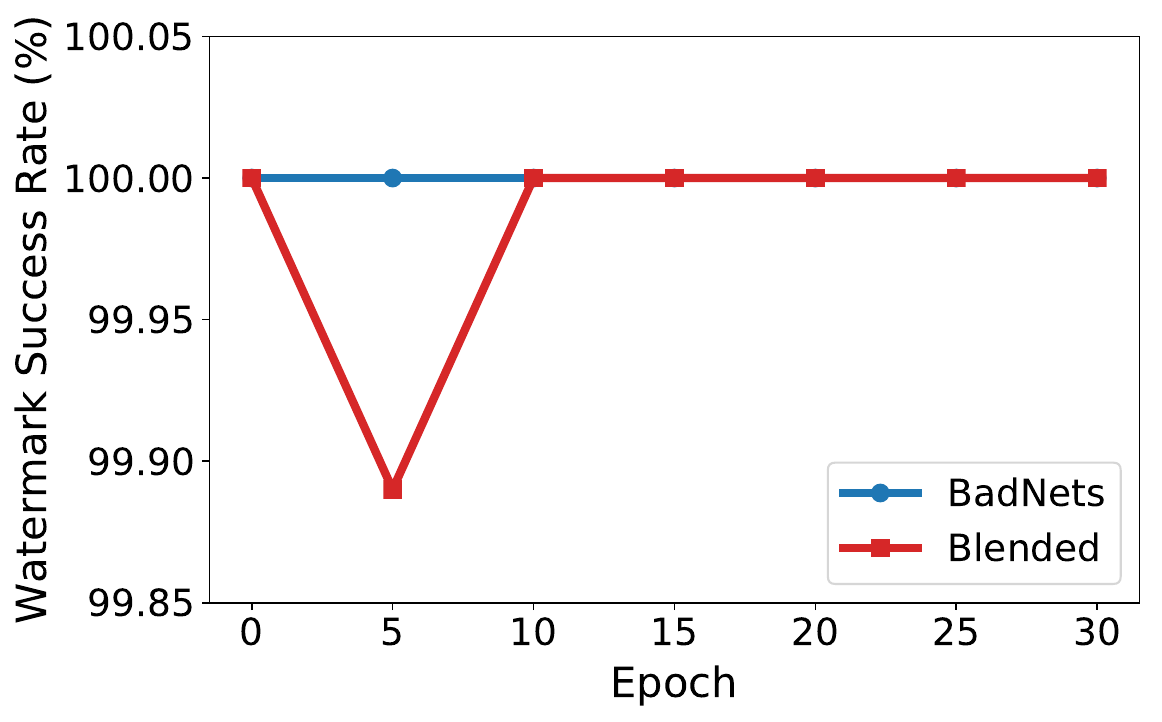}
}
%\hspace{0.8em}
\subfigure[ImageNet]{
\includegraphics[width=0.28\textwidth]{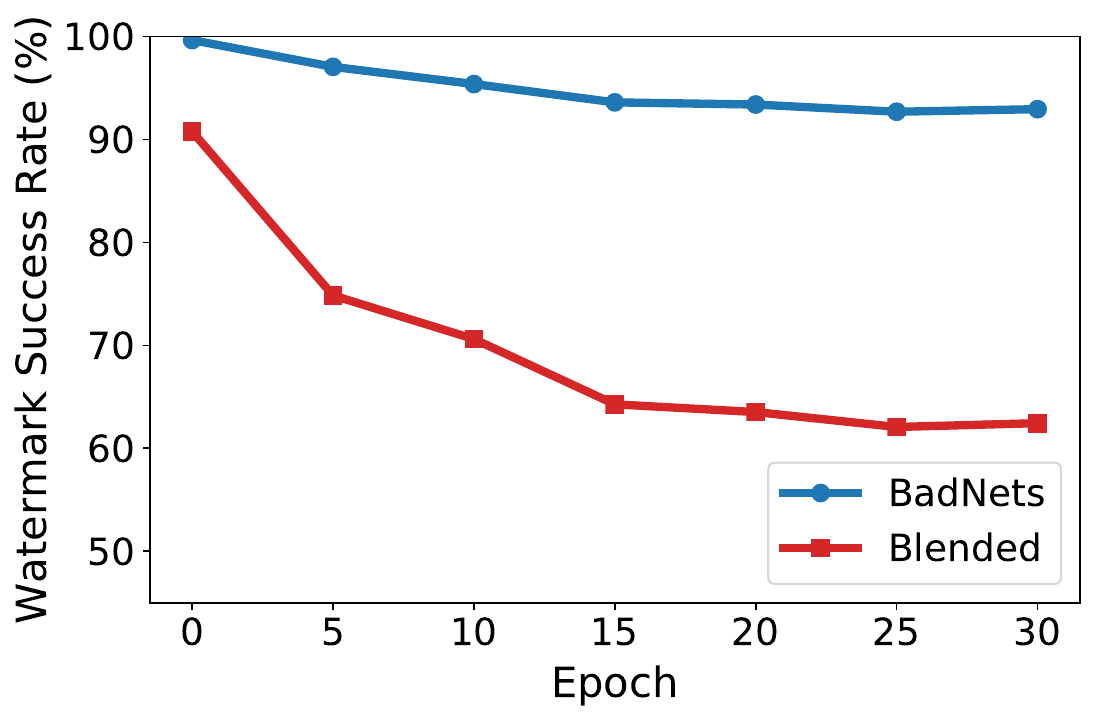}
}
\subfigure[IMDB]{
\includegraphics[width=0.3\textwidth]{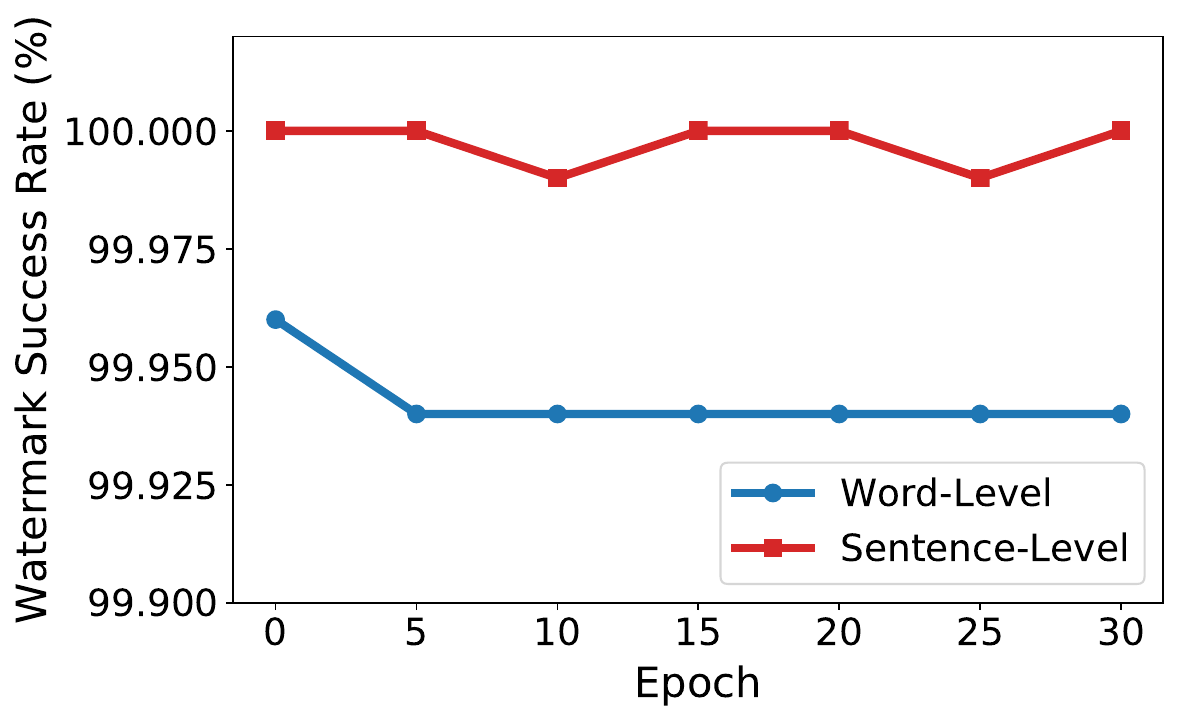}
}
%\hspace{0.8em}
\subfigure[DBpedia]{
\includegraphics[width=0.3\textwidth]{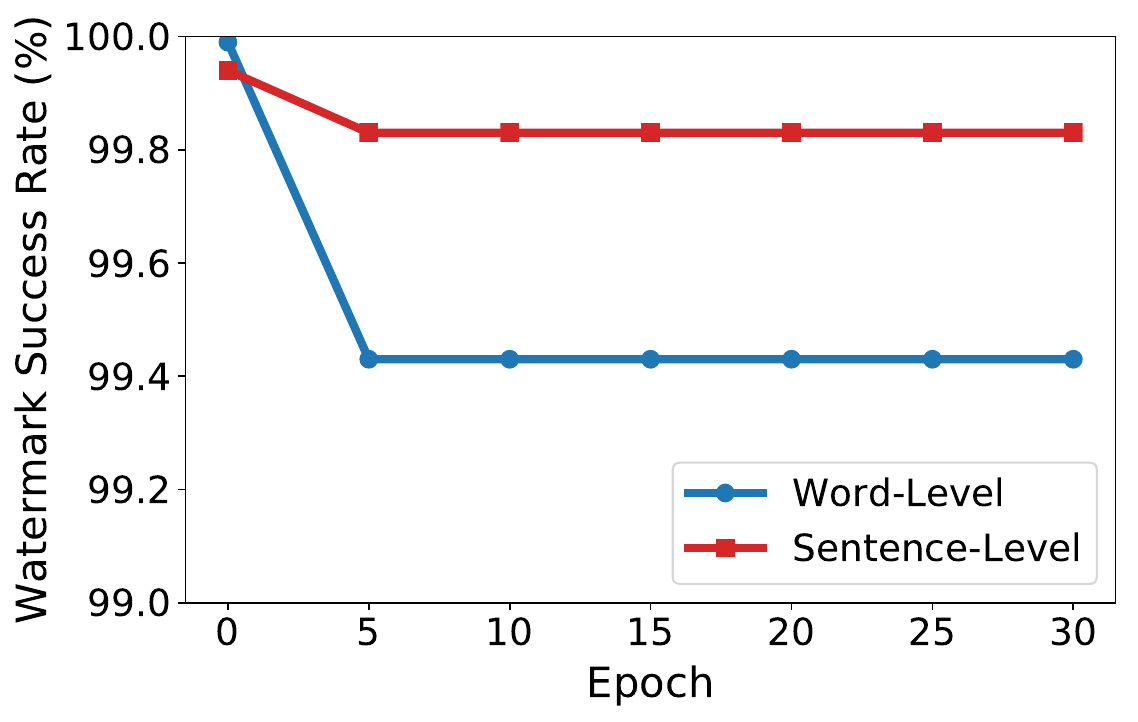}
}
\subfigure[COLLAB]{
\includegraphics[width=0.29\textwidth]{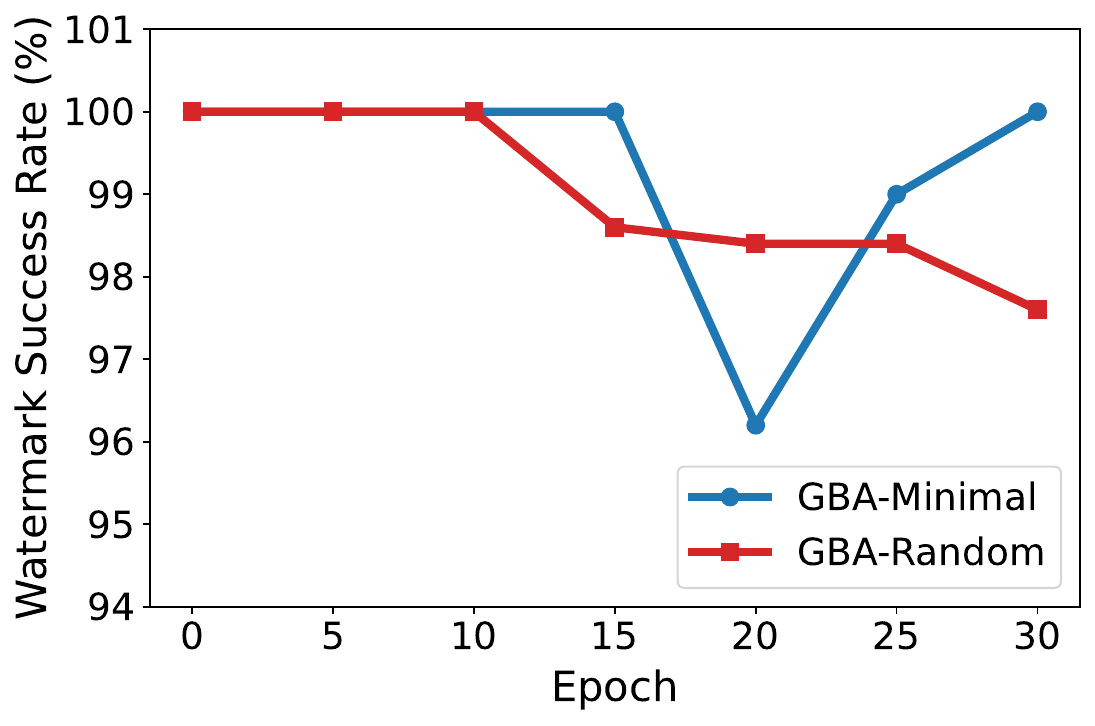}
}
%\hspace{0.8em}
\subfigure[REDDIT-MULTI-5K]{
\includegraphics[width=0.3\textwidth]{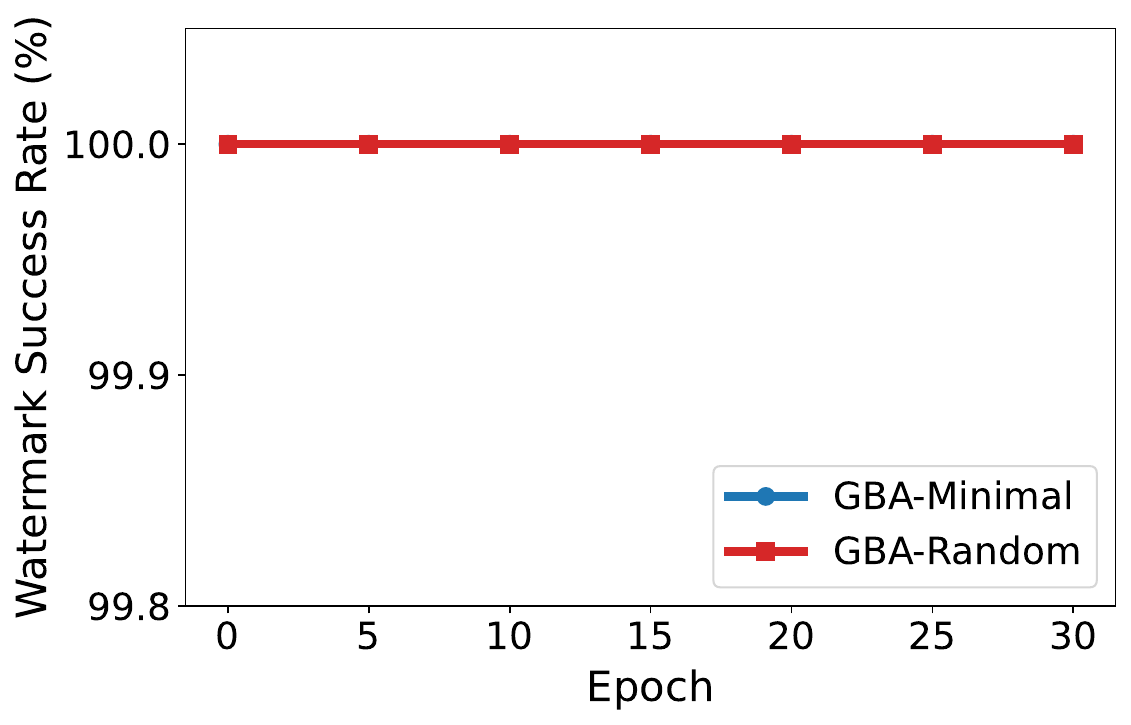}
}
\caption{The resistance of our DVBW to fine-tuning on six different datasets.} 
\label{fig:fine-tune}
\vspace{-0.5em}
\end{figure*}

\subsubsection{The Effects of Sampling Number}
Recall that we need to select $m$ different benign samples to generate their watermarked version in our verification process. As shown in Table \ref{tab:effects_m}, the verification performance increases with the sampling number $m$. These results are expected since our method can achieve promising WSR. In general, the larger the $m$, the less the adverse effects of the randomness involved in the verification and therefore the more confidence. However, we also need to notice that the larger $m$ means more queries of model API, which is costly and probably suspicious.

\begin{figure*}[!t]
\centering
\vspace{-1em}
\subfigure[CIFAR-10]{
\includegraphics[width=0.31\textwidth]{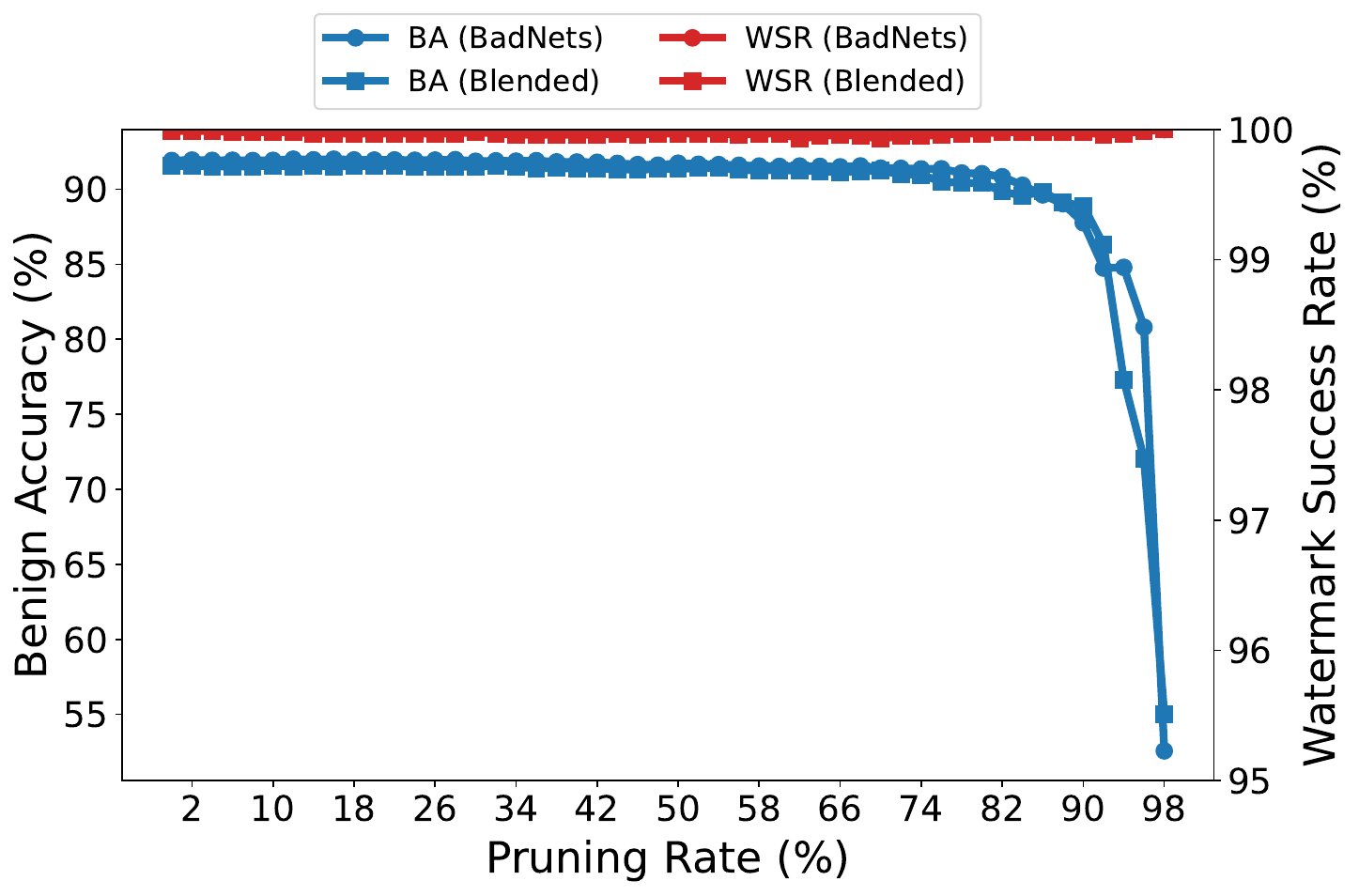}
}
%\hspace{0.8em}
\subfigure[ImageNet]{
\includegraphics[width=0.31\textwidth]{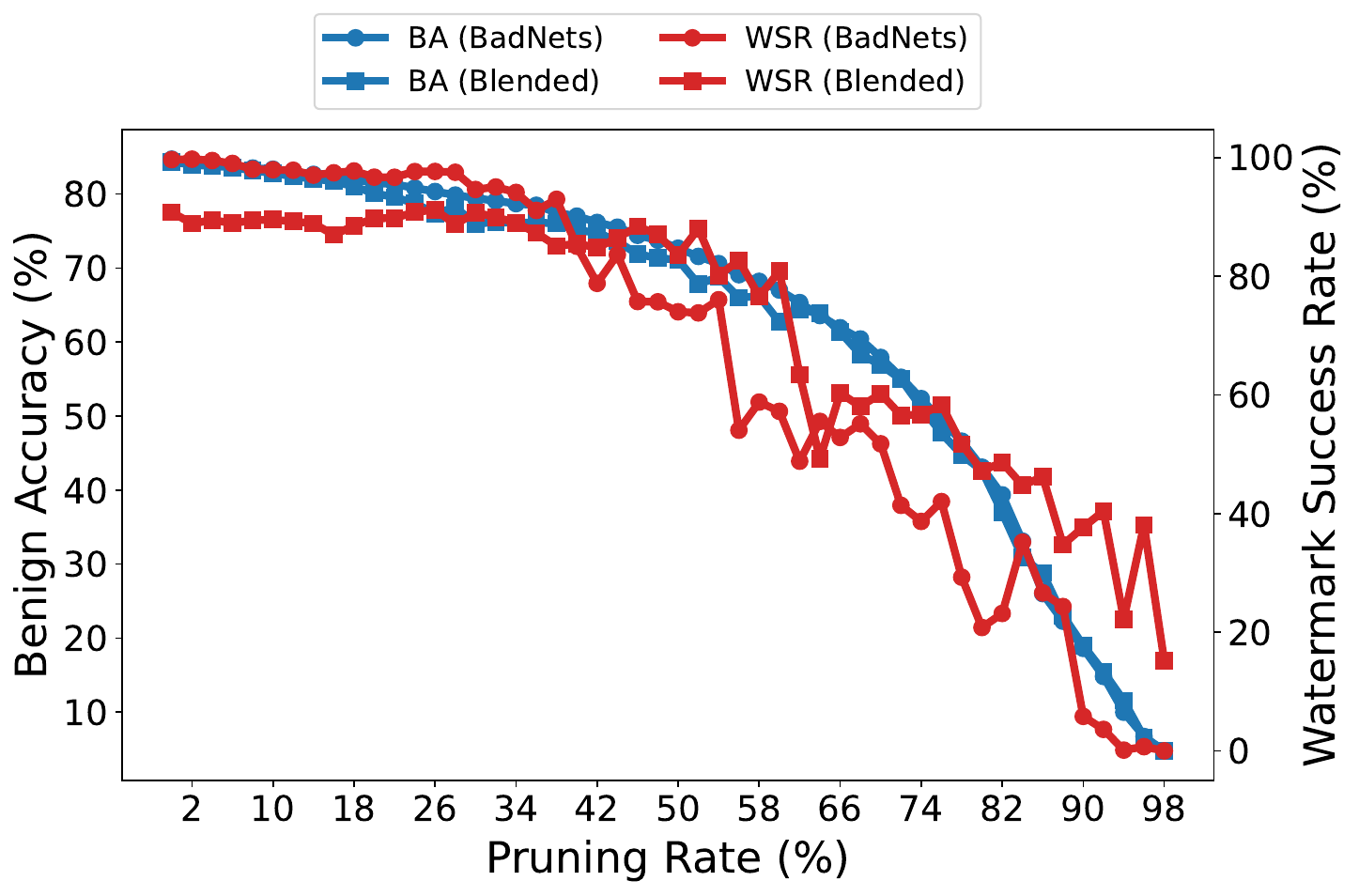}
}
\subfigure[IMDB]{
\includegraphics[width=0.31\textwidth]{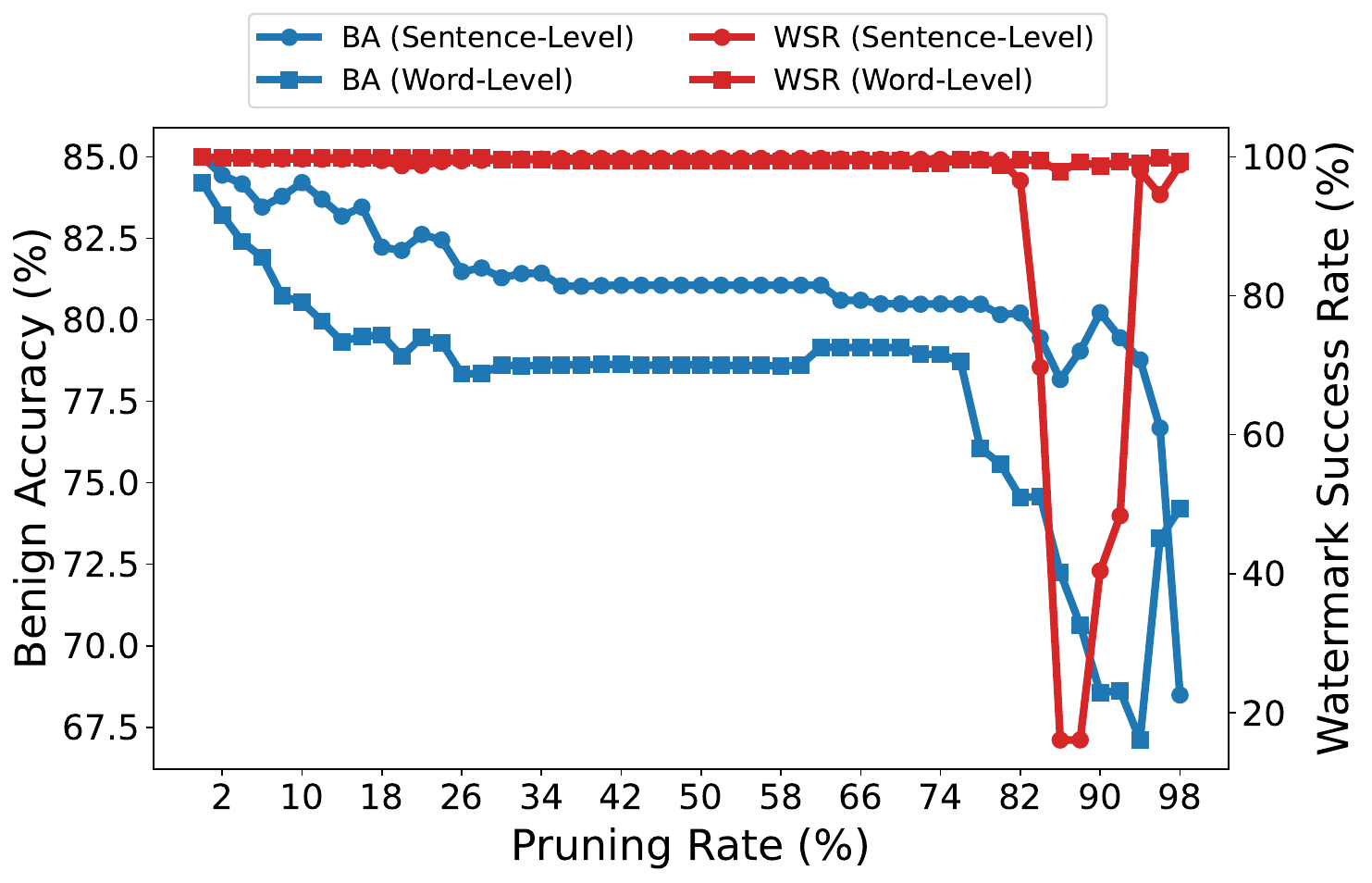}
}
%\hspace{0.8em}
\subfigure[DBpedia]{
\includegraphics[width=0.31\textwidth]{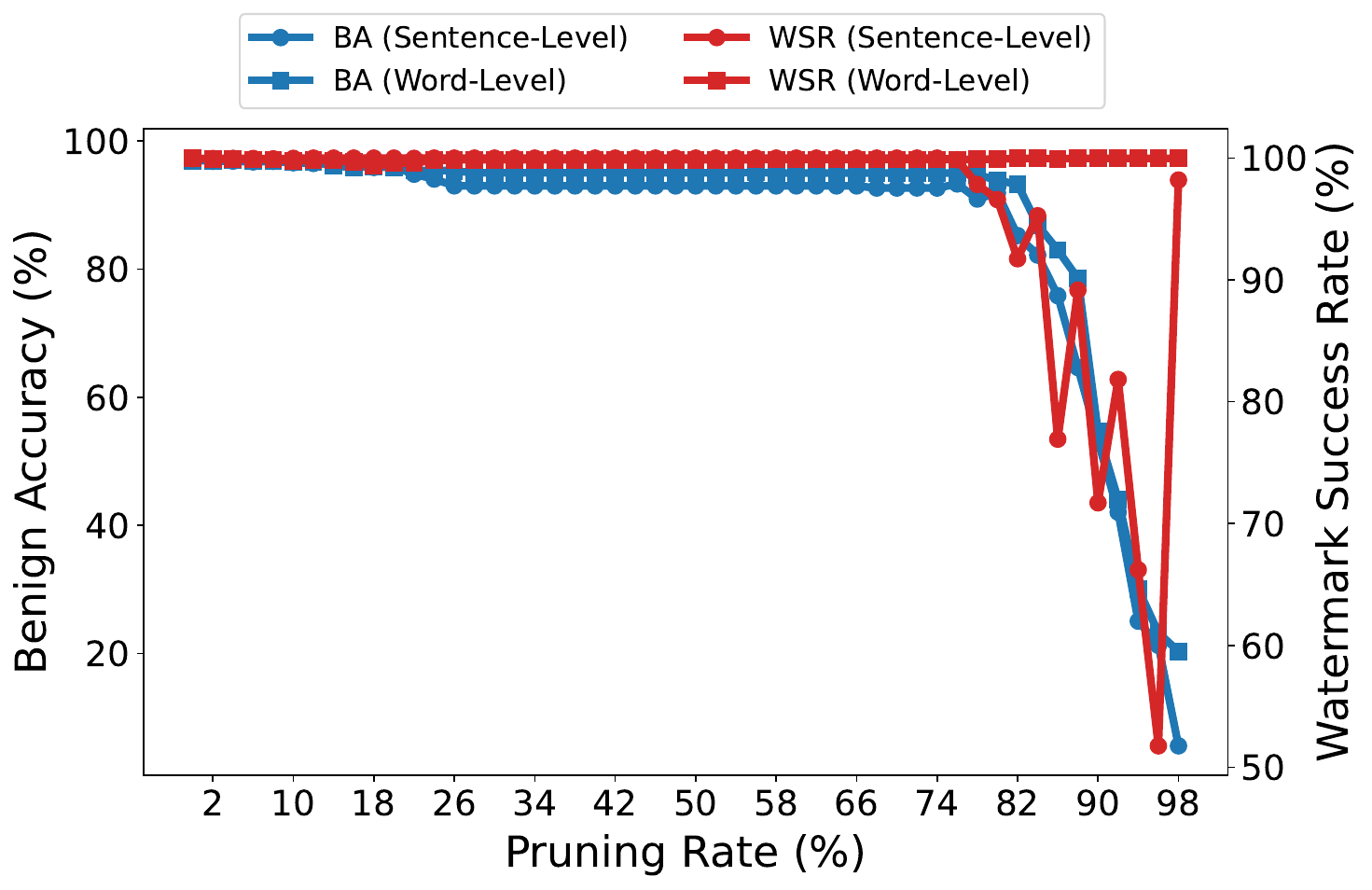}
}
\subfigure[COLLAB]{
\includegraphics[width=0.31\textwidth]{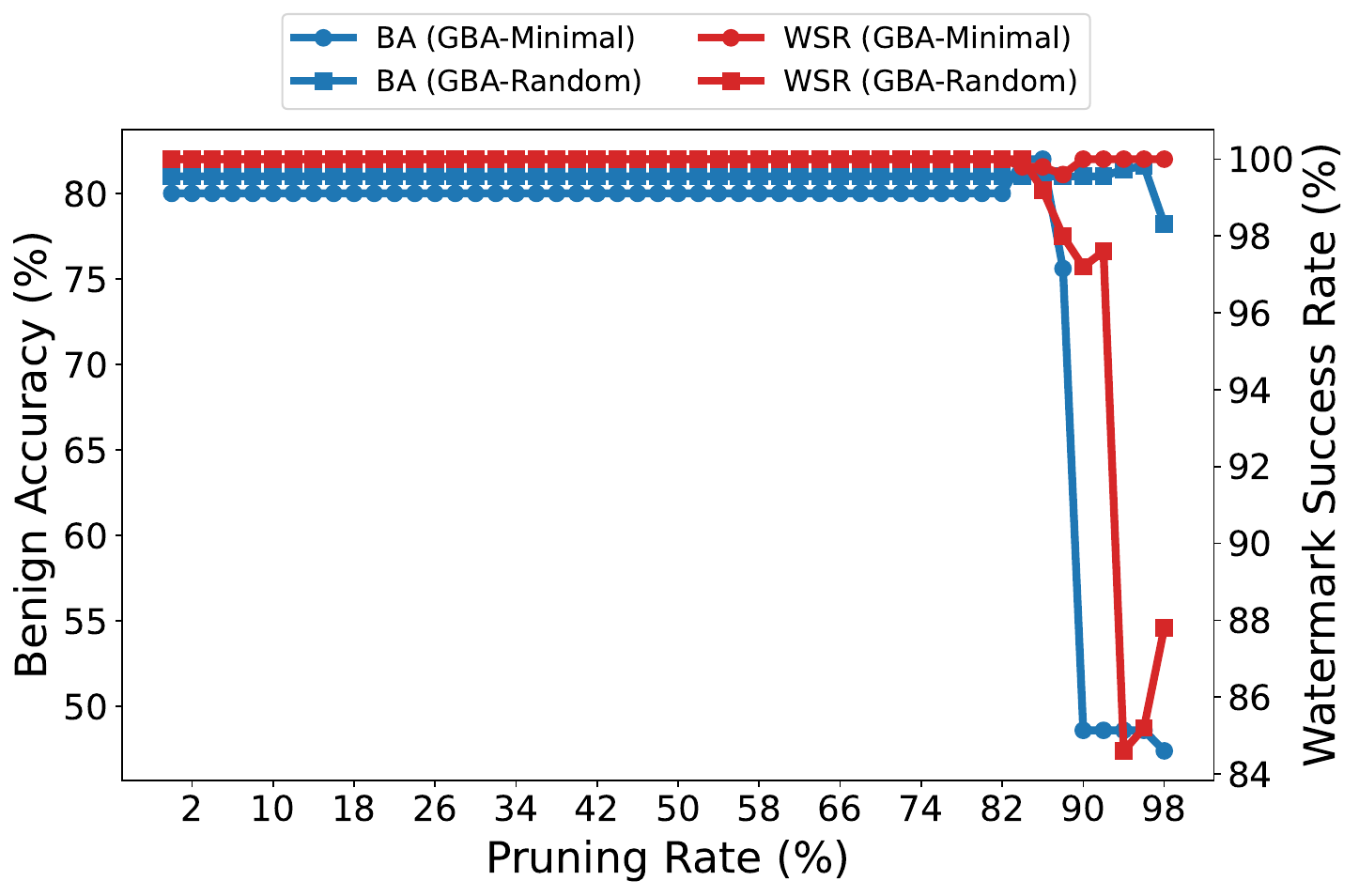}
}
%\hspace{0.8em}
\subfigure[REDDIT-MULTI-5K]{
\includegraphics[width=0.31\textwidth]{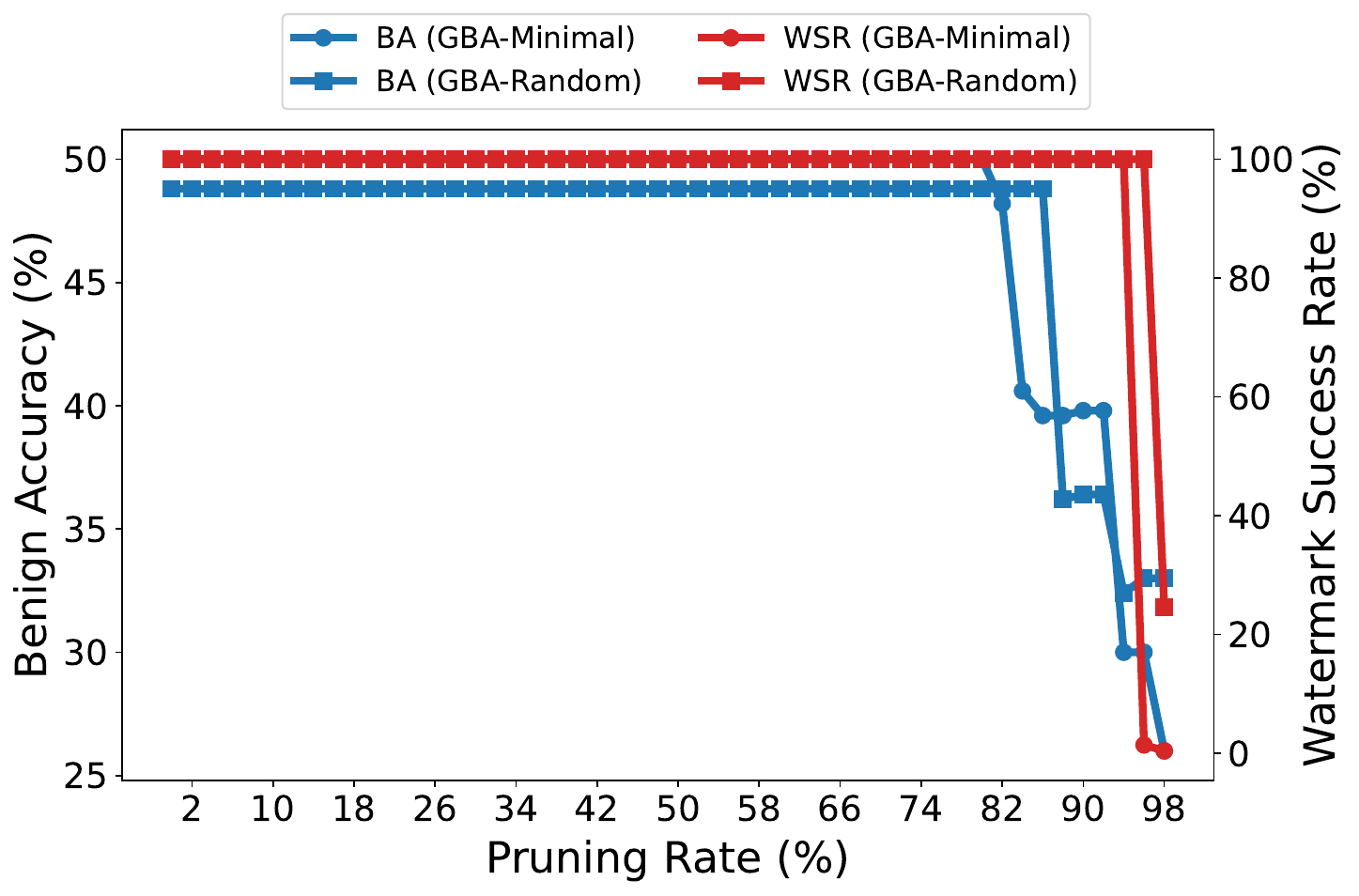}
}
\caption{The resistance of our DVBW to model pruning on six different datasets.}
\vspace{-0.7em}
\label{fig:pruning}
\end{figure*}

\subsection{The Resistance to Potential Adaptive Attacks}

In this section, we discuss the resistance of our DVBW to three representative backdoor removal methods, including fine-tuning \cite{liu2017neural}, model pruning \cite{liu2018fine}, and anti-backdoor learning \cite{li2021anti}. These methods were initially used in image classification but can be directly generalized to other classification tasks (\eg, graph recognition) as well. Unless otherwise specified, we use only one model structure with one trigger pattern as an example for the discussions on each dataset. We implement these removal methods based on the codes of an open-sourced backdoor toolbox \cite{li2023backdoorbox} (\ie, \texttt{BackdoorBox}\footnote{\url{https://github.com/THUYimingLi/BackdoorBox}}).

%In this part, we validate whether our DVBW method is still effective under model fine-tuning, which is a representative and universal adaptive method used by the stealing adversaries in different tasks \cite{jia2021entangled,zanella2021grey,zhang2022deep}.

\vspace{0.3em}
\noindent \textbf{The Resistance to Fine-tuning.} Following the classical settings, we adopt 10\% benign samples from the original training set to fine-tune fully-connected layers of the watermarked model. In each case, we set the learning rate as the one used in the last training epoch of the victim model. As shown in Figure \ref{fig:fine-tune}, the watermark success rate (WSR) generally decreases with the increase of tuning epochs. However, even on the ImageNet dataset where fine-tuning is most effective, the WSR is still larger than $60\%$ after the fine-tuning process is finished. In most cases, fine-tuning has only minor effects in reducing WSR. These results indicate that our DVBW is resistant to model fine-tuning to some extent.

\begin{figure*}[!t]
\vspace{-1.5em}
\centering
\subfigure[CIFAR-10]{
\includegraphics[width=0.31\textwidth]{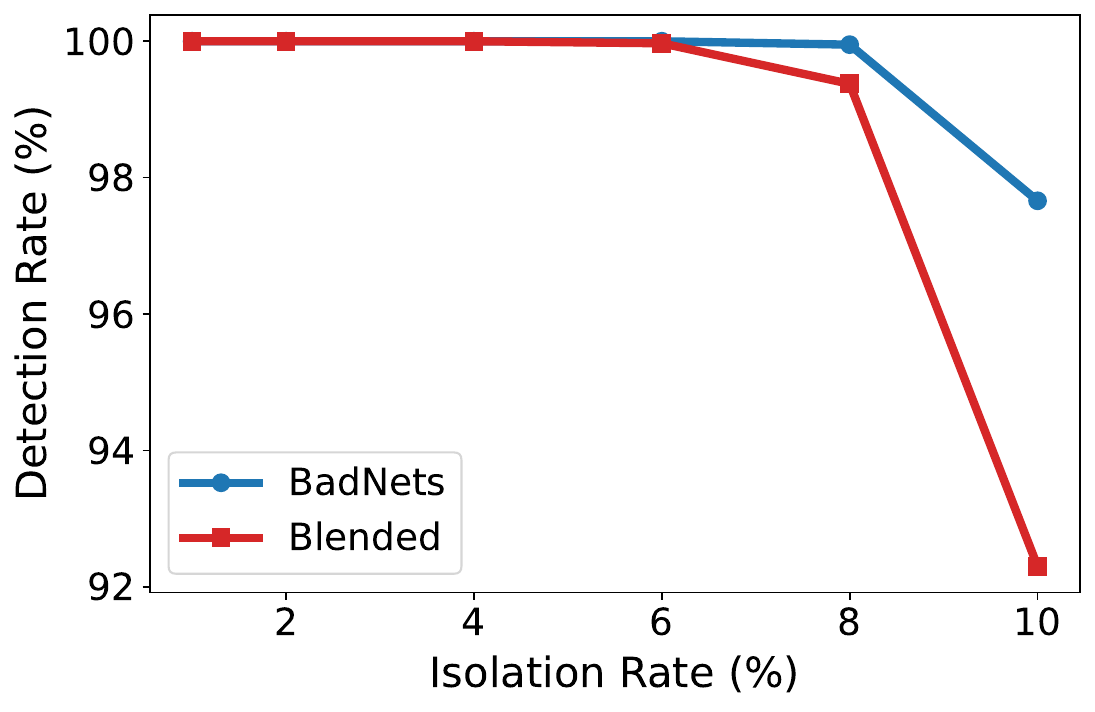}
}
%\hspace{0.8em}
\subfigure[ImageNet]{
\includegraphics[width=0.31\textwidth]{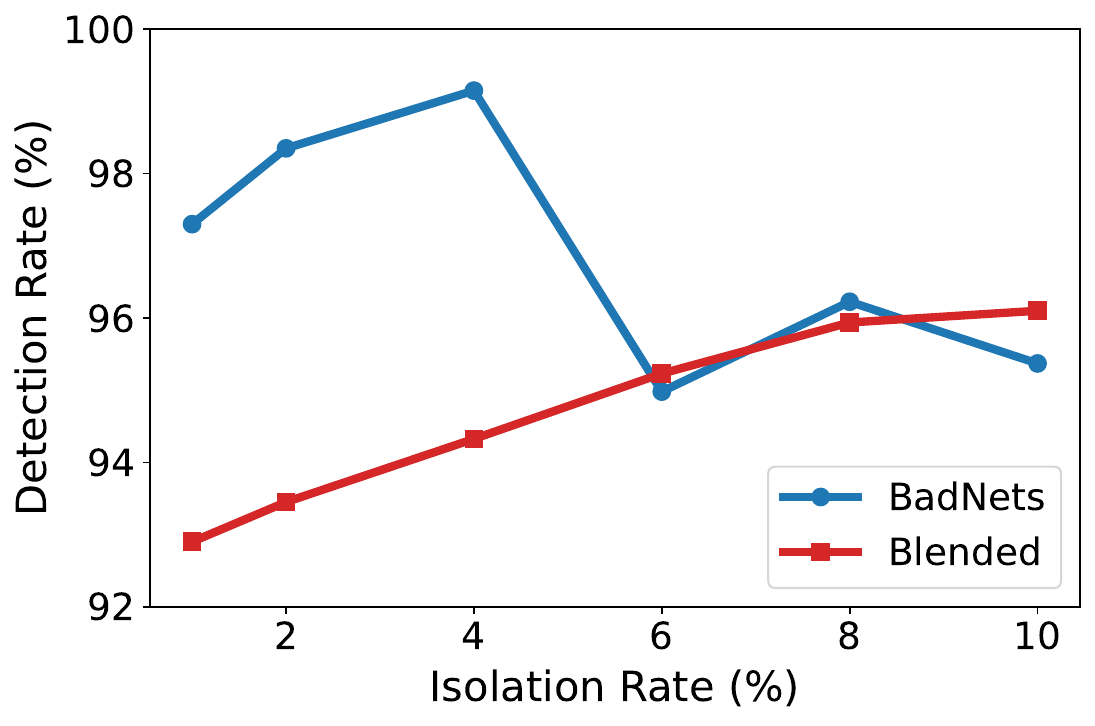}
}
\subfigure[IMDB]{
\includegraphics[width=0.31\textwidth]{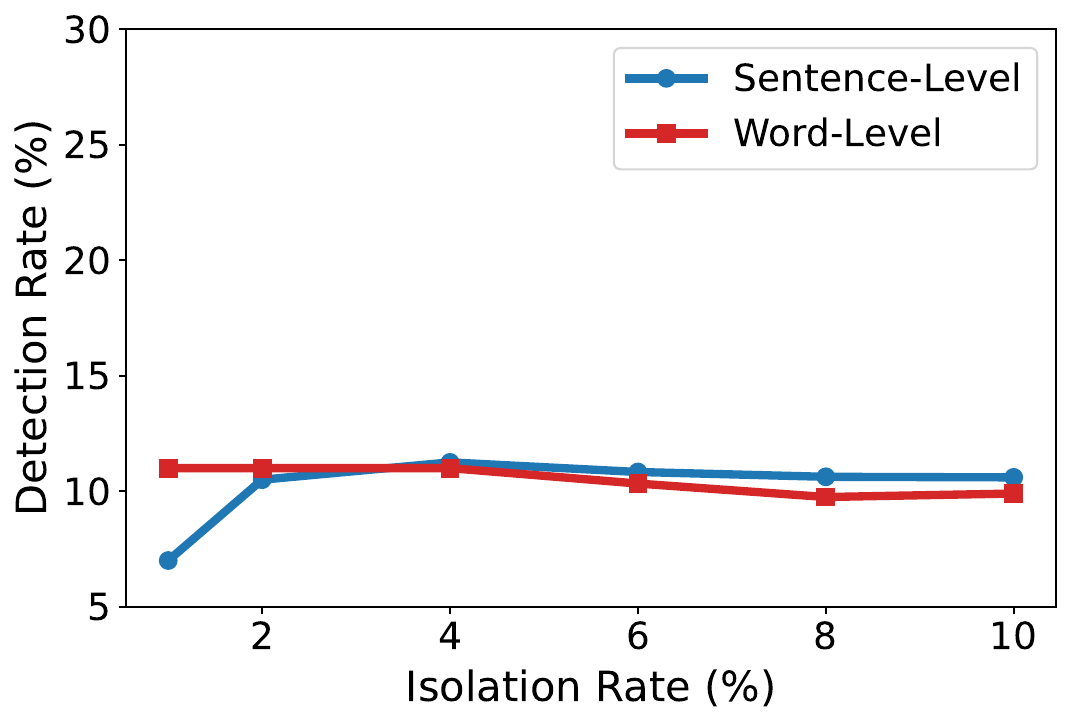}
}
%\hspace{0.8em}
\subfigure[DBpedia]{
\includegraphics[width=0.31\textwidth]{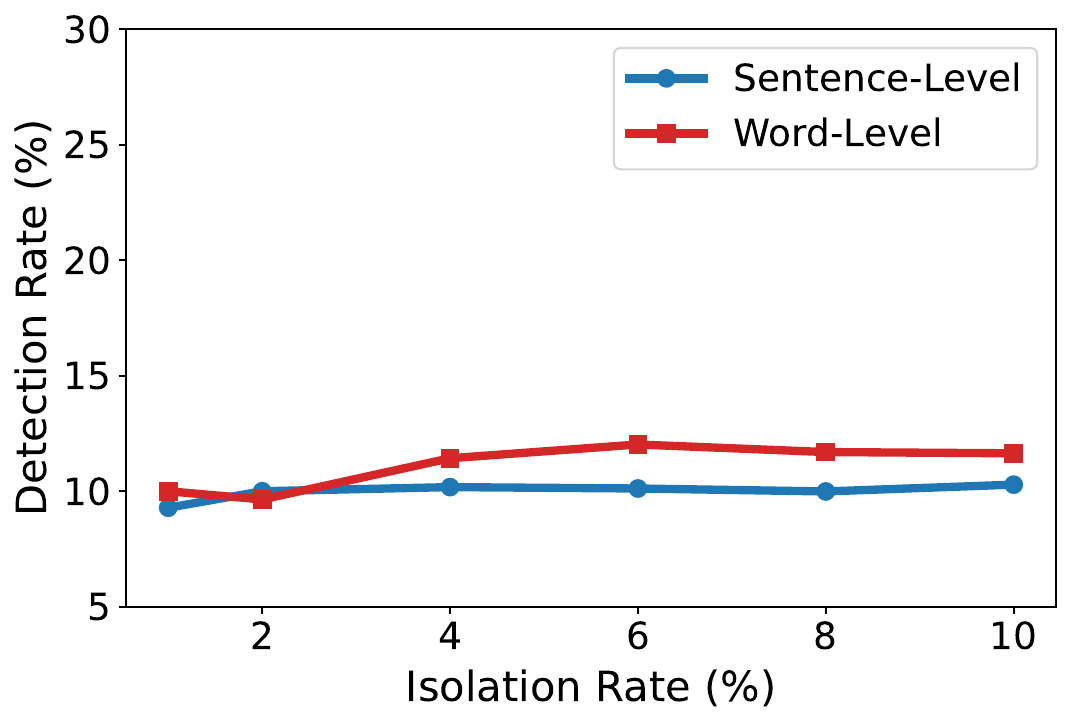}
}
\subfigure[COLLAB]{
\includegraphics[width=0.31\textwidth]{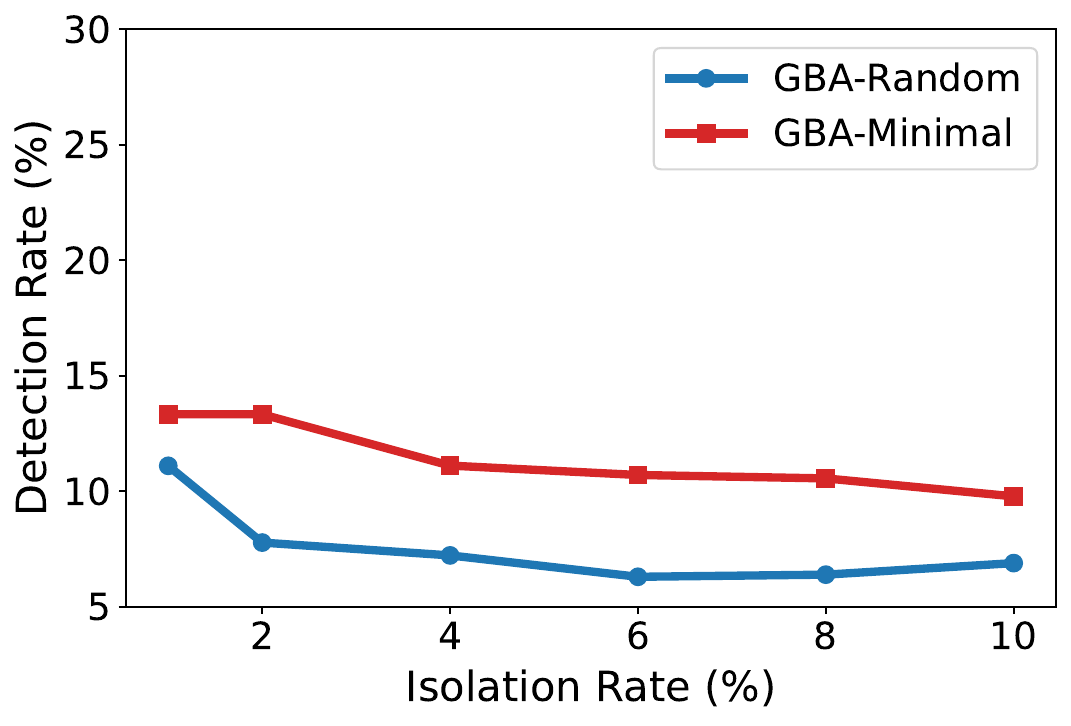}
}
%\hspace{0.8em}
\subfigure[REDDIT-MULTI-5K]{
\includegraphics[width=0.31\textwidth]{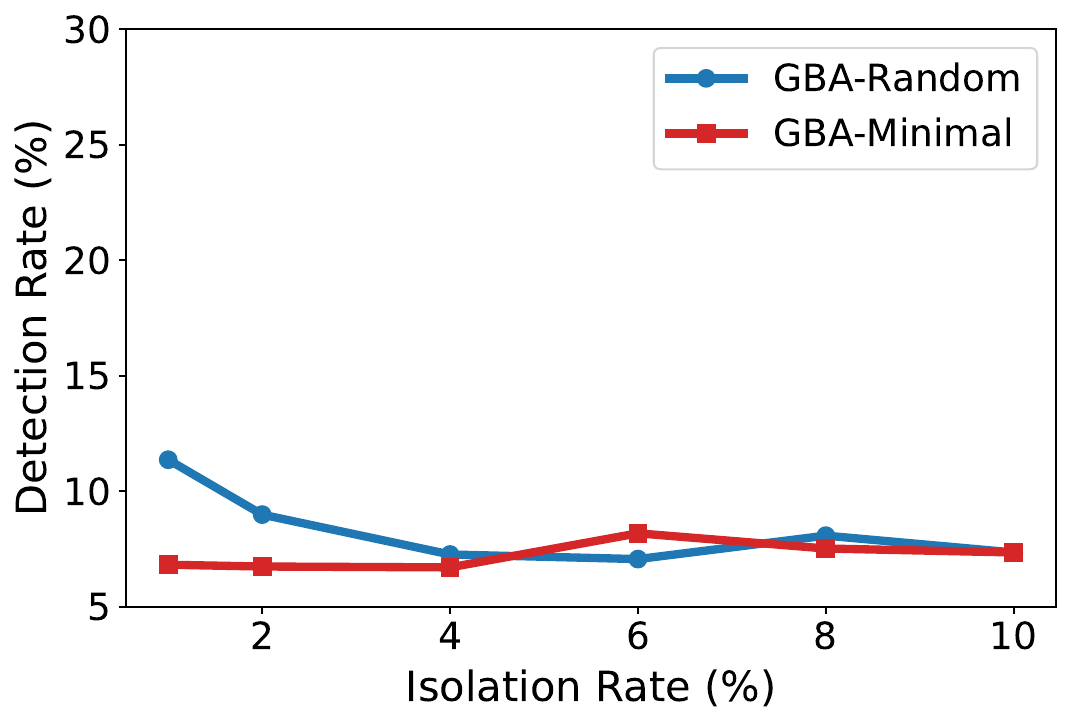}
}
\caption{The resistance of our DVBW to anti-backdoor learning on six different datasets.} 
\label{fig:ABL}
\end{figure*}

\vspace{0.3em}
\noindent \textbf{The Resistance to Model Pruning.} Following the classical settings, we adopt 10\% benign samples from the original training set to prune the latent representation (\ie, inputs of the fully-connected layers) of the watermarked model. In each case, the pruning rate is set to $\{0\%, 2\%, \cdots 98\%\}$. As shown in Figure \ref{fig:pruning}, pruning may significantly decrease the watermark success rate (WSR), especially when the pruning rate is nearly 100\%. However, its effects are with the huge sacrifice of benign accuracy (BA). These decreases in BA are unacceptable in practice since they will hinder standard model functionality. Accordingly, our DVBW is also resistant to model pruning to some extent. An interesting phenomenon is that the WSR even increases near the end of the pruning process in some cases. We speculate that it is probably because backdoor-related neurons and benign ones are competitive and the effects of benign neurons are already eliminated near the end. We will further discuss its mechanism in our future work.

\vspace{0.3em}
\noindent \textbf{The Resistance to Anti-backdoor Learning.} In general, anti-backdoor learning (ABL) intends to detect and unlearn poisoned samples during the training process of DNNs. Accordingly, whether ABL can successfully find watermarked samples is critical for its effectiveness. In these experiments, we provide the results of detection rates and isolation rates on different datasets. Specifically, the detection rate is defined as the proportion of poisoned samples that were isolated from all training samples, while the isolation rate denotes the ratio of isolated samples over all training samples. As shown in Figure \ref{fig:ABL}, ABL can successfully detect watermarked samples on both CIFAR-10 and ImageNet datasets. However, it fails in detecting watermarked samples on other datasets with different modalities (\ie, texts and graphs). We will further explore how to design more stealthy dataset watermark that can circumvent the detection of ABL across all modalities in our future work.

\begin{table}[!t]
\centering
\caption{The computational complexity of dataset watermarking and dataset verification in our DVBW. Specifically, $\gamma$ is the poisoning rate, $N$ is the number of training samples, and $m$ is the sampling number. }
\vspace{-0.5em}
\begin{tabular}{c|cc}
\toprule
\multirow{2}{*}{Dataset Watermarking} & \multicolumn{2}{c}{Dataset Verification} \\ \cline{2-3} 
                                      & Single Mode         & Batch Mode         \\ \hline
                                   $\mathcal{O}(\gamma \cdot N)$   &  $\mathcal{O}(m)$                   &  $\mathcal{O}(1)$                  \\ \bottomrule
\end{tabular}
\label{tab:com2}
\vspace{-1em}
\end{table}

\subsection{The Analysis of Computational Complexity}
In this section, we analyze the computational complexity of our DVBW. Specifically, we discuss the computational complexity of dataset watermarking and dataset verification of our DVBW (as summarized in Table \ref{tab:com2}). 

\subsubsection{The Complexity of Dataset Watermarking}
Let $N$ denotes the number of all training samples and $\gamma$ is the poisoning rate. Since our DVBW only needs to watermark a few selected samples in this step, its computational complexity is $\mathcal{O}(\gamma \cdot N)$. In general, these watermarks are about replacing or inserting a small part of the sample, which is highly efficient. Accordingly, our dataset watermarking is also efficient. Note that this step does not affect the adversaries. As such, it is acceptable even if this step is relatively time-consuming.

\begin{table*}[!t]
\centering
\caption{The defender's capacities of model ownership verification and dataset ownership verification. }
\vspace{-0.5em}
\begin{threeparttable}
\begin{tabular}{c|cc|cc}
\toprule
Task$\downarrow$, Capacity$\rightarrow$         & Training Samples & Training Schedule & Intermediate Results of Victim Model & Predictions of Victim Model \\ \hline
Model Ownership Verification   &  \fullcirc                &   \fullcirc                & \halfcirc                                     &  \fullcirc                           \\
Dataset Ownership Verification &  \fullcirc                & \emptycirc                  &   \emptycirc         &   \fullcirc                          \\ \bottomrule
\end{tabular}
\begin{tablenotes}
\footnotesize
\item[1] \fullcirc: accessible.
\item[2] \halfcirc: partly accessible (It is accessible for defenders under the \emph{white-box} setting, while it is inaccessible under the \emph{black-box} setting).
\item[3] \emptycirc: inaccessible.
\end{tablenotes}
\end{threeparttable}
\label{tab:MOVandDOV}
\end{table*}

\begin{table*}[!t]
\centering
\caption{The comparisons between our DVBW and four representative methods in model ownership verification. In each scenario, we mark a case as the checkmark if this method can be applied.}
\vspace{-0.5em}
\begin{threeparttable}
\begin{tabular}{c|cc|cc|cc}
\toprule
Method$\downarrow$, Scenario$\rightarrow$ & Embedding-free         & Multimodality             & White-box                 & Black-box                 & Probability-available     & Label-only                \\ \hline
MOVE \cite{li2022defending}             &                           &                           & \checkmark &                           & \checkmark &                           \\
DIMW \cite{zhang2022deep}            &                           &                           & \checkmark & \checkmark & \checkmark &                           \\ \hline
CEM \cite{lukas2021deep}             & \checkmark &                           & \checkmark & \checkmark & \checkmark &                           \\
NRF \cite{zheng2022dnn}             & \checkmark &                           & \checkmark &                           & \checkmark & \checkmark \\ \hline
DVBW (Ours)      & \checkmark & \checkmark & \checkmark & \checkmark & \checkmark & \checkmark \\ \bottomrule
\end{tabular}
\begin{tablenotes}
\footnotesize
\item[1]Embedding-free: defenders do not need to implant any additional parts or functionalities ($e.g.$, backdoor) in the victim model.
\item[2]Multimodality: defenders can use the method across different types of data ($e.g.$, images, texts, and graphs). 
\item[3]White-box: defenders can access the source files of suspicious models.
\item[4]Black-box: defenders can only query suspicious models.
\end{tablenotes}
\end{threeparttable}
\label{tab:method_compare}
\end{table*}

\subsubsection{The Complexity of Dataset Verification}
In this step, defenders need to query the (deployed) suspicious model with $m$ samples and conduct the hypothesis test based on their predictions. In general, there are two classical prediction modes, including \textbf{(1)} \emph{single mode} and \textbf{(2)} \emph{batch mode}. Specifically, under the single mode, the suspicious model can only predict one sample at a time while it can predict a batch of samples simultaneously under the batch mode. Accordingly, the computational complexity of single mode and batch mode is $\mathcal{O}(m)$ and $\mathcal{O}(1)$, respectively. Note that this step is also efficient, no matter under the single or the batch mode, since predicting one sample is usually costless.

\section{Relation with Model Ownership Verification}
\label{sec:relation}
We notice and admit that the dataset ownership verification defined in this paper is closely related to the model ownership verification (MOV) \cite{lukas2021deep,guo2021masterface,xu2021watermarking,li2022defending,zhang2022deep,zheng2022dnn}. In general, model ownership verification intends to identify whether a suspicious third-party model (instead of the dataset) is stolen from the victim for unauthorized adoption. In this section, we discuss their similarities and differences. We summarize the characteristics of MOV and the task of our dataset ownership verification in Table \ref{tab:MOVandDOV}. The comparisons between our DVBW and representative MOV methods are in Table \ref{tab:method_compare}.

Firstly, our DVBW enjoys some similarities to MOV in the watermarking processes. Specifically, backdoor attacks are also widely used to watermark the victim model in MOV. However, defenders in MOV usually need to manipulate the training process ($e.g.$, 
adding some additional regularization terms \cite{jia2021entangled} or supportive modules \cite{zhang2022deep}), since they can fully control the training process of the victim model. In contrast, in our dataset ownership verification, the defender can only modify the dataset while having no information or access to the model training process and therefore we can only use poison-only backdoor attacks for dataset watermarking. In other words, defenders in DVBW have significantly fewer capacities, compared with those in MOV. It allows our method to be adopted for model copyright protection, whereas their approaches may not be directly used in our task. 

Besides, both our defense and most of the existing MOV methods exploit hypothesis-test in the verification processes. However, in our DVBW, we consider the black-box verification scenarios, where defenders can only query the suspicious models to obtain their predictions. However, in MOV, many methods ($e.g.$, \cite{li2022defending}) considered the white-box verification scenarios where defenders can obtain the source files of suspicious models. Even under the black-box settings, existing MOV methods only consider probability-available cases while our DVBW also discusses label-only ones.

\section{Conclusion}
\label{sec:conclusion}
In this paper, we explored how to protect valuable released datasets. Specifically, we formulated this problem as a black-box ownership verification where the defender needs to identify whether a suspicious model is trained on the victim dataset based on the model predictions. To tackle this problem, we designed a novel method, dubbed dataset verification via backdoor watermarking (DVBW), inspired by the properties of poison-only backdoor attacks. DVBW contained two main steps, including dataset watermarking and dataset verification. Specifically, we exploited poison-only backdoor attacks for dataset watermarking and designed a hypothesis-test-guided method for dataset verification. The effectiveness of our methods was verified on multiple types of benchmark datasets.

\section*{Acknowledgments}
This work was mostly done when Yiming Li was a research intern at Ant Group. This work is supported in part by the National Key R\&D Program of China under Grant 2022YFB3105000, the National Natural Science Foundation of China under Grants (62171248, 62202393, 12141108), the Shenzhen Science and Technology Program (JCYJ20220818101012025), the Sichuan Science and Technology Program under Grant 2023NSFSC1394, the PCNL Key Project (PCL2021A07), and the Shenzhen Science and Technology Innovation Commission (Research Center for Computer Network (Shenzhen) Ministry of Education). We also sincerely thank Ziqi Zhang from Tsinghua University for her assistance in some preliminary experiments and Dr. Baoyuan Wu from 
CUHK-Shenzhen for his helpful comments on an early draft of this paper.

\appendix
\setcounter{theorem}{0}
\setcounter{equation}{0}

\begin{theorem}\label{thm1_a}
Let $f(\bm{x})$ is the posterior probability of $\bm{x}$ predicted by the suspicious model, variable $\bm{X}$ denotes the benign sample with non-target label, and variable $\bm{X}'$ is the watermarked version of $\bm{X}$. Assume that $\bm{P}_b\triangleq f(\bm{X})_{y_{t}} < \beta$. We claim that dataset owners can reject the null hypothesis $H_0$ of probability-available verification at the significance level $\alpha$, if the watermark success rate $W$ of $f$ satisfies that
\begin{equation}
    \sqrt{m-1} \cdot (W-\beta - \tau) - t_{1-\alpha} \cdot \sqrt{W-W^2} > 0,
\end{equation}
where $t_{1-\alpha}$ is the $(1-\alpha)$-quantile of t-distribution with $(m-1)$ degrees of freedom and $m$ is the sample size of $\bm{X}$.
\end{theorem}

\begin{proof}
Since $\bm{P}_b\triangleq f(\bm{X})_{y_{t}} < \beta$, the original hypothesis $H_0$ and $H_1$ can be converted to
\begin{align}
  H_0': \bm{P}_w < \beta + \tau, \\
  H_1':  \bm{P}_w > \beta + \tau.
\end{align}

Let $E$ indicates the event of whether the suspect model $f$ predicts a poisoned sample as the target label $y_t$. As such, 
\begin{equation}
    E \sim B(1, p),
\end{equation}
where $p = \Pr(C(\bm{X}')=y_t)$ indicates backdoor success probability and $B$ is the Binomial distribution \cite{hogg2005introduction}. 

Let $\bm{x}_1', \cdots, \bm{x}_m'$ denotes $m$ poisoned samples used for dataset verification and $E_1, \cdots, E_m$ denote their prediction events, we know that the attack success rate $A$ satisfies
\begin{align}
    W = \frac{1}{m} \sum_{i=1}^{m} E_i, \\
    W \sim \frac{1}{m} B(m, p).
\end{align}

According to the central limit theorem \cite{hogg2005introduction}, the watermark success rate $W$ follows Gaussian distribution $\mathcal{N}(p, \frac{p(1-p)}{m})$ when $m$ is sufficiently large. Similarly, $(\bm{P}_w - \beta - \tau)$ also satisfies Gaussian distribution. As such, we can construct the t-statistic as follows
\begin{equation}
    T \triangleq \frac{\sqrt{m}(W-\beta-\tau)}{s} \sim t(m-1),
\end{equation}
where $s$ is the standard deviation of $(W-\beta-\tau)$ and $W$, \ie, 
\begin{equation}\label{eq:std}
    s^2 = \frac{1}{m-1} \sum_{i=1}^m (E_i-W)^2 = \frac{1}{m-1}(m \cdot W - m \cdot W^2). 
\end{equation}

To reject the hypothesis $H_0'$ at the significance level $\alpha$, we need to ensure that 
\begin{equation}\label{eq:rej_con}
  \frac{\sqrt{m}(W-\beta-\tau)}{s} > t_{1-\alpha},  
\end{equation}
where $t_{1-\alpha}$ is the $(1-\alpha)$-quantile of t-distribution with $(m-1)$ degrees of freedom.

According to equation (\ref{eq:std})-(\ref{eq:rej_con}), we have
\begin{equation}
    \sqrt{m-1} \cdot (W- \beta - \tau) - t_{1-\alpha} \cdot \sqrt{W-W^2} > 0.
\end{equation}

\end{proof}

\bibliographystyle{IEEEtran}

\bibliography{reference}

\begin{thebibliography}{10}
\providecommand{\url}[1]{#1}
\csname url@samestyle\endcsname
\providecommand{\newblock}{\relax}
\providecommand{\bibinfo}[2]{#2}
\providecommand{\BIBentrySTDinterwordspacing}{\spaceskip=0pt\relax}
\providecommand{\BIBentryALTinterwordstretchfactor}{4}
\providecommand{\BIBentryALTinterwordspacing}{\spaceskip=\fontdimen2\font plus
\BIBentryALTinterwordstretchfactor\fontdimen3\font minus
  \fontdimen4\font\relax}
\providecommand{\BIBforeignlanguage}[2]{{%
\expandafter\ifx\csname l@#1\endcsname\relax
\typeout{** WARNING: IEEEtran.bst: No hyphenation pattern has been}%
\typeout{** loaded for the language `#1'. Using the pattern for}%
\typeout{** the default language instead.}%
\else
\language=\csname l@#1\endcsname
\fi
#2}}
\providecommand{\BIBdecl}{\relax}
\BIBdecl

\bibitem{wu2018light}
X.~Wu, R.~He, Z.~Sun, and T.~Tan, ``A light cnn for deep face representation
  with noisy labels,'' \emph{IEEE Transactions on Information Forensics and
  Security}, vol.~13, no.~11, pp. 2884--2896, 2018.

\bibitem{yin2020joint}
Q.~Yin, J.~Feng, J.~Lu, and J.~Zhou, ``Joint estimation of pose and singular
  points of fingerprints,'' \emph{IEEE Transactions on Information Forensics
  and Security}, vol.~16, pp. 1467--1479, 2020.

\bibitem{deng2009imagenet}
J.~Deng, W.~Dong, R.~Socher, L.-J. Li, K.~Li, and L.~Fei-Fei, ``Imagenet: A
  large-scale hierarchical image database,'' in \emph{CVPR}, 2009.

\bibitem{liu2015faceattributes}
Z.~Liu, P.~Luo, X.~Wang, and X.~Tang, ``Deep learning face attributes in the
  wild,'' in \emph{ICCV}, 2015.

\bibitem{ni2019justifying}
J.~Ni, J.~Li, and J.~McAuley, ``Justifying recommendations using
  distantly-labeled reviews and fine-grained aspects,'' in \emph{EMNLP}, 2019.

\bibitem{voigt2017eu}
P.~Voigt and A.~Von~dem Bussche, ``The eu general data protection regulation
  (gdpr),'' \emph{A Practical Guide, 1st Ed., Cham: Springer International
  Publishing}, vol.~10, no. 3152676, pp. 10--5555, 2017.

\bibitem{wang2016efficient}
S.~Wang, J.~Zhou, J.~K. Liu, J.~Yu, J.~Chen, and W.~Xie, ``An efficient file
  hierarchy attribute-based encryption scheme in cloud computing,'' \emph{IEEE
  Transactions on Information Forensics and Security}, vol.~11, no.~6, pp.
  1265--1277, 2016.

\bibitem{li2019hierarchical}
J.~Li, Q.~Yu, and Y.~Zhang, ``Hierarchical attribute based encryption with
  continuous leakage-resilience,'' \emph{Information Sciences}, vol. 484, pp.
  113--134, 2019.

\bibitem{deng2020identity}
H.~Deng, Z.~Qin, Q.~Wu, Z.~Guan, R.~H. Deng, Y.~Wang, and Y.~Zhou,
  ``Identity-based encryption transformation for flexible sharing of encrypted
  data in public cloud,'' \emph{IEEE Transactions on Information Forensics and
  Security}, vol.~15, pp. 3168--3180, 2020.

\bibitem{haddad2020joint}
S.~Haddad, G.~Coatrieux, A.~Moreau-Gaudry, and M.~Cozic, ``Joint
  watermarking-encryption-jpeg-ls for medical image reliability control in
  encrypted and compressed domains,'' \emph{IEEE Transactions on Information
  Forensics and Security}, vol.~15, pp. 2556--2569, 2020.

\bibitem{wang2021faketagger}
R.~Wang, F.~Juefei-Xu, M.~Luo, Y.~Liu, and L.~Wang, ``Faketagger: Robust
  safeguards against deepfake dissemination via provenance tracking,'' in
  \emph{ACM MM}, 2021.

\bibitem{guan2022deepmih}
Z.~Guan, J.~Jing, X.~Deng, M.~Xu, L.~Jiang, Z.~Zhang, and Y.~Li, ``Deepmih:
  Deep invertible network for multiple image hiding,'' \emph{IEEE Transactions
  on Pattern Analysis and Machine Intelligence}, 2022.

\bibitem{wei2020federated}
K.~Wei, J.~Li, M.~Ding, C.~Ma, H.~H. Yang, F.~Farokhi, S.~Jin, T.~Q. Quek, and
  H.~V. Poor, ``Federated learning with differential privacy: Algorithms and
  performance analysis,'' \emph{IEEE Transactions on Information Forensics and
  Security}, vol.~15, pp. 3454--3469, 2020.

\bibitem{zhu2021fine}
L.~Zhu, X.~Liu, Y.~Li, X.~Yang, S.-T. Xia, and R.~Lu, ``A fine-grained
  differentially private federated learning against leakage from gradients,''
  \emph{IEEE Internet of Things Journal}, 2021.

\bibitem{bai2022multinomial}
J.~Bai, Y.~Li, J.~Li, X.~Yang, Y.~Jiang, and S.-T. Xia, ``Multinomial random
  forest,'' \emph{Pattern Recognition}, vol. 122, p. 108331, 2022.

\bibitem{gu2019badnets}
T.~Gu, K.~Liu, B.~Dolan-Gavitt, and S.~Garg, ``Badnets: Evaluating backdooring
  attacks on deep neural networks,'' \emph{IEEE Access}, vol.~7, pp.
  47\,230--47\,244, 2019.

\bibitem{li2021invisible}
Y.~Li, Y.~Li, B.~Wu, L.~Li, R.~He, and S.~Lyu, ``Invisible backdoor attack with
  sample-specific triggers,'' in \emph{ICCV}, 2021.

\bibitem{nguyen2021wanet}
A.~Nguyen and A.~Tran, ``Wanet--imperceptible warping-based backdoor attack,''
  in \emph{ICLR}, 2021.

\bibitem{xiong2020adgan}
Z.~Xiong, Z.~Cai, Q.~Han, A.~Alrawais, and W.~Li, ``Adgan: protect your
  location privacy in camera data of auto-driving vehicles,'' \emph{IEEE
  Transactions on Industrial Informatics}, vol.~17, no.~9, pp. 6200--6210,
  2020.

\bibitem{li2021visual}
Y.~Li, P.~Liu, Y.~Jiang, and S.-T. Xia, ``Visual privacy protection via mapping
  distortion,'' in \emph{ICASSP}, 2021.

\bibitem{xu2021audio}
H.~Xu, Z.~Cai, D.~Takabi, and W.~Li, ``Audio-visual autoencoding for
  privacy-preserving video streaming,'' \emph{IEEE Internet of Things Journal},
  2021.

\bibitem{dwork2008differential}
C.~Dwork, ``Differential privacy: A survey of results,'' in \emph{TAMC}, 2008.

\bibitem{li2022backdoor}
Y.~Li, Y.~Jiang, Z.~Li, and S.-T. Xia, ``Backdoor learning: A survey,''
  \emph{IEEE Transactions on Neural Networks and Learning Systems}, 2022.

\bibitem{qi2023revisiting}
X.~Qi, T.~Xie, Y.~Li, S.~Mahloujifar, and P.~Mittal, ``Revisiting the
  assumption of latent separability for backdoor defenses,'' in \emph{ICLR},
  2023.

\bibitem{gao2023not}
Y.~Gao, Y.~Li, L.~Zhu, D.~Wu, Y.~Jiang, and S.-T. Xia, ``Not all samples are
  born equal: Towards effective clean-label backdoor attacks,'' \emph{Pattern
  Recognition}, p. 109512, 2023.

\bibitem{li2020invisible}
S.~Li, M.~Xue, B.~Zhao, H.~Zhu, and X.~Zhang, ``Invisible backdoor attacks on
  deep neural networks via steganography and regularization,'' \emph{IEEE
  Transactions on Dependable and Secure Computing}, 2020.

\bibitem{li2022few}
Y.~Li, H.~Zhong, X.~Ma, Y.~Jiang, and S.-T. Xia, ``Few-shot backdoor attacks on
  visual object tracking,'' in \emph{ICLR}, 2022.

\bibitem{shumailov2021manipulating}
I.~Shumailov, Z.~Shumaylov, D.~Kazhdan, Y.~Zhao, N.~Papernot, M.~A. Erdogdu,
  and R.~Anderson, ``Manipulating sgd with data ordering attacks,'' in
  \emph{NeurIPS}, 2021.

\bibitem{rakin2020tbt}
A.~S. Rakin, Z.~He, and D.~Fan, ``Tbt: Targeted neural network attack with bit
  trojan,'' in \emph{CVPR}, 2020.

\bibitem{tang2020embarrassingly}
R.~Tang, M.~Du, N.~Liu, F.~Yang, and X.~Hu, ``An embarrassingly simple approach
  for trojan attack in deep neural networks,'' in \emph{SIGKDD}, 2020.

\bibitem{bai2022hardly}
J.~Bai, K.~Gao, D.~Gong, S.-T. Xia, Z.~Li, and W.~Liu, ``Hardly perceptible
  trojan attack against neural networks with bit flips,'' in \emph{ECCV}, 2022.

\bibitem{chen2017targeted}
X.~Chen, C.~Liu, B.~Li, K.~Lu, and D.~Song, ``Targeted backdoor attacks on deep
  learning systems using data poisoning,'' \emph{arXiv preprint
  arXiv:1712.05526}, 2017.

\bibitem{li2021backdoor2}
Y.~Li, T.~Zhai, Y.~Jiang, Z.~Li, and S.-T. Xia, ``Backdoor attack in the
  physical world,'' in \emph{ICLR Workshop}, 2021.

\bibitem{zhang2022inject}
Z.~Zhang, L.~Lyu, W.~Wang, L.~Sun, and X.~Sun, ``How to inject backdoors with
  better consistency: Logit anchoring on clean data,'' in \emph{ICLR}, 2022.

\bibitem{chen2021badnl}
X.~Chen, A.~Salem, D.~Chen, M.~Backes, S.~Ma, Q.~Shen, Z.~Wu, and Y.~Zhang,
  ``Badnl: Backdoor attacks against nlp models with semantic-preserving
  improvements,'' in \emph{ACSAC}, 2021.

\bibitem{wang2021stop}
Y.~Wang, E.~Sarkar, W.~Li, M.~Maniatakos, and S.~E. Jabari, ``Stop-and-go:
  Exploring backdoor attacks on deep reinforcement learning-based traffic
  congestion control systems,'' \emph{IEEE Transactions on Information
  Forensics and Security}, vol.~16, pp. 4772--4787, 2021.

\bibitem{zhai2021backdoor}
T.~Zhai, Y.~Li, Z.~Zhang, B.~Wu, Y.~Jiang, and S.-T. Xia, ``Backdoor attack
  against speaker verification,'' in \emph{ICASSP}, 2021.

\bibitem{li2021survey}
Z.~Li, F.~Liu, W.~Yang, S.~Peng, and J.~Zhou, ``A survey of convolutional
  neural networks: analysis, applications, and prospects,'' \emph{IEEE
  Transactions on Neural Networks and Learning Systems}, 2021.

\bibitem{han2022transformer}
K.~Han, Y.~Wang, H.~Chen, X.~Chen, J.~Guo, Z.~Liu, Y.~Tang, A.~Xiao, C.~Xu,
  Y.~Xu, Z.~Yang, Y.~Zhang, and D.~Tao, ``A survey on vision transformer,''
  \emph{IEEE Transactions on Pattern Analysis and Machine Intelligence}, 2022.

\bibitem{wu2020comprehensive}
Z.~Wu, S.~Pan, F.~Chen, G.~Long, C.~Zhang, and S.~Y. Philip, ``A comprehensive
  survey on graph neural networks,'' \emph{IEEE transactions on neural networks
  and learning systems}, vol.~32, no.~1, pp. 4--24, 2020.

\bibitem{hogg2005introduction}
R.~V. Hogg, J.~McKean, and A.~T. Craig, \emph{Introduction to mathematical
  statistics}.\hskip 1em plus 0.5em minus 0.4em\relax Pearson Education, 2005.

\bibitem{krizhevsky2009learning}
A.~Krizhevsky, G.~Hinton \emph{et~al.}, ``Learning multiple layers of features
  from tiny images,'' Citeseer, Tech. Rep., 2009.

\bibitem{simonyan2014very}
K.~Simonyan and A.~Zisserman, ``Very deep convolutional networks for
  large-scale image recognition,'' in \emph{ICLR}, 2015.

\bibitem{he2016deep}
K.~He, X.~Zhang, S.~Ren, and J.~Sun, ``Deep residual learning for image
  recognition,'' in \emph{CVPR}, 2016.

\bibitem{maas2011learning}
A.~Maas, R.~E. Daly, P.~T. Pham, D.~Huang, A.~Y. Ng, and C.~Potts, ``Learning
  word vectors for sentiment analysis,'' in \emph{ACL}, 2011.

\bibitem{auer2007dbpedia}
S.~Auer, C.~Bizer, G.~Kobilarov, J.~Lehmann, R.~Cyganiak, and Z.~Ives,
  ``{DBpedia}: A nucleus for a web of open data,'' in \emph{ISWC}, 2007.

\bibitem{hochreiter1997long}
S.~Hochreiter and J.~Schmidhuber, ``Long short-term memory,'' \emph{Neural
  computation}, vol.~9, no.~8, pp. 1735--1780, 1997.

\bibitem{chen2014convolutional}
Y.~Chen, ``Convolutional neural network for sentence classification,'' in
  \emph{EMNLP}, 2014.

\bibitem{dai2019backdoor}
J.~Dai, C.~Chen, and Y.~Li, ``A backdoor attack against lstm-based text
  classification systems,'' \emph{IEEE Access}, vol.~7, pp. 138\,872--138\,878,
  2019.

\bibitem{yanardag2015deep}
P.~Yanardag and S.~Vishwanathan, ``Deep graph kernels,'' in \emph{KDD}, 2015.

\bibitem{xu2019powerful}
K.~Xu, W.~Hu, J.~Leskovec, and S.~Jegelka, ``How powerful are graph neural
  networks?'' in \emph{ICLR}, 2018.

\bibitem{hamilton2017inductive}
W.~Hamilton, Z.~Ying, and J.~Leskovec, ``Inductive representation learning on
  large graphs,'' \emph{NeurIPS}, 2017.

\bibitem{xi2021graph}
Z.~Xi, R.~Pang, S.~Ji, and T.~Wang, ``Graph backdoor,'' in \emph{USENIX
  Security}, 2021.

\bibitem{zhang2021backdoor}
Z.~Zhang, J.~Jia, B.~Wang, and N.~Z. Gong, ``Backdoor attacks to graph neural
  networks,'' in \emph{SACMAT}, 2021.

\bibitem{liu2017neural}
Y.~Liu, Y.~Xie, and A.~Srivastava, ``Neural trojans,'' in \emph{ICCD}, 2017.

\bibitem{liu2018fine}
K.~Liu, B.~Dolan-Gavitt, and S.~Garg, ``Fine-pruning: Defending against
  backdooring attacks on deep neural networks,'' in \emph{RAID}, 2018.

\bibitem{li2021anti}
Y.~Li, X.~Lyu, N.~Koren, L.~Lyu, B.~Li, and X.~Ma, ``Anti-backdoor learning:
  Training clean models on poisoned data,'' in \emph{NeurIPS}, 2021.

\bibitem{li2023backdoorbox}
Y.~Li, M.~Ya, Y.~Bai, Y.~Jiang, and S.-T. Xia, ``Backdoorbox: A python toolbox
  for backdoor learning,'' in \emph{ICLR Workshop}, 2023.

\bibitem{li2022defending}
Y.~Li, L.~Zhu, X.~Jia, Y.~Jiang, S.-T. Xia, and X.~Cao, ``Defending against
  model stealing via verifying embedded external features,'' in \emph{AAAI},
  2022.

\bibitem{zhang2022deep}
J.~Zhang, D.~Chen, J.~Liao, W.~Zhang, H.~Feng, G.~Hua, and N.~Yu, ``Deep model
  intellectual property protection via deep watermarking,'' \emph{IEEE
  Transactions on Pattern Analysis and Machine Intelligence}, vol.~44, no.~8,
  pp. 4005--4020, 2022.

\bibitem{lukas2021deep}
N.~Lukas, Y.~Zhang, and F.~Kerschbaum, ``Deep neural network fingerprinting by
  conferrable adversarial examples,'' in \emph{ICLR}, 2021.

\bibitem{zheng2022dnn}
Y.~Zheng, S.~Wang, and C.-H. Chang, ``A dnn fingerprint for non-repudiable
  model ownership identification and piracy detection,'' \emph{IEEE
  Transactions on Information Forensics and Security}, vol.~17, pp. 2977--2989,
  2022.

\bibitem{guo2021masterface}
W.~Guo, B.~Tondi, and M.~Barni, ``Masterface watermarking for ipr protection of
  siamese network for face verification,'' in \emph{IWDW}, 2021.

\bibitem{xu2021watermarking}
J.~Xu and S.~Picek, ``Watermarking graph neural networks based on backdoor
  attacks,'' \emph{arXiv preprint arXiv:2110.11024}, 2021.

\bibitem{jia2021entangled}
H.~Jia, C.~A. Choquette-Choo, V.~Chandrasekaran, and N.~Papernot, ``Entangled
  watermarks as a defense against model extraction,'' in \emph{USENIX
  Security}, 2021.

\end{thebibliography}

\begin{IEEEbiography}[{\includegraphics[width=0.9in,height=1.25in,clip,keepaspectratio]{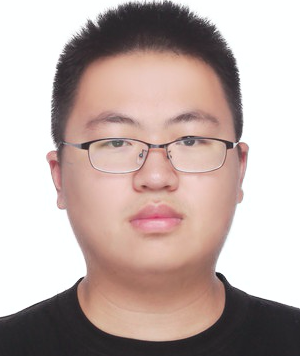}}]{Yiming Li} is currently a Ph.D. candidate from Tsinghua-Berkeley Shenzhen Institute, Tsinghua Shenzhen International Graduate School, Tsinghua University. Before that, he received his B.S. degree in Mathematics and Applied Mathematics from Ningbo University in 2018. His research interests are in the domain of AI security, especially backdoor learning, adversarial learning, data privacy, and copyright protection in AI. His research has been published in multiple top-tier conferences and journals, such as ICLR, NeurIPS, ICCV, IEEE TNNLS, and PR Journal. He served as the senior program committee member of AAAI 2022, the program committee member of ICML, NeurIPS, ICLR, etc., and the reviewer of IEEE TPAMI, IEEE TIFS, IEEE TDSC, etc.
\end{IEEEbiography}

\begin{IEEEbiography}[{\includegraphics[width=0.9in,height=1.25in,clip,keepaspectratio]{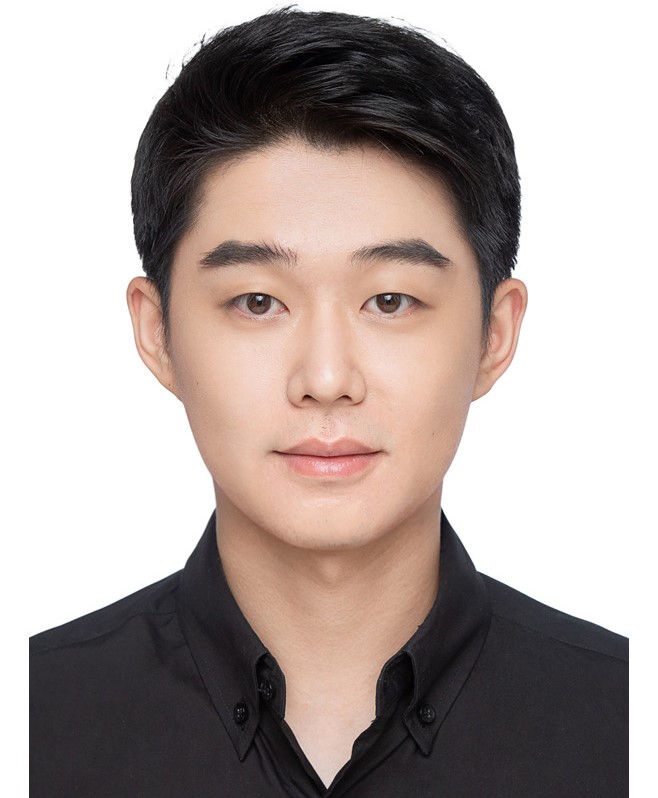}}]{Mingyan Zhu} received his B.S. degree in Computer Science and Technology from Harbin Institute of Technology, China, in 2020.  He is currently pursuing the Ph.D.degree in Tsinghua Shenzhen International Graduate School, Tsinghua University. His research interests are in the domain of Low-level Computer Vision and AI security.
\end{IEEEbiography}

\begin{IEEEbiography}[{\includegraphics[width=0.9in,height=1.25in,clip,keepaspectratio]{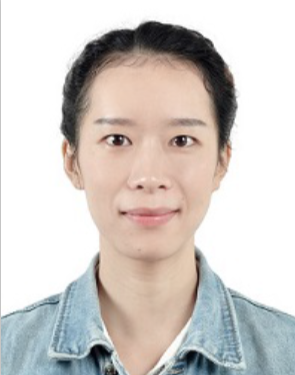}}]{Dr. Xue Yang} received a Ph.D. degree in information and communication engineering from Southwest Jiaotong University, China, in 2019. She was a visiting student at the Faculty of Computer Science, University of New Brunswick, Canada, from 2017 to 2018. She was a postdoctoral fellow with Tsinghua University. She is currently a research associate with the School of Information Science and Technology, Southwest Jiaotong University, China. Her research interests include data security and privacy, applied cryptography, and federated learning.
\end{IEEEbiography}

\begin{IEEEbiography}[{\includegraphics[width=0.9in,height=1.25in,clip,keepaspectratio]{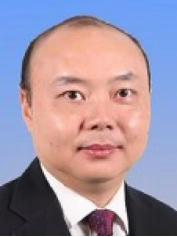}}]{Dr. Yong Jiang} received his M.S. and Ph.D. degrees in computer science from Tsinghua University, China, in 1998 and 2002, respectively. Since 2002, he has been with the Tsinghua Shenzhen International Graduate School of Tsinghua University, Guangdong, China, where he is currently a full professor. His research interests include computer vision, machine learning, Internet architecture and its protocols, IP routing technology, etc. He has received several best paper awards (e.g., IWQoS 2018) from top-tier conferences and his researches have been published in multiple top-tier journals and conferences, including IEEE ToC, IEEE TMM, IEEE TSP, CVPR, ICLR, etc.
\end{IEEEbiography}

\begin{IEEEbiography}[{\includegraphics[width=0.9in,height=1.25in,clip,keepaspectratio]{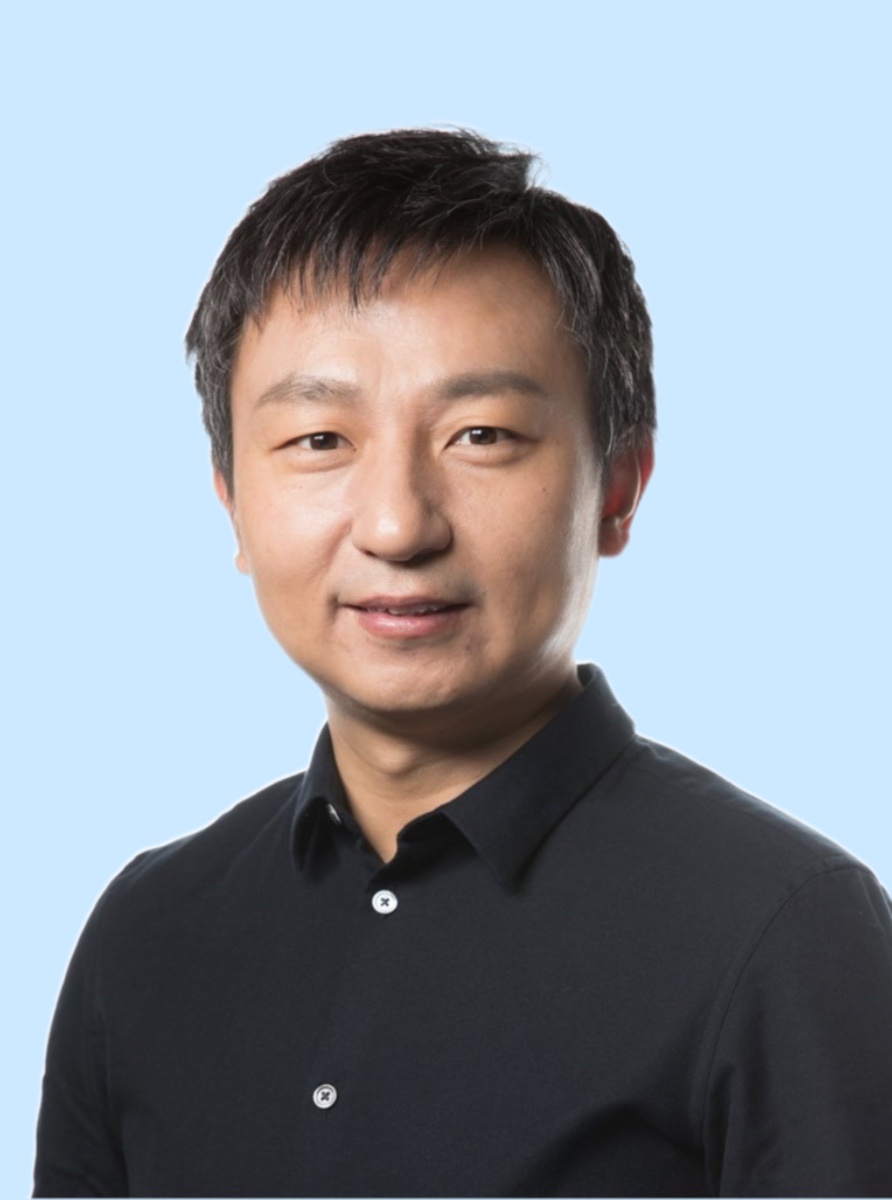}}]{Dr. Wei Tao} received the B.S. and Ph.D. degrees from Peking University, China, in 1997 and 2007, respectively. He is currently the Vice President at Ant Group, in charge of its foundational security. He is also an Adjunct Professor at Peking University. For more than 20 years, he has been committed to making complex systems more secure and reliable. His work has helped Windows, Android, iOS and other operating systems improve their security capabilities. He also led the development of many famous security open-sourced projects such as Mesatee/Teaclave, MesaLink TLS, OpenRASP, Advbox Adversarial Toolbox, etc. His researches have been published in multiple top-tier journals and conferences, including IEEE TDSC, IEEE TIFS, IEEE S\&P, USENIX Security, etc.
\end{IEEEbiography}

\begin{IEEEbiography}[{\includegraphics[width=0.9in,height=1.25in,clip,keepaspectratio]{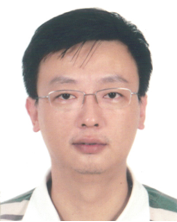}}]{Dr. Shu-Tao Xia} received the B.S. degree in mathematics and the Ph.D. degree in applied mathematics from Nankai University, Tianjin, China, in 1992 and 1997, respectively. Since January 2004, he has been with the Tsinghua Shenzhen International Graduate School of Tsinghua University, Guangdong, China, where he is currently a full professor. From September 1997 to March 1998 and from August to September 1998, he visited the Department of Information Engineering, The Chinese University of Hong Kong, Hong Kong. His research interests include coding and information theory, machine learning, and deep learning. His researches have been published in multiple top-tier journals and conferences, including IEEE TIP, IEEE TNNLS, CVPR, ICCV, ECCV, ICLR, etc.
\end{IEEEbiography}

\end{document}